\documentclass[]{Liebert_Author_revised}

\title{A spatial-correlated multitask linear mixed-effects model for imaging genetics} 

\author{Zhibin Pu $^{1}$ and Shufei Ge $^{1\ast}$\\
{$^{1}$Institute of Mathematical Sciences, ShanghaiTech University}\\
{393 Middle Huaxia Road, Shanghai, 201210, China}\\
{$^\ast$To whom correspondence should be addressed;}\\
{E-mail:  geshf@shanghaitech.edu.cn.}
}

\begin{document} 

\maketitle


\begin{abstract}
  Imaging genetics aims to uncover the hidden relationship between imaging quantitative traits (QTs) and genetic markers (\emph{e.g.} single nucleotide polymorphism (SNP)), and brings valuable insights into the pathogenesis of complex diseases, such as cancers and cognitive disorders (\emph{e.g.} the Alzheimer's Disease). However, most linear models in imaging genetics didn't explicitly model the inner relationship among QTs, which might miss some potential efficiency gains from information borrowing across brain regions. In this work, we developed a novel Bayesian regression framework for identifying significant associations between QTs and genetic markers while explicitly modelling spatial dependency between QTs,  with the main contributions as follows. Firstly, we developed a spatial-correlated multitask linear mixed-effects model (LMM) to account for dependencies between QTs. We incorporated a population-level mixed-effects term into the model, taking full advantage of the dependent structure of brain imaging-derived QTs. Secondly, we implemented the model in the Bayesian framework and derived a Markov chain Monte Carlo (MCMC) algorithm to achieve the model inference. Further, we incorporated the MCMC samples with the Cauchy combination test (CCT) to examine the association between SNPs and QTs, which avoided computationally intractable multi-test issues. The simulation studies indicated improved power of our proposed model compared to classic models where inner dependencies of QTs were not modelled. We also applied the new spatial model to an imaging dataset obtained from the  Alzheimer's Disease Neuroimaging Initiative (ADNI) database (\url{https://adni.loni.usc.edu}). The implementation of our method is available at \url{https://github.com/ZhibinPU/spatialmultitasklmm.git}.
\end{abstract}

\section{Introduction}

\label{sec:intro}
 Imaging genetics uncovers potential risk genes by analysing their effects on organisms (e.g. brain structure and lung cancer), and further explains for the pathogenic mechanism of target disorders \citep{gerber2009imaging}. The past decade has witnessed discovery of numerous gene markers in this research field. The single nucleotide polymorphism (SNP) $rs1344706$ was shown to impact gray matter volumes in several brain regions and further contribute to schizophrenia \citep{lencz2010schizophrenia}; $rs2365715$, $rs3762515$ and $rs67827860$ were found associated with white matter lesions causing main depressive disorders \citep{elliott2018genome}; 
$rs4348791$ ranked the top in the relationship analysis with phenotypes derived from the left caudate nucleus, which accounted for impaired neural development \citep{wang2022phenotypic}.  Genome-wise association studies (GWAS) has been widely used in imaging genetics by replacing the one-hot encoding (case-control status) with quantitative traits (QTs) derived from images. Unlike univariate phenotype in classical GWAS,  QTs derived from images usually are multi-dimensional and spatial correlated. Some studies applied voxel-wise approaches which treat imaging traits separately, such as vGWAS and its faster version - a univariate linear model for critical correlations between each SNP and voxel \citep{stein2010voxelwise, hibar2011voxelwise,  huang2015fvgwas}. However, QTs in these models were considered independent, potentially overlooking interactive connections among them. Besides, QTs in most imaging genetic studies were summarised from a voxel-wised level into a coarse level, e.g. region of interest (ROI)\citep{wang2012identifying, greenlaw2017bayesian, zhu2014bayesian}. This ‘trick’ reduces the computational burden by sacrificing partial spatial information in the whole-brain images to some extent.

Besides the massive univariate analysis, the multivariate analysis has been also widely used in genetic studies to explore the joint effect of correlated SNPs on a single phenotype. These models usually require a significant reduction in the dimensionality of data, thus sparse regression technologies are introduced. One line of research is adding regularization to select variables of interest: only a relatively small subset of massive genetic variants has significant effects on the phenotype of interest. This is usually achieved by posing $L_1$ or $L_2$ norm constraints on the regression model, such as the Group-Lasso \citep{vounou2010discovering, yuan2006model}.
The sparsity in these models is often controlled by a hyper-parameter named regularization factor. 

Another line of research performs decomposition techniques and then iterates in a greedy way to obtain the coefficient matrix, such as the L2RM \citep{kong2019l2rm}. To be precise, sparse constraints are imposed on the coefficient matrix, where sparsity is implicitly determined by singular values instead of regularization. These multivariate methods for GWAS offers new benefits over univariate framework: they have not only brought enhanced predictive performance due to correlated genotypes, but also alleviated the multiple testing problem to avoid the loss of efficiency in handling high put-through datasets.

Despite these advantages, a limitation of most association studies in imaging genetics is only furnishing a point estimate of the coefficient matrix but dismissing the uncertainty. \cite{little2019statistical} proposed a Bayesian multivariate normal model (MVN), which obtained high efficiency by fitting covariance between QTs. However, the performance suffers in datasets with high relatedness due to the ignorance of correlations between samples. \cite{dahl2016multiple} proposed a Bayesian multiple-phenotype mixed model (MPMM), which uses matrix normal (MN) distributions with the Kronecker products of row and column covariance matrices. Though this approach considered the correlations between both samples and traits and achieved better performance over other approaches on five simulated datasets, it mainly focused on the imputation of missing phenotypes rather than on the efficiency of computing effect sizes. Besides, \cite{song2022bayesian} proposed a Bayesian spatial model to accommodate the correlation structures in brain images by a proper bivariate conditional autoregressive process. The model was designed for assessing the association between a moderate number of QTs and SNPs (a few hundred to a few thousand), it may lose its efficiency in handling large-scale input data. Nevertheless, these models are valuable attempts to perform Bayesian inference in imaging genetic studies.

Association tests often identify significant loci associated with given phenotypes from a statistical perspective. The test statistics usually vary from model to model and depend on the model assumptions. Univariate approaches obtain a standard \textit{P-}value concerning effect size for each SNP-QT pair, where the effect size is 0 under the null hypothesis \citep{stein2010voxelwise, hibar2011voxelwise}.  Test statistics are carefully designed to avoid computationally intractable multi-test issues in a multivariate background. For example, the sequence kernel association test only fits the null model where phenotypes are regressed on the covariates alone \citep[SKAT,][]{wu2011rare}. Furthermore, \cite{cao2022gene} proposed a test named ‘overall’ to boost its power by aggregating the information from three types of traditional association tests including SKAT. In the Bayesian framework, each effect size is considered as a random variable rather than a fixed scalar, the association tests are quite different. One feasible choice is to use the credible interval of each effect size approximated from the posterior distribution, and significance is determined by whether ``0" is within the confidence interval~\citep{hespanhol2019understanding, lu2012analysis, eberly2003estimating}. The difficulty in dealing with the \textit{P-}values in the Bayesian framework is that the samples obtained from the posterior distribution of each effect size are usually dependent, not satisfying the independent assumptions commonly used in the frequentist hypothesis tests. Recently, \cite{liu2024ensemble} used the Cauchy combination test (CCT) to aggregate a set of individual \textit{P-}values into a single test statistic via Cauchy transformations, and these \textit{P-}values are not necessarily independent.  Inspired by this, we adopted the CCT into our model, and used the credible interval approach as a benchmark. 

In this work, we developed a novel Bayesian regression framework for identifying significant associations between QTs and genetic markers while modeling the spatial dependency among QTs explicitly. Our work has three-fold primary contributions as follows. Firstly, we incorporated a population-level mixed-effects term into the linear model, taking full advantage of the dependent structure of brain imaging QTs. This fits the fact that a SNP is interlinked to multiple QTs by pleiotropy - a common biological phenomenon. The population-level mixed-effects term avoids the unidentifiable issue triggered by the individual-level mixed-effects term. Secondly, we implemented the model in the Bayesian framework and derived a Markov chain Monte Carlo (MCMC) algorithm to do the model inference. Further, we incorporated the MCMC samples with the CCT in an ensemble framework to explore the significant association between SNPs and QTs, which avoided computationally intractable multi-test issues.   With this model, we can perform the association analysis between a set of QTs and a given SNP simultaneously instead of testing one QT at a time.   The spatial model can effectively utilize spatial information and boost its power in association studies.  Our simulation studies indicated improved power of our method concerning the metric area under the receiver operating characteristics curve (AUC, \citealp{ling2003auc,huang2005using}) compared to the traditional linear model. In addition, though we validated our method in a simulation study containing moderate QTs and SNPs, our method can be easily extended to a large-scale of SNP dataset by parallel allocating disjoint SNP sub-datasets to multiple compute servers.

\section{Methods}
\label{sec: 2}
\subsection{A Bayesian spatial-correlated multitask regression model}
A linear model (LM) explains how the $\textit{outcome}$ variable varies over \textit{predictors} by a linear function. Denote $y_{i}$ the observed response value of individual $i$ ($i=1,\ldots,n$).  
 Let $x_i$ be the 1$-$dimensional predictor of individual $i$, the distribution of outcomes $y_{i}$ given the predictor $x_i$ ($i=1, \ldots, n$) and coefficients $\beta_{0},\beta_{1}$, is normally distributed with variance $\sigma^2$ and mean
\begin{equation}
   \label{eqlm}
    E(y_{i}|\beta_{1}, \beta_{0}, x_i) =  x_i\beta_{1} + \beta_{0},
\end{equation}
where $\beta_{1}$ is a scalar, representing the effect size of predictor $x_i$ concerning the response variable, and  $\beta_{0}$ is the interception term.  This univariate LM implicitly assumes that $y_{i},~i=1,\ldots,n$ are independent. Thus, Eq. (\ref{eqlm}) can be equalized as 
\begin{equation}
   \label{eqlm_m}
   y_i|\beta_{1}, \beta_{0}, x_i \sim {N(  x_i\beta_{1} + \beta_{0}, \sigma^2}).
\end{equation}

However, grouping factors, such as populations, species and regions, exist widely in biological data, which cause the data points not to be truly independent. Thus, a linear mixed model (LMM) is developed to deal with structured data by modeling the relationship of outcomes within groups:
\begin{equation}
    E(y_i|{\mathbf{\beta}_1}, \beta_0, {x_i}, {b_i}) =  {x_i}\beta_1 + \beta_0 + {b_i}, 
\end{equation}
where {$b_i$} is a mixed-effects term, modelling the potential inner linkage of $y_1,\ldots,y_n$.  Compared to a LM, a LMM models the dependency among $y_i$ explicitly, which boosts its power in association studies. 

In this work, we extended the traditional univariate LMM by proposing a multi-task univariate spatial regression model in the Bayesian framework to accommodate the dependency among phenotypes. To ensure the completeness, we used the linear model as a comparison. To be specific, let {$\mathbf{y}_i \in \mathbb{R}^{p}$} be the vector of $p$ phenotypes of  $i$th individual, let $x_{i}$ be the observed value of individual $i$ given the specified predictor. Start with a simple linear model,
\begin{align}
\label{con:lm}
    {\mathbf{y}_i} = x_{i}{\bm{\beta}_1} + {\bm{\beta}_0} + {\bm{\epsilon_i}}
\end{align}
where ${\bm{\epsilon_i}} \sim MVN(0,{\sigma^2} I_p)$ is the noise term, ${\bm{\beta}_1} =(\beta_{11},\ldots,\beta_{1p})^T$ and ${\bm{\beta}_0}=(\beta_{01},\ldots,\beta_{0p})^T$ with $\beta_{1j}$ and $ \beta_{0j}$ representing the effect size and intercept term of the predictor concerning phenotype $j$ {($j=1,\ldots,p$)}, respectively. The equivalent expression of Eq. (\ref{con:lm}) is 
\begin{align}
\label{con:lm_2}
    {\mathbf{y}_i} |{\sigma^2}, {\bm{\beta}_1}, {\bm{\beta}_0} \sim{MVN(x_{i}{\bm{\beta}_1}+{\bm{\beta}_0}, {\sigma^2}I_p)}.
\end{align}

To accommodate the dependency among phenotypes, we incorporated a mixed-effects term $b_i$ in the linear model,
\begin{align}
\label{con:lm33}
    {\mathbf{y}_i} = x_{i}{\bm{\beta}_1} + {\bm{\beta}_0} + {\mathbf{b}_i}+{\bm{\epsilon_i}}, 
\end{align}
where ${\mathbf{b}_i}\sim MVN(0,\Sigma)$, where $\Sigma$ is an unknown positive definite matrix representing the dependency among phenotypes. However, ${\mathbf{b}_i}$ would intrigue unidentifiable issue in the model inference without further constrains. As ${\mathbf{b}_i}$ represents the dependency among phenotypes, for individuals from the same population, it's reasonable to assume the dependency of all individuals are the same at the population level, thus, the Eq. (\ref{con:lm33}) can be adjusted to 
\begin{align}
\label{con:lm3}
    {\mathbf{y}_i} = x_{i}{\bm{\beta}_1} + {\bm{\beta}_0} + {\mathbf{h}}+{\bm{\epsilon_i}}, 
\end{align}
where ${\mathbf{h}}\sim MVN(0,\Sigma)$ represents the population-level phenotype dependency, and theoretically $\Sigma$ can take any positive definite matrix. To further simplify the structure, in this work, we set $\Sigma=\sigma_p^2 G$, $\sigma^2_p$ is unknown, $G$ is known and can be given by the sample covariance of ${\mathbf{y}_1},\ldots, {\mathbf{y}_n}$ since $\Sigma$ representing the group-level phenotype dependency. Further, if $y_i$ are $p/2$ paired phenotypes, $G$ can be given by the Kronecker product of $A\otimes B$, where $A$ and $B$ represent the correlation within paired QTs and between paired QTs, respectively.

In the Bayesian framework, for the LMM described in Eq. (\ref{con:lm3}), we used multivariate normal distributions as prior distributions for parameters $\bm{\beta}_0,\bm{\beta}_1$ and Inv-Gamma distributions as priors for $\sigma_e^2,\sigma_p^2$. The fully Bayesian LMM model is then given as below,
\begin{align*}
 {\mathbf{y}_i} | {\bm{\beta}_1},{\bm{\beta}_0},{\mathbf{h}},\sigma_e^2 & \sim MVN( x_{i}{\bm{\beta}_1} + {\bm{\beta}_0} + {\mathbf{h}}, \sigma_e^2I_p),\\
   {\bm{\beta}_1} & \sim{MVN({\mathbf{0}}, I_p)},\\
   {\bm{\beta}_0} & \sim{MVN({\bm{\mu}_0}, I_p)},\\
    {\mathbf{h}}|\sigma^2_p &\sim{MVN({\mathbf{0}}, \sigma_p^2G_{p\times p})}, \\
    \sigma_p^2& \sim{IG(a, b)}, \\
    \sigma_e^2& \sim{IG(c, d)}, 
\end{align*}
where $a,b,c,d$ and $\mu_0$ are hyperparameters. Similarly, the fully Bayesian LM model described in Eq. (\ref{con:lm}) is given as follows,
\begin{align*}
 {\mathbf{y}_i} | {\bm{\beta}_1},{\bm{\beta}_0},{\sigma^2} & \sim MVN( x_{i}{\bm{\beta}_1} + {\bm{\beta}_0},{\sigma^2}I_p),\\
   {\bm{\beta}_1} & \sim{MVN({\mathbf{0}}, I_p)},\\
   {\bm{\beta}_0} & \sim{MVN({\bm{\mu}_0}, I_p)},\\
    \sigma_e^2& \sim{IG(c, d)}.
\end{align*}

\subsection{MCMC Inference}
We implemented an MCMC algorithm, \emph{i.e.} Gibbs sampling in our content, to approximate the posterior of parameters. The basic idea of MCMC is to generate a Markov chain with equilibrium distribution to be the target.  Precisely, with the data denoted as $y$ and parameters to be represented by $\theta$, Gibbs sampling is performed to approximate the posterior distribution $p(\theta|y)$ by iterative sampling from every single parameter's full conditional distribution $p(\theta_{i}^{(t)} | \theta_{1}^{(t+1)}, \ldots, \theta_{i-1}^{(t+1)},  
\theta_{i+1}^{(t)}
,\cdots, \theta_{d}^{(t)}, y)$. Given the current samples $\theta^{(t)} = (\theta^{(t)}_1, \ldots, \theta^{(t)}_d)$ at time $t$, each parameter at time $t+1$ can be drawn from the full conditional distribution iteratively:
\begin{equation}
    \theta^{(t+1)}_i \sim p(\theta^{(t)}_i| \theta^{(t+1)}_1, \ldots, \theta^{(t+1)}_{i-1}, \theta^{(t)}_{i+1}, \ldots, \theta^{(t)}_{d}, y)
\end{equation}
   Denote the remaining samples after the burn-in stage as $\{\theta^{(t)}: t= 1, \ldots, m\}$, the posterior distribution of $\theta_i$ ($i = 1, \ldots, d$) can be approximated by $P(\theta_i)=\frac{1}{m}\sum_{i=1}^{m}I(\theta_i=\theta^{(t)}_{i})$, and its point estimate can be estimated by taking the sample average: $ \Hat{\theta}_{i} = \frac{1}{m}\sum_{t=1}^m \theta_{i}^{(t)}$.

\subsubsection{Gibbs sampling}
The joint posterior distribution for all unknown parameters of our proposed LMM is given by:
 \begin{equation}
 \begin{aligned}
 & p({\bm{\beta}_1}, \sigma_e^2, \sigma_p^2, {\mathbf{h}}, {\bm{\beta}_0}| y_1, \cdots, y_n) 
 \propto \prod_{i=1}^n |\sigma_e^2I_p|^{-1/2} \\
 &\times 
 \exp\{-\frac{1}{2\sigma_e^2}||{\mathbf{y}_i}-x_i{\bm{\beta}_1}-{\mathbf{h}}- {\bm{\beta}_0}||^2 -\frac{1}{2}{\bm{\beta}_1}^T{\bm{\beta}_1}\} \\
 &\times  |\sigma_p^2G|^{-1/2} 
 \exp\{-\frac{1}{2\sigma_p^2}{\mathbf{h}}^TG^{-1}{\mathbf{h}}\} \\
 &\times (\sigma_p^2)^{-a-1}e^{-\frac{b}{\sigma_p^2}}  (\sigma_e^2)^{-c-1}e^{-\frac{d}{\sigma_e^2}} \\ 
 &\times \exp\{-\frac{1}{2}({\bm{\beta}_0}-{\bm{\mu}_0})^T({\bm{\beta}_0}-{\bm{\mu}_0})\}.
 \end{aligned}
 \end{equation}
The full conditional distribution of each parameter is derived as follows:
\begin{align}
    \sigma_e^2| rest &\sim IG(c+\frac{np}{2}, \frac{1}{2}\sum_{i=1}^n||{\mathbf{y}_i}-x_i{\bm{\beta}_1}-{\mathbf{h}}- {\bm{\beta}_0}||^2+d), \label{con: lmm1}\\
        \sigma_p^2| rest &\sim IG(a+\frac{p}{2}, \frac{1}{2}{\mathbf{h}}^TG^{-1}{\mathbf{h}}+b), \label{con: lmm2}\\
    {\bm{\beta}_1}| rest &\sim 
    MVN({\bm{\mu}}_{{\bm{\beta}_1}}, \Sigma_{{\bm{\beta}_1}}),
    \label{con: lmm3}\\
    {\bm{\beta}_0}| rest &\sim 
    MVN({\bm{\mu}}_{{\bm{\beta}_0}}
    , \Sigma_{{\bm{\beta}_0}}),
    \label{con: lmm4} \\
    {\mathbf{h}}| rest &\sim 
    MVN({\bm{\mu}}_{{\mathbf{h}}}, \Sigma_{{\mathbf{h}}}). 
    \label{con: lmm5}
\end{align}
where ${\bm{\mu}}_{{\mathbf{h}}}=(\frac{n}{\sigma_e^2}I_p+\frac{1}{\sigma_p^2}G^{-1})^{-1}\frac{\sum_i({\mathbf{y}_i}-x_i{\bm{\beta}_1}- {\bm{\beta}_0})}{\sigma_e^2}$, $\Sigma_{{\mathbf{h}}}=(\frac{n}{\sigma_e^2}I_p+\frac{1}{\sigma_p^2}G^{-1})^{-1}$, ${\bm{\mu}_{\bm{\beta}_1}}=\frac{\frac{1}{\sigma_e^2}\sum_i({\mathbf{y}_i}-{\mathbf{h}}-
    {\bm{\beta}_0})x_i}{\frac{1}{\sigma_e^2}\sum_ix_i^2+1} $, $\Sigma_{{\bm{\beta}_1}}= (\frac{\sum_ix_i^2}{\sigma_e^2}+1)^{-1}I_p$, ${\bm{\mu}}_{{\bm{\beta}_0}}=(\frac{n}{\sigma_e^2}+1)^{-1}(\frac{\sum_i({\mathbf{y}_i}-x_i{\bm{\beta}_1}- {\mathbf{h}})}{\sigma_e^2}+{\bm{\mu}_0})
    $, and $\Sigma_{{\bm{\beta}_0}}=(\frac{n}{\sigma_e^2}+1)^{-1}I_p$.
    
    Algorithm \ref{alg:lmm} depicts the Gibbs sampling algorithm of the LMM. We refer readers to Appendix A for the derivation details. 
     The Gibbs sampling algorithm for the full Bayesian LM and the derivation details are also given in Appendix A. The implementation of our method is available at \url{https://github.com/ZhibinPU/spatialmultitasklmm.git}.

\begin{algorithm}
    \caption{Gibbs sampling algorithm for the LMM}
    \label{alg:lmm}
  {\bfseries Input:} $\textbf{X}=(x_{ij}) \in \mathbb{R}^{n \times d}$
        , $\textbf{Y}=\{\textbf{y}_1, \textbf{y}_2, \cdots, \textbf{y}_n\} \in \mathbb{R}^{p \times n}$, total number of iteration $m$. \\
 {\bfseries Output:} Gibbs Samples $\{\sigma_e^{2(t)}, \sigma_p^{2(t)}, {\bm{\beta}_1}^{(t)}, {\mathbf{h}}^{(t)}, {\bm{\beta}_0}^{(t)}\}_{t=1}^{m}$\\
  {\bfseries Initialization:}  Set $t \leftarrow 0$, and initialize $\sigma_e^{2(0)}$, $\sigma_p^{2(0)}$, ${\bm{\beta}_1}^{(0)}$, ${\bm{\beta}_0}^{(0)}$,
        ${\mathbf{h}}^{(0)}$.\\
\While{$t \le m$} {
 Set $t \leftarrow t + 1$.\\
  Sample
        $\sigma_e^{2(t+1)}$ according to Eq. (\ref{con: lmm1}) conditional on $\sigma_p^{2(t)}$, 
        ${\bm{\beta}_1}^{(t)}$, 
        ${\mathbf{h}}^{(t)}$,
        ${\bm{\beta}_0}^{(t)}$.\\
  Sample $\sigma_p^{2(t+1)} $ according to Eq. (\ref{con: lmm2}) conditional on $\sigma_e^{2(t+1)}$, ${\bm{\beta}_1}^{(t)}$, ${\bm{\beta}_0}^{(t)}$, ${\mathbf{h}}^{(t)}$.\\
Sample ${\bm{\beta}_1}^{(t+1)}$ according to Eq. (\ref{con: lmm3}) conditional on $\sigma_e^{2(t+1)}$, 
        $\sigma_p^{2(t+1)}$, 
        ${\bm{\beta}_0}^{(t)}$,
        ${\mathbf{h}}^{(t)}$.\\
Sample ${\bm{\beta}_0}^{(t+1)}$ according to Eq. ({\ref{con: lmm4}}) conditional on 
        $\sigma_e^{2(t+1)}$, 
        $\sigma_p^{2(t+1)}$, 
        ${\bm{\beta}_1}^{(t+1)}$,
        ${\mathbf{h}}^{(t)}$.\\
Sample ${\mathbf{h}}^{(t+1)}$ according to Eq. (\ref{con: lmm5}) conditional on
        $\sigma_e^{2(t+1)}$, 
        $\sigma_p^{2(t+1)}$, 
        ${\bm{\beta}_1}^{(t+1)}$,
        ${\bm{\beta}_0}^{(t+1)}$.\\
} 
\end{algorithm}

{Recall that $n$ was the sample size, $m$ was the total number of SNPs, and $p$ was the total number of phenotypes. The computational complexity of a single step iteration of MCMC was $\mathcal{O}(nmp^2)$ for the LM,  and was $\max\{\mathcal{O}(nmp^2), \mathcal{O}(mT_{inv})\}$ for the LMM, where $T_{inv}$ represented the computational complexity for computing the inverse of a matrix of size ${p \times p }$. \cite{tveit2003complexity} showed that the lower bound of $T_{inv}$ was $O(p^{2}log(p))$, and $T_{inv}$ at most took $\mathcal{O}(p^3)$. When $n < log(p)$, the LMM would be more theoretically computationally expensive than the LM. However, in practice, $p$ (\emph{i.e.} the number of phenotypes) is often less than $n$ (\emph{i.e.} the sample size).}

\section{Simulation}
In this section, we simulated data in different scenarios to evaluate the performance of our method.  We generated  $d$ independent SNPs of sample size $n$ with software \textit{gG2P}, a GWAS simulation tool \citep{tang2019g2p}. Equation (\ref{con: sim}) was used to generate $p$-dimensional phenotypes with an underlying trait covariance as well as the random noisy, which were controlled by parameters $\sigma^2_p$, $G$ and $\sigma^2_e$.  
  \begin{equation}
    {\mathbf{y}_i} = \sum_{j=1}^d x_{ij}{\bm{\beta}_j} + {\mathbf{h}} + {\bm{\epsilon}_i}, ~~~i=1,\ldots,n.
    \label{con: sim}
\end{equation}

 We considered a SNP size of $d=100$, with effect sizes ${\bm{\beta}_j}$ ($j=1,\ldots,d$) being a $p$-dimensional random vector that follows a mixture of a normal distribution and a Dirac distribution, ${\bm{\beta}_j} \sim \pi_1\delta(\textbf{0})+\pi_2MVN(0,I_p)$, where $\delta(\textbf{0})$ means the probability mass being zero at every point except at $0$,  $\pi_1 = 0.95$ and $\pi_0 = 0.05$. In other words, among total $d$ SNPs, only about $5\%$ were significant (non-zeros). Besides, the mixed-effects $h$ and random noise $\epsilon_{ij}$ were generated from their prior distributions described in Section \ref{sec: 2}, with $\sigma_e^2=0.1^2$ and $\sigma_p^2=0.2^2$.

\begin{figure}[h]
  \centering
  \begin{subfigure}{0.45\linewidth}
    \includegraphics[width=\linewidth]{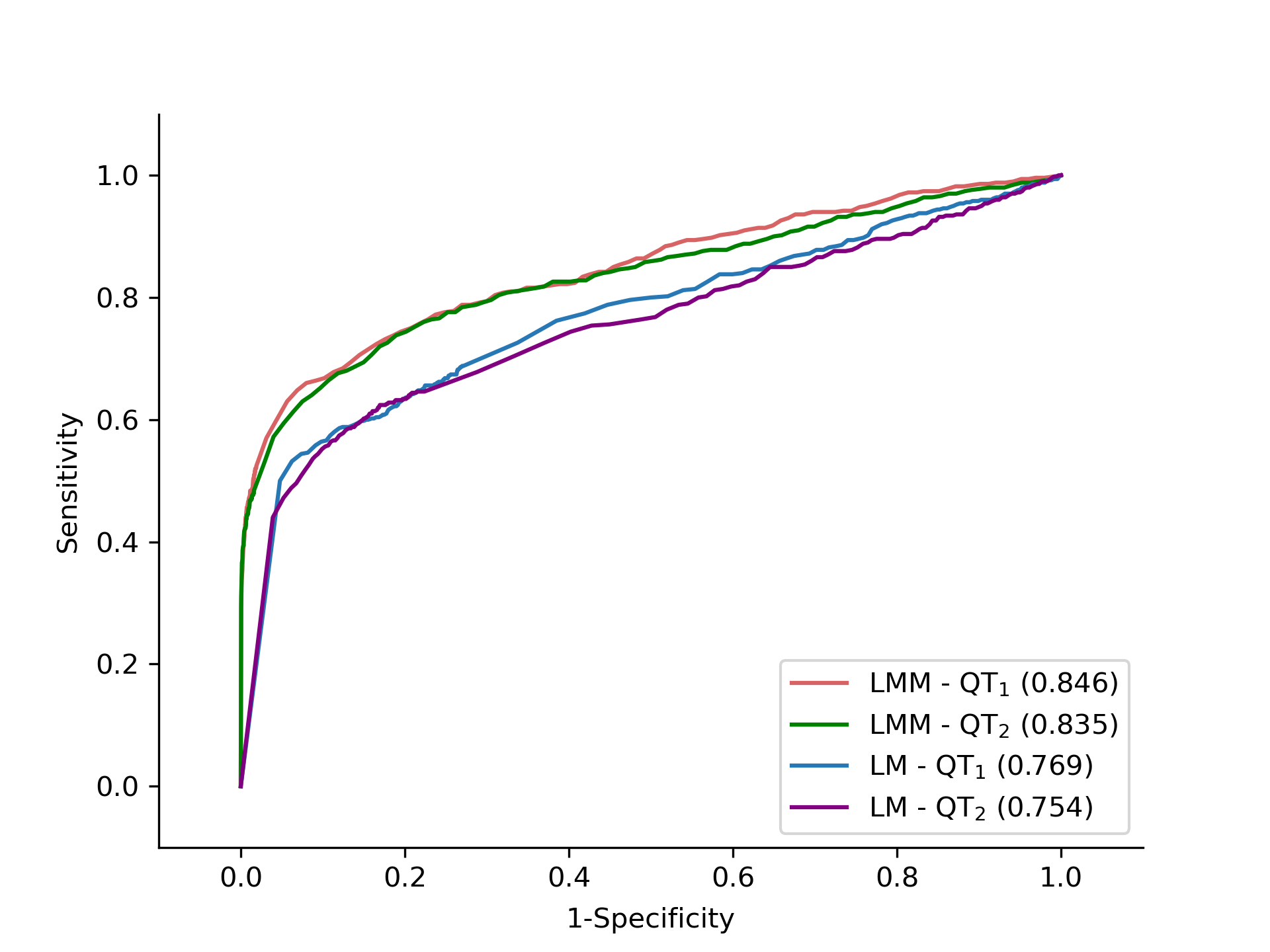}
    \label{subfig:n=50_average_i}
    \subcaption{$n=50$}
  \end{subfigure}
  \hfill
  \begin{subfigure}{0.45\linewidth}
    \includegraphics[width=\linewidth]{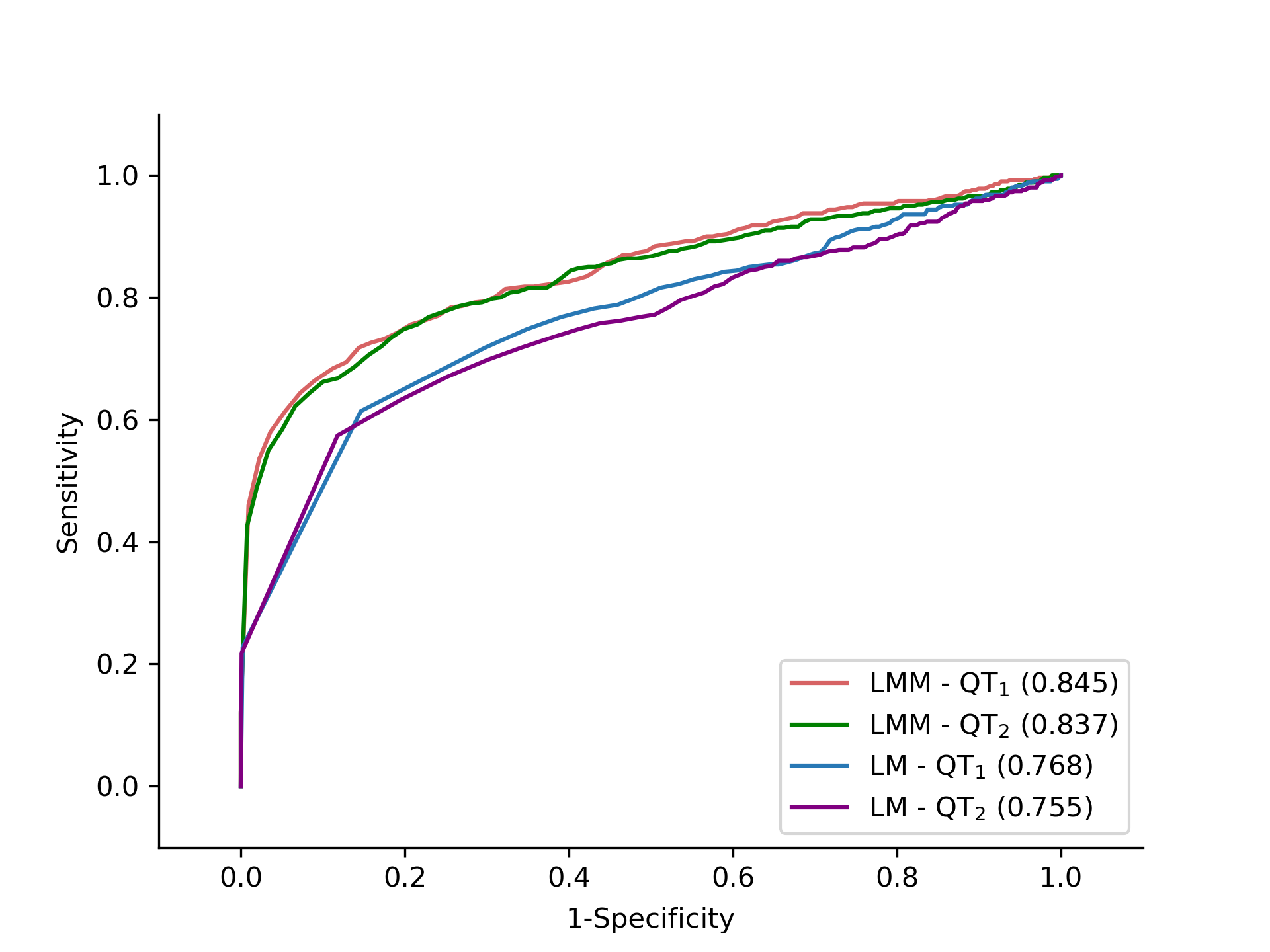}
    \label{subfig:n=50_average_p}
    \subcaption{$n=50$}
  \end{subfigure}
\par
\begin{subfigure}{0.45\linewidth}
    \includegraphics[width=\linewidth]{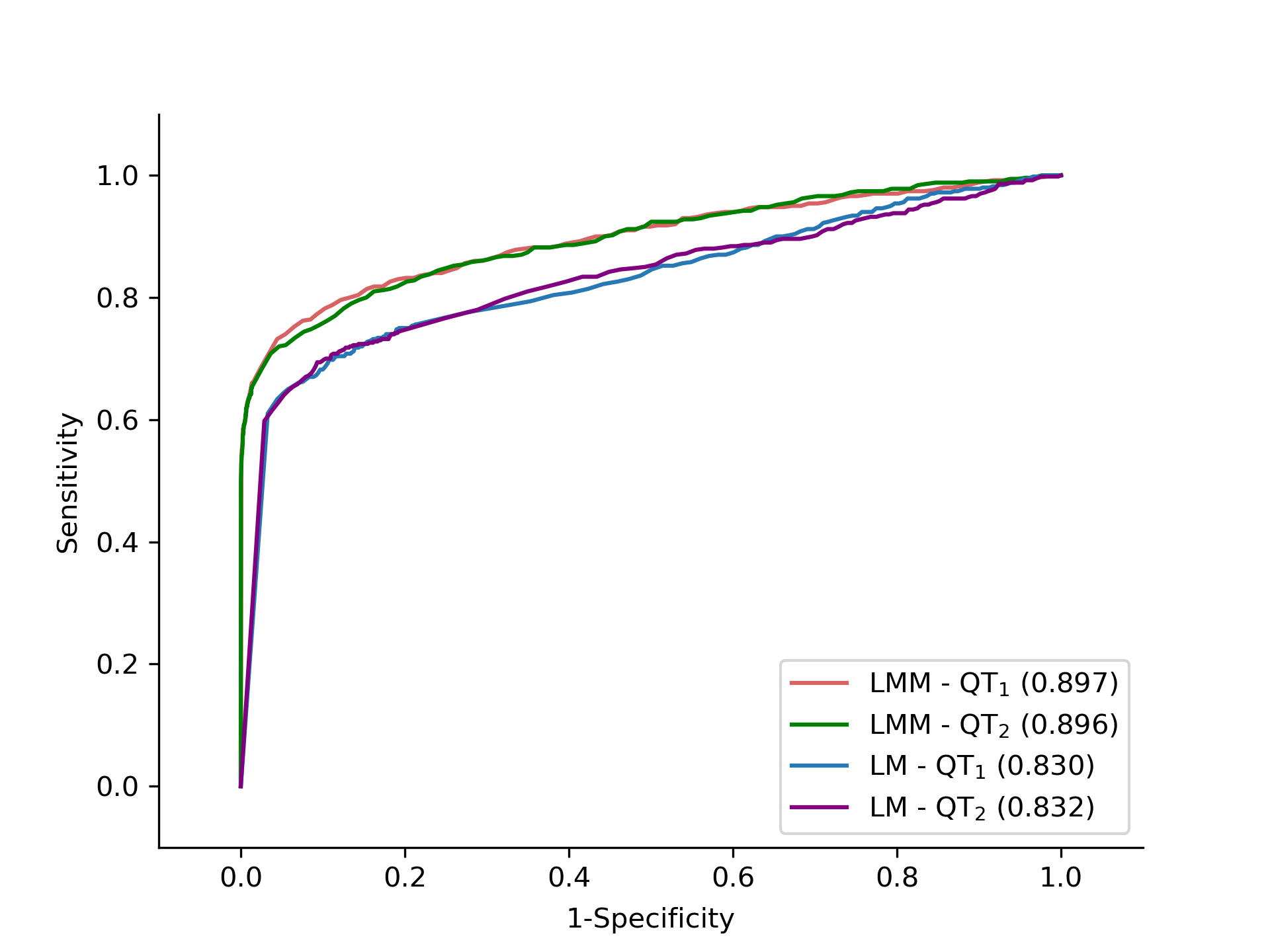}
    \label{subfig:n=100_average_i}
    \subcaption{$n=100$}
  \end{subfigure}
  \hfill
  \begin{subfigure}{0.45\linewidth}
    \includegraphics[width=\linewidth]{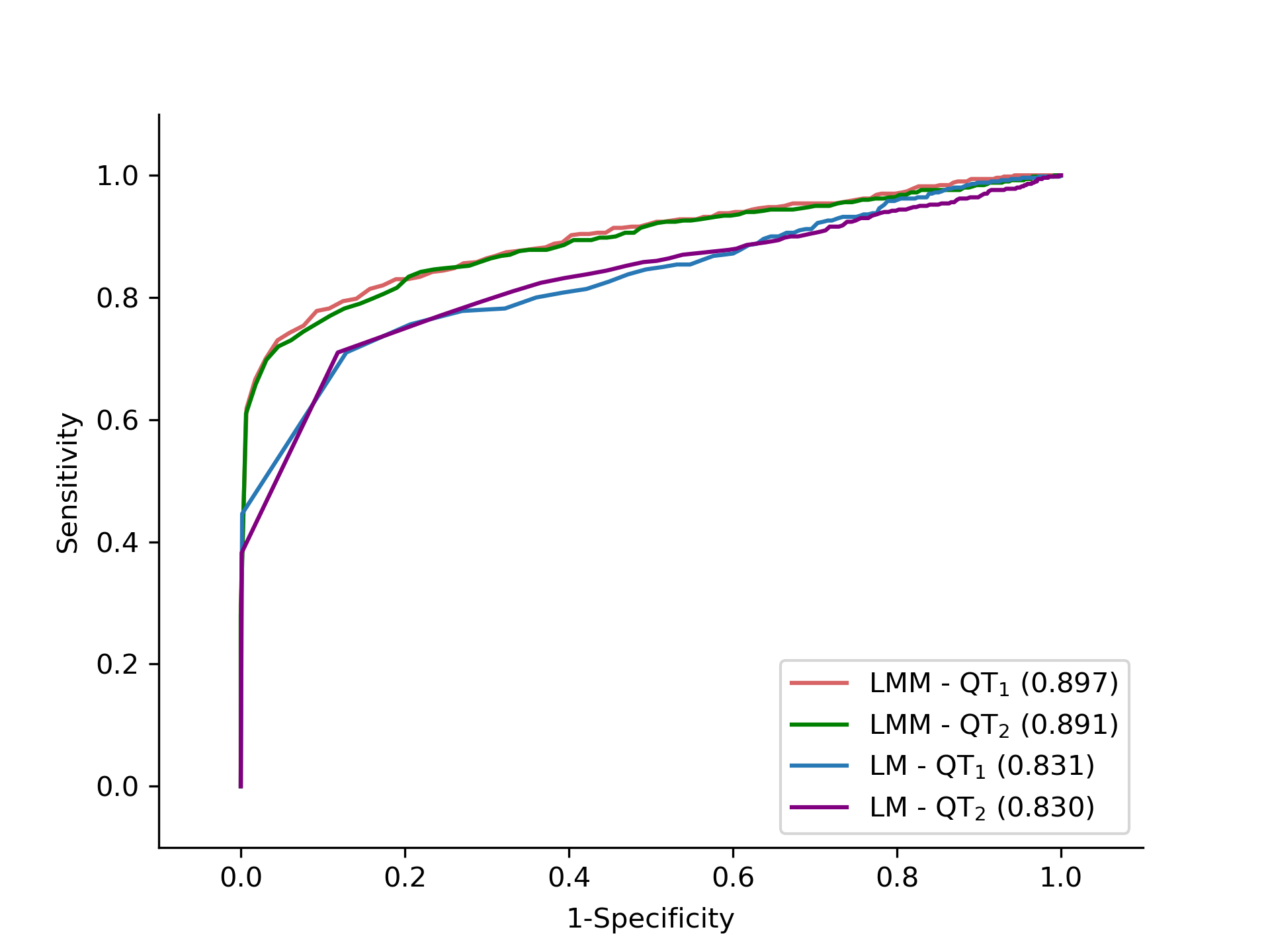}
    \label{subfig:n=100_average_p}
    \subcaption{$n=100$}
  \end{subfigure}
\caption{Case 1. ROC curves of LMM and LM based on credible intervals (a,c) and the CCT ({b},d) when there was a moderate dependency presented among phenotypes with sample size varies.  Curves of QT$_1$, QT$_2$ represent the ROC curves concerning QT$_1$ and QT$_2$.  The figures in  brackets indicated the corresponding AUC for each curve. }  
  \label{fig:sp1}
\end{figure}

\begin{figure}[h]
  \centering
  \begin{subfigure}{0.45\linewidth}
    \includegraphics[width=\linewidth]{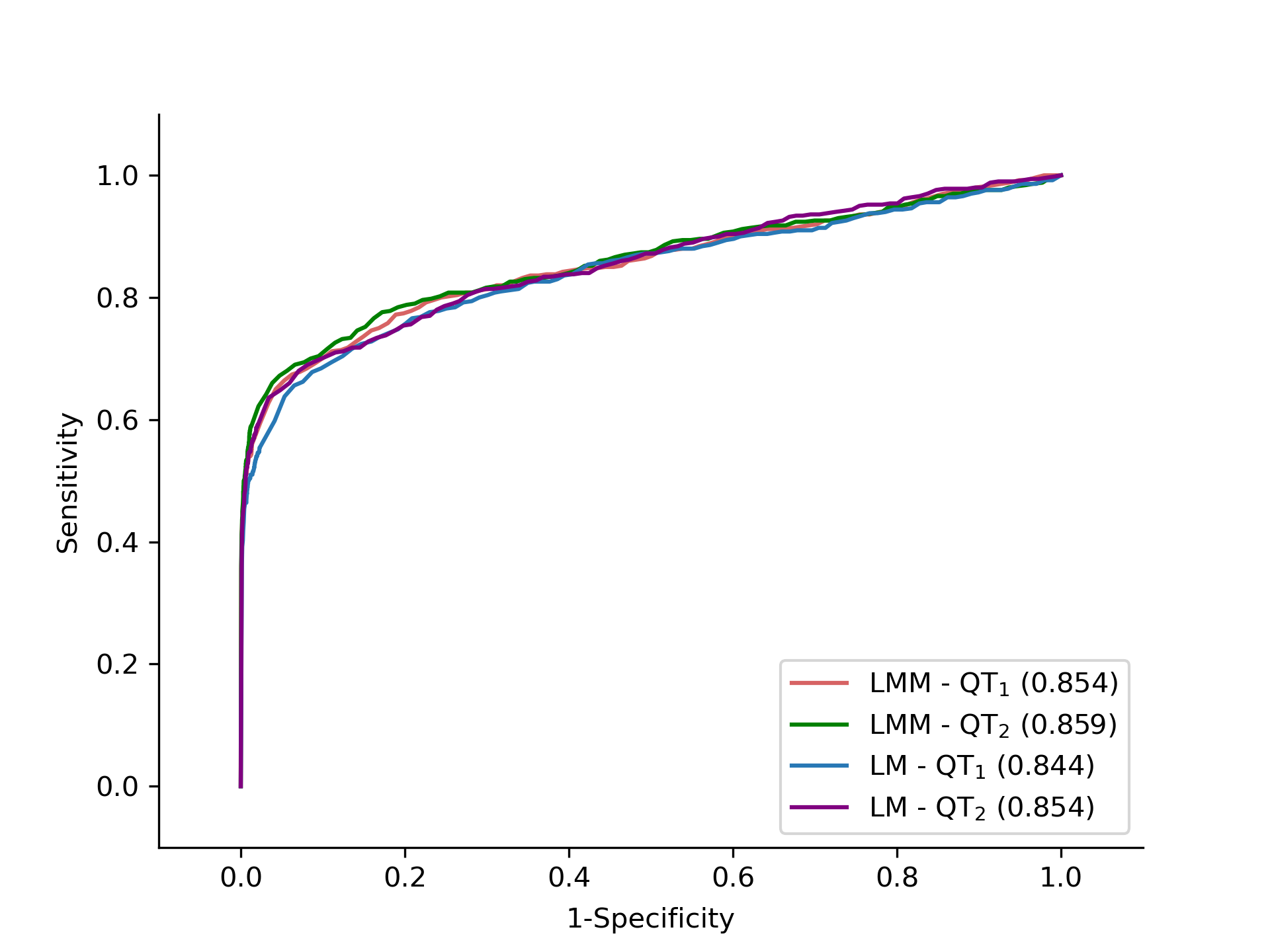}
    \label{subfig:np_n=50_average_i}
    \subcaption{$n=50$ }
  \end{subfigure}
  \hfill
  \begin{subfigure}{0.45\linewidth}
    \includegraphics[width=\linewidth]{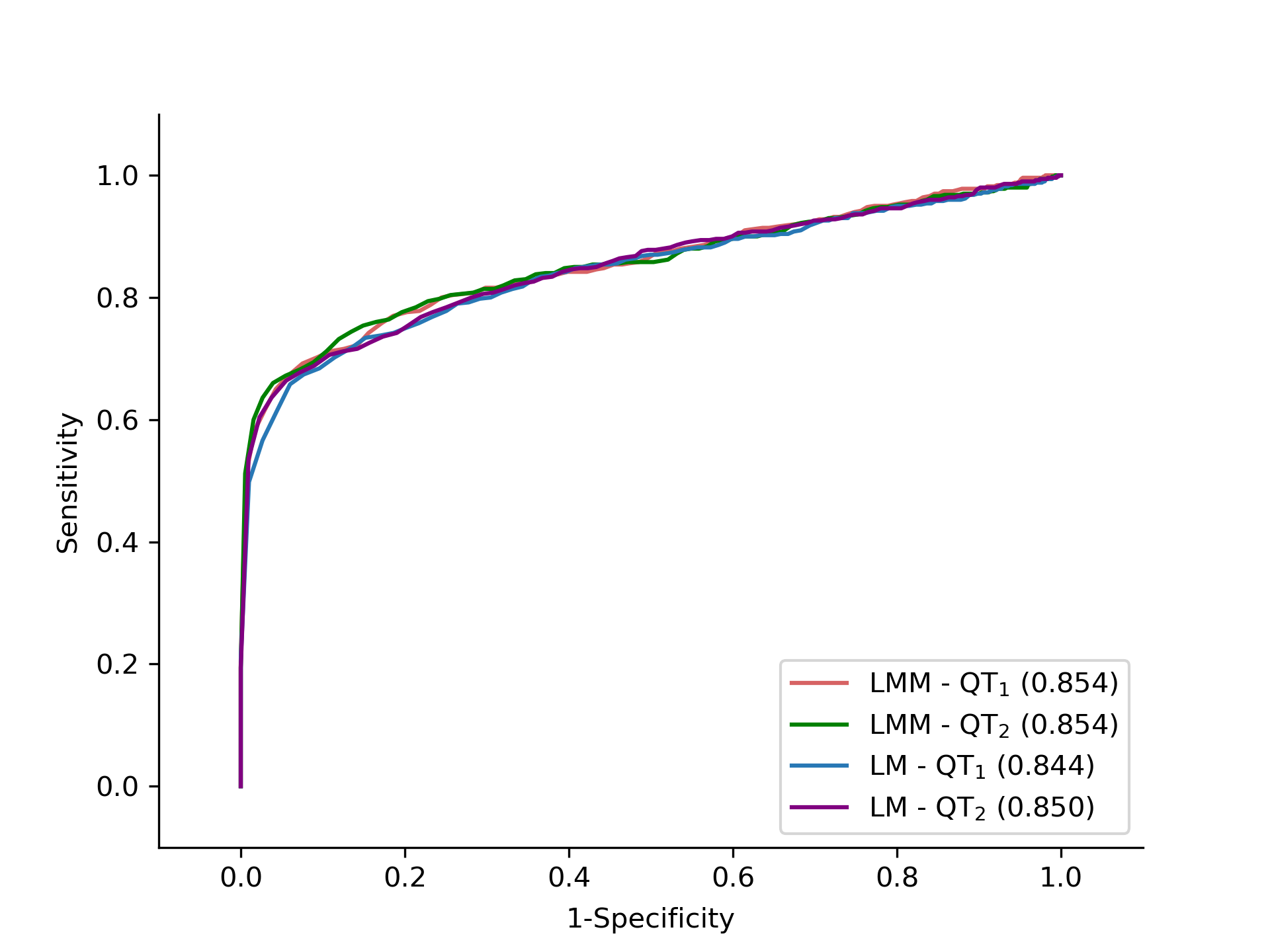}
    \label{subfig:np_n=50_average_p}
    \subcaption{$n=50$ }
  \end{subfigure}
\par
\begin{subfigure}{0.45\linewidth}
    \includegraphics[width=\linewidth]{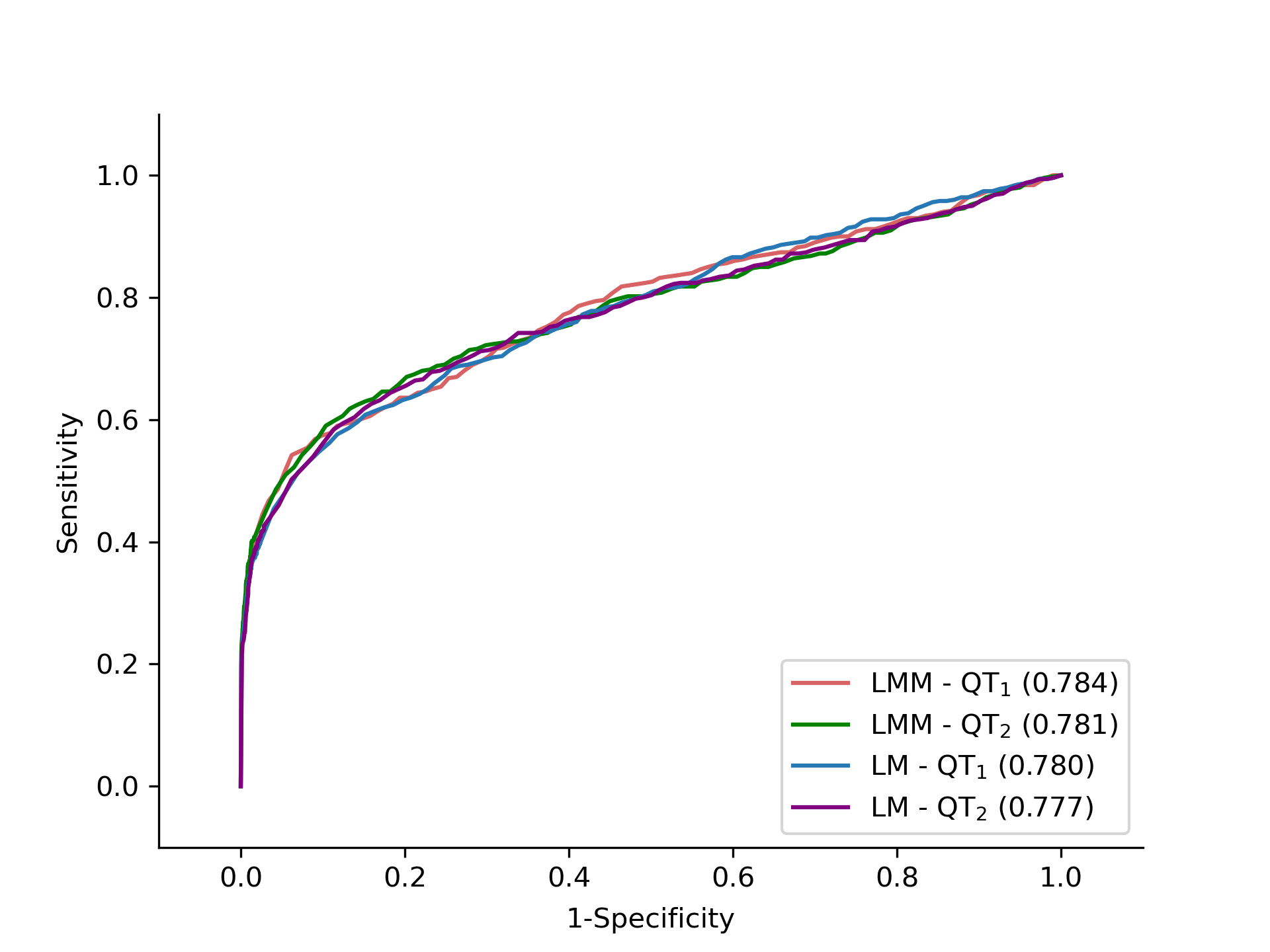}
    \label{subfig:np_n=100_average_i}
    \subcaption{$n=100$ }
  \end{subfigure}
  \hfill
  \begin{subfigure}{0.45\linewidth}
    \includegraphics[width=\linewidth]{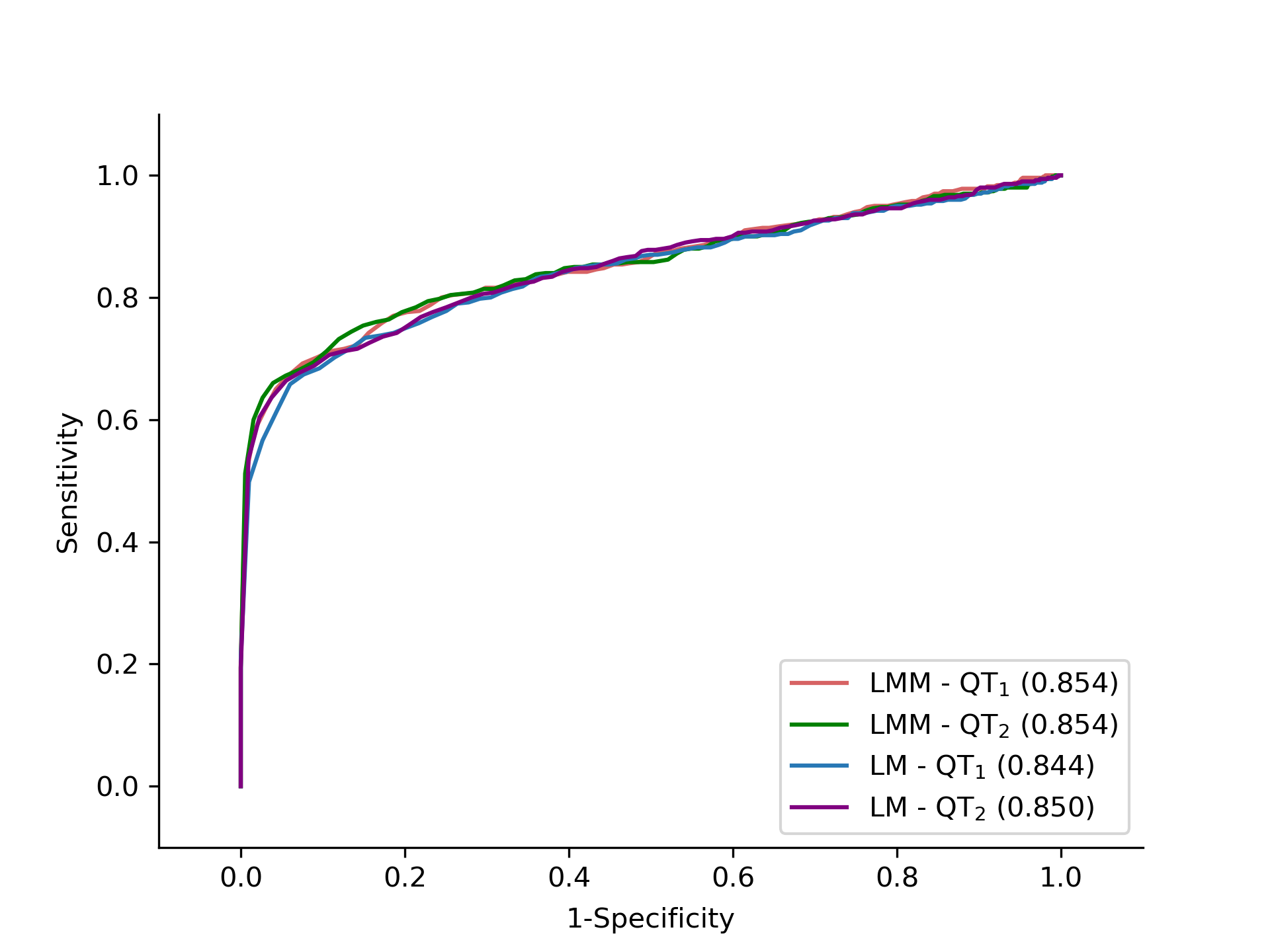}
    \label{subfig:np_n=100_average_p}
    \subcaption{$n=100$ }
  \end{subfigure}
\caption{Case 2. ROC curves of LMM and LM based on credible intervals (a,c) and the CCT ({b},d) when there was no dependence presented among phenotypes with sample size varies. Curves of QT$_1$, QT$_2$ represent the ROC curves concerning QT$_1$ and QT$_2$.  The figures in  brackets indicated the corresponding AUC for each curve.}
  \label{fig:sp2}
\end{figure}

 We set the sample size to $n = 50$ (less than the SNP size),  and $n = 100$ (equal to the SNP size) respectively.  $G$ was set to different values to demonstrate the robustness of our method corresponding to scenarios with independent, weakly, moderately, and strongly dependent phenotypes.  We also compared our method to a basic LM with respect to the metric AUC based on credible intervals and CCT {\textit{P-}values} described in Section \ref{sec:intro}. 

In \textbf{Cases 1} and \textbf{2}, we set QTs $p=2$, and the dependency matrix $G$ to be 
$\begin{bmatrix}
    1 & 0.5 \\
    0.5 & 1 \\
\end{bmatrix}$  (moderate), 
$\begin{bmatrix}
    1 & 0 \\
    0 & 1 \\
\end{bmatrix}$ (zero), respectively. We applied the LMM and LM to the simulated data and we repeated the experiment $100$ times. We computed the averaged point-wise sensitivity and specificity, resulting in Receiver Operating Characteristic (ROC) curves as shown in Figure~\ref{fig:sp1} (moderate) and Figure \ref{fig:sp2} (zero). Figure~\ref{fig:sp1} showed improved performance of the LMM over the LM concerning the AUC based on both the credible intervals and aggregated {\textit{P-}values} of the LMMs were better when there was a moderate dependency among phenotypes. When there was no spatial dependency among QTs, as shown in Figure \ref{fig:sp2}, the performance of both models was similar as LMMs gained no spatial information in this case compared to the LMs.

\begin{figure}
  \centering
  \begin{subfigure}{0.49\linewidth}
    \includegraphics[width=\linewidth]{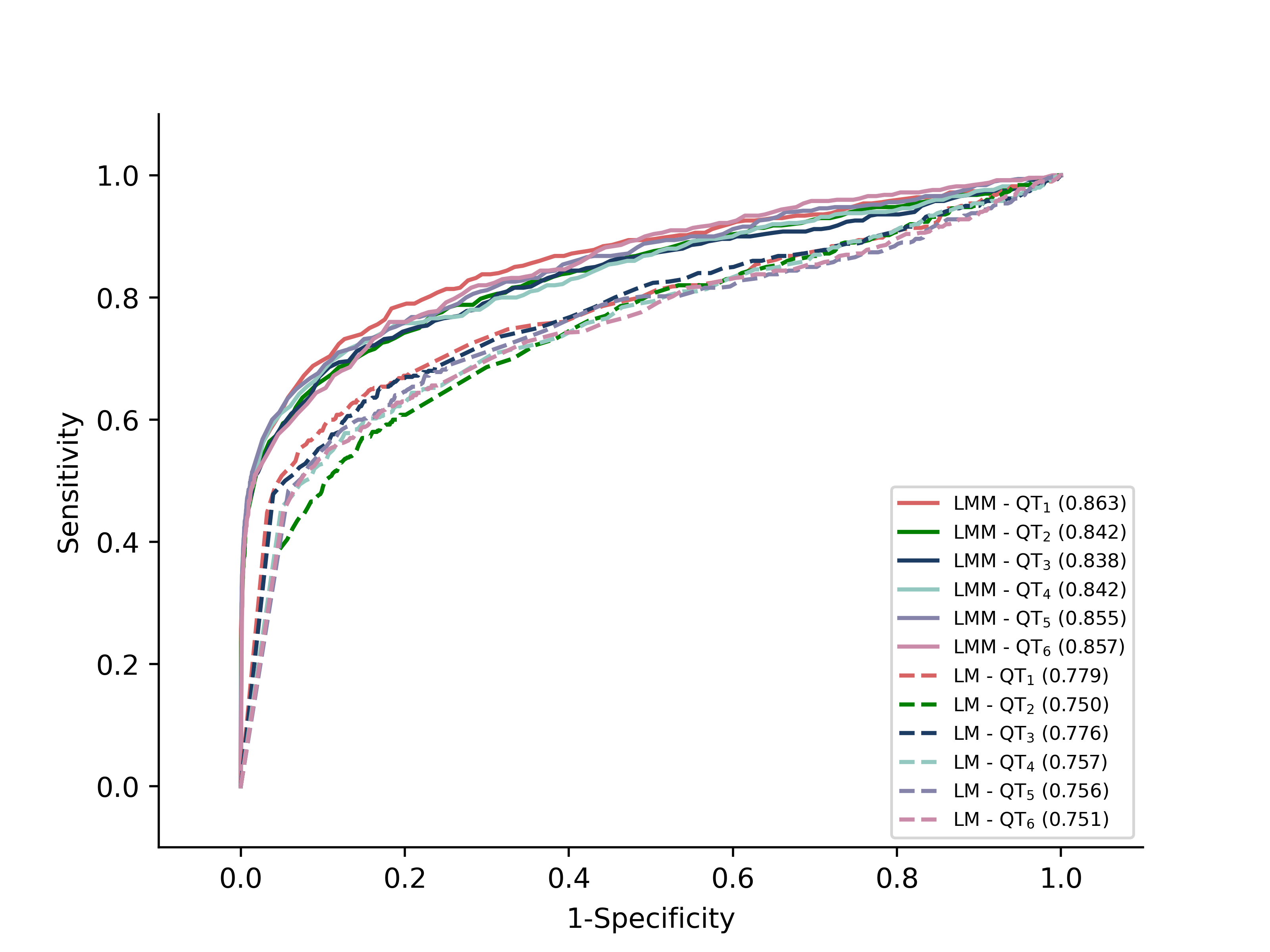}
    \label{subfig:case5_n=50_average_i}
    \subcaption{$n=50$}
  \end{subfigure}
  \hfill
  \begin{subfigure}{0.49\linewidth}
    \includegraphics[width=\linewidth]{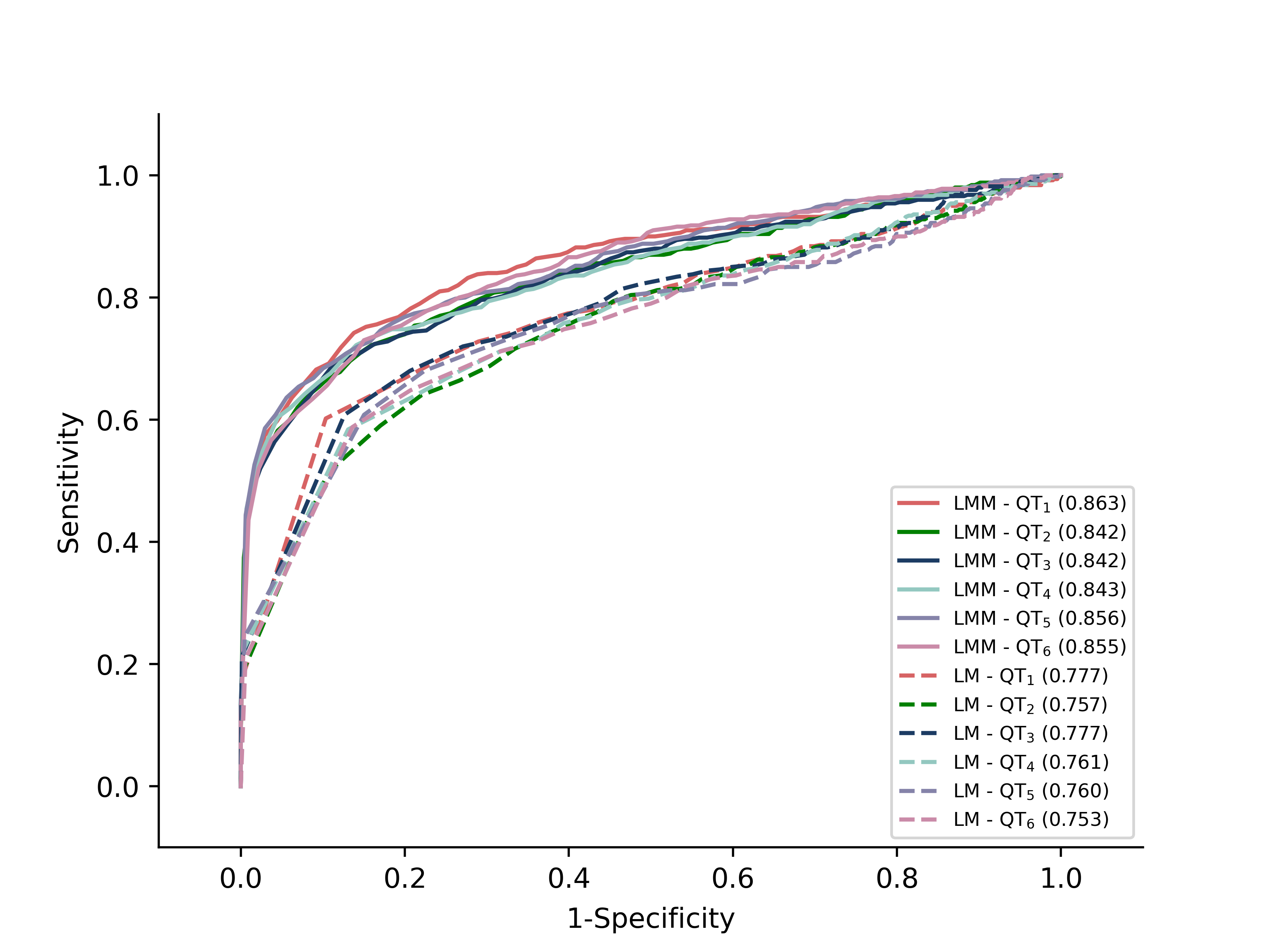}
    \label{subfig:case5_n=50_average_p}
    \subcaption{$n=50$ }
  \end{subfigure}
\par
\begin{subfigure}{0.49\linewidth}
    \includegraphics[width=\linewidth]{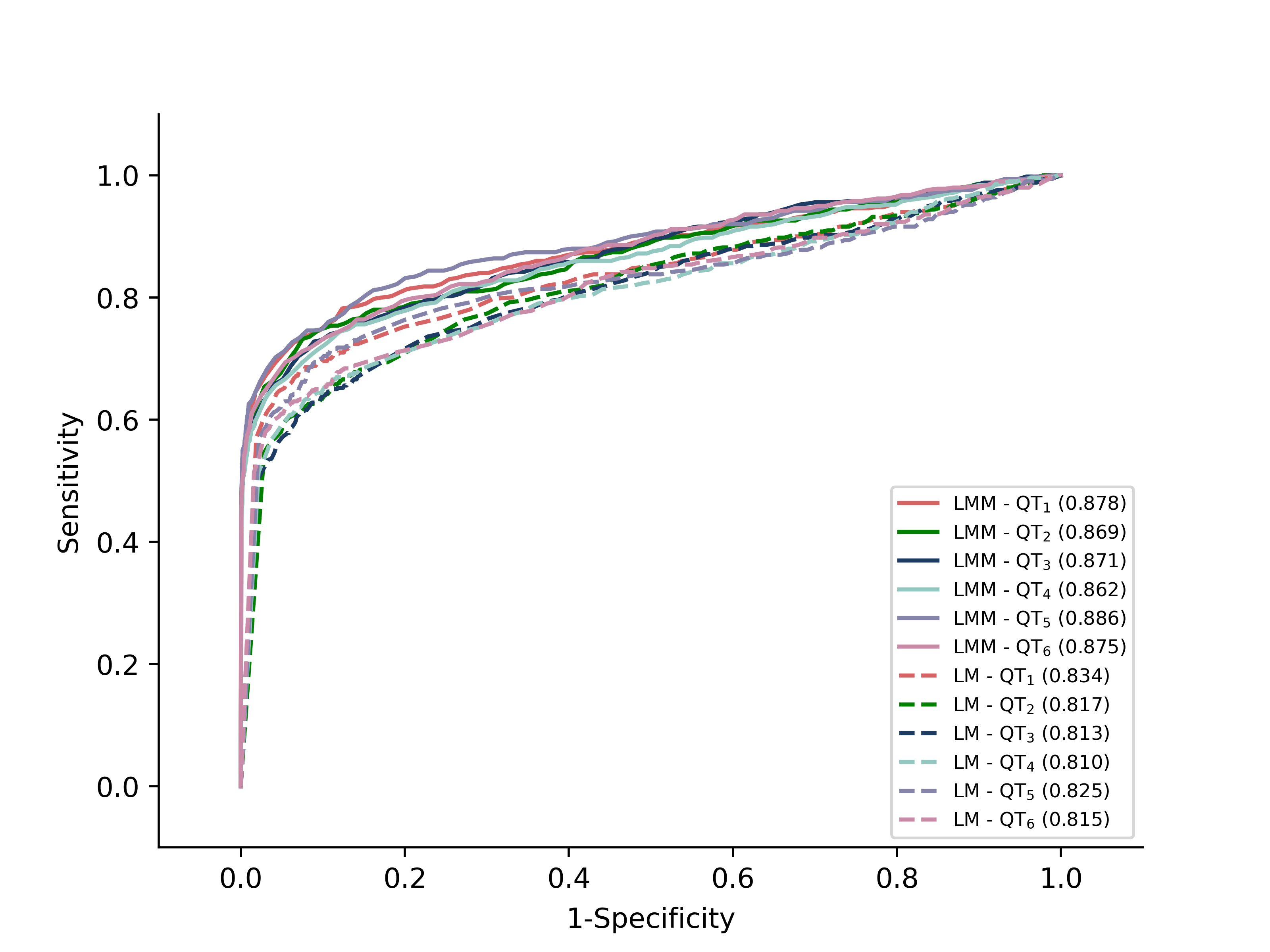}
    \label{subfig:case5_n=100_average_i}
    \subcaption{$n=100$ }
  \end{subfigure}
  \hfill
  \begin{subfigure}{0.49\linewidth}
    \includegraphics[width=\linewidth]{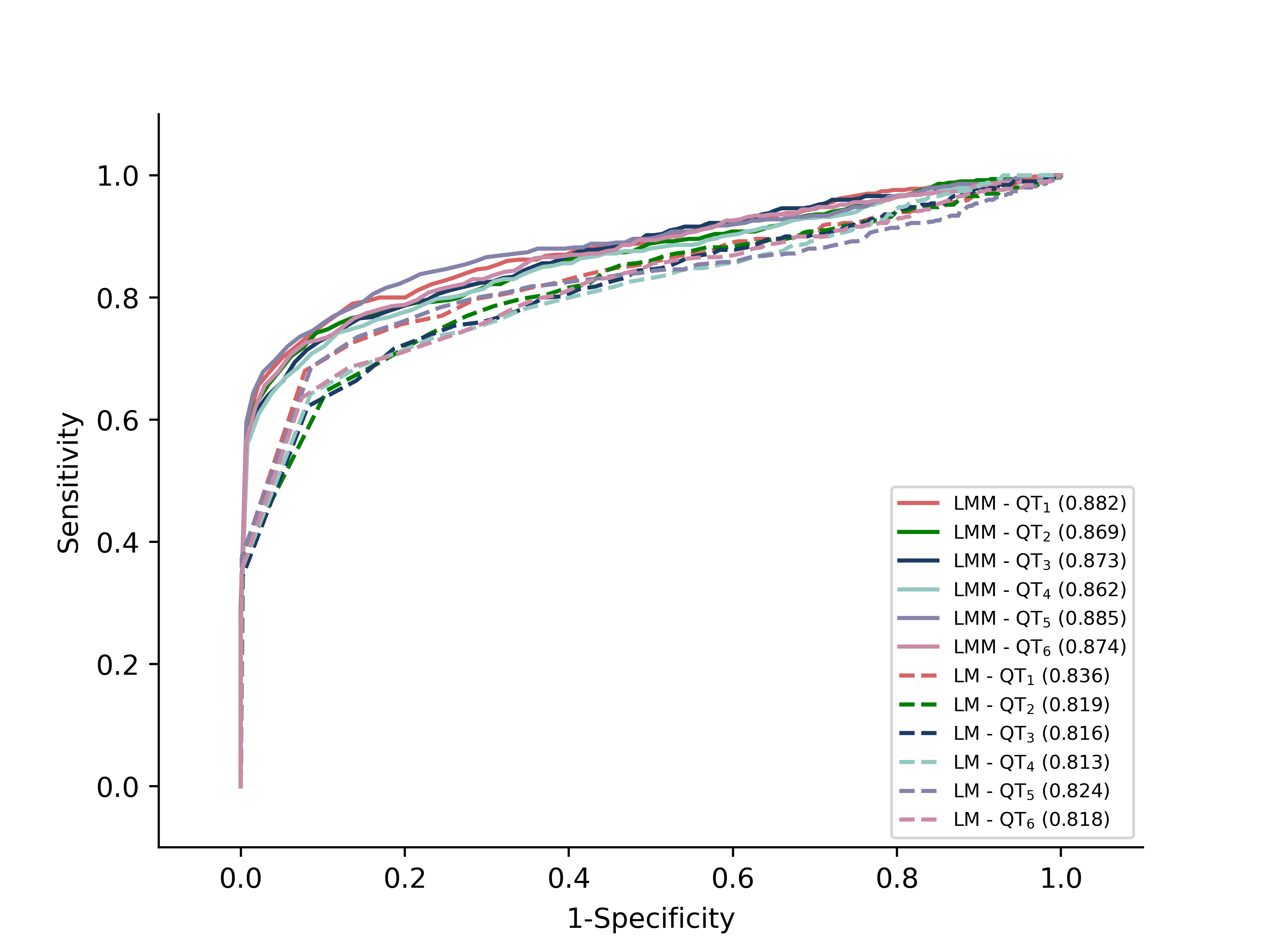}
    \label{subfig:case5_n=100_average_p}
    \subcaption{$n=100$ }
  \end{subfigure}
\caption{Case 3. ROC curves of both the LMM and the LM based on credible intervals (a,c) and the CCT ({b},d) when there was a complex dependency presented among phenotypes with sample size varies. Curves of QT$_1,\ldots$, QT$_6$ represents the ROC curves concerning QT$_1, \ldots$, QT$_6$. The figures in  brackets indicated the corresponding AUC for each curve.}
  \label{fig:spc4}
\end{figure}

In \textbf{Case 3}, we simulated a more sophisticated scenario. The phenotypes could be divided into $3$ pairs, we used a Kronecker product to represent the dependence among these $6$ phenotypes. The dependency matrix $G$ was set to  
$\begin{bmatrix}
    1 & 0.6 \\
    0.6 & 1 \\
\end{bmatrix}
\otimes
\begin{bmatrix}
    1 & 0.1 & 0.4 \\
    0.1 & 0.6 & 0.1 \\
    0.4 & 0.1 & 0.5 \\
\end{bmatrix}$.  We repeated this experiment $100$ times and computed the averaged point-wise sensitivity and specificity, and obtained the ROC curves as shown in Figure \ref{fig:spc4}. As indicated in Figure \ref{fig:spc4},  the LMM outperformed the LM with concerning metric AUC when complex dependency structures of QTs were presented.  

In \textbf{Cases 4, 5}, we set QTs $p=2$, and the dependency matrix $G$ to be 
$\begin{bmatrix}
    1 & 0.2 \\
    0.2 & 1 \\
\end{bmatrix}$ (weak),
and 
$\begin{bmatrix}
    1 & 0.8 \\
    0.8 & 1 \\
\end{bmatrix}$ (strong), respectively. In these two cases, we repeated each experiment $100$ and recorded the AUC of both LMM and LM in each experiment.

\begin{figure}
  \centering
  \begin{subfigure}{0.49\linewidth}
    \includegraphics[width=\linewidth]{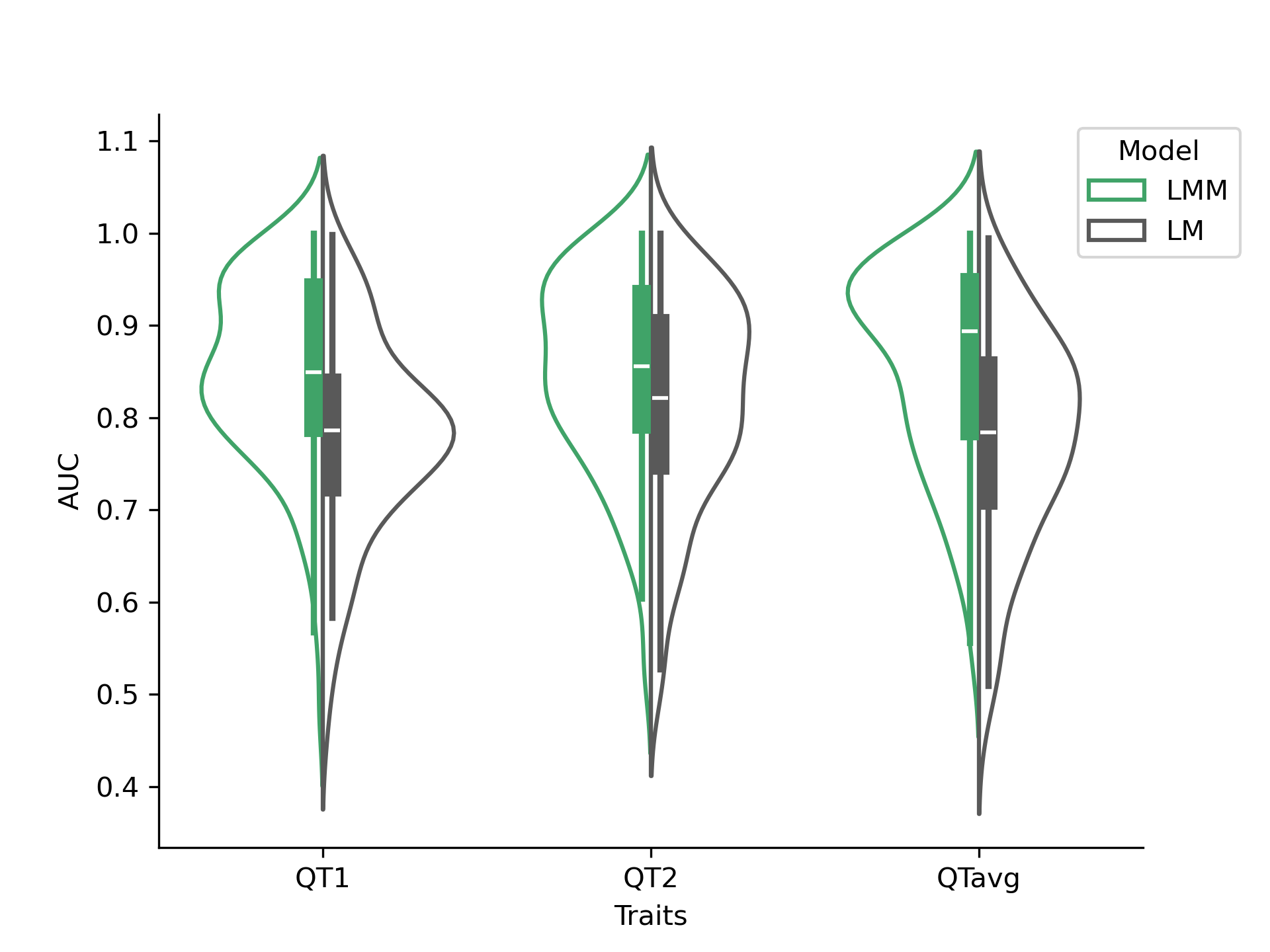}
    \label{subfig:case3_n=100_average_i}
    \subcaption{based on credible interval}
  \end{subfigure}
  \hfill
  \begin{subfigure}{0.49\linewidth}
    \includegraphics[width=\linewidth]{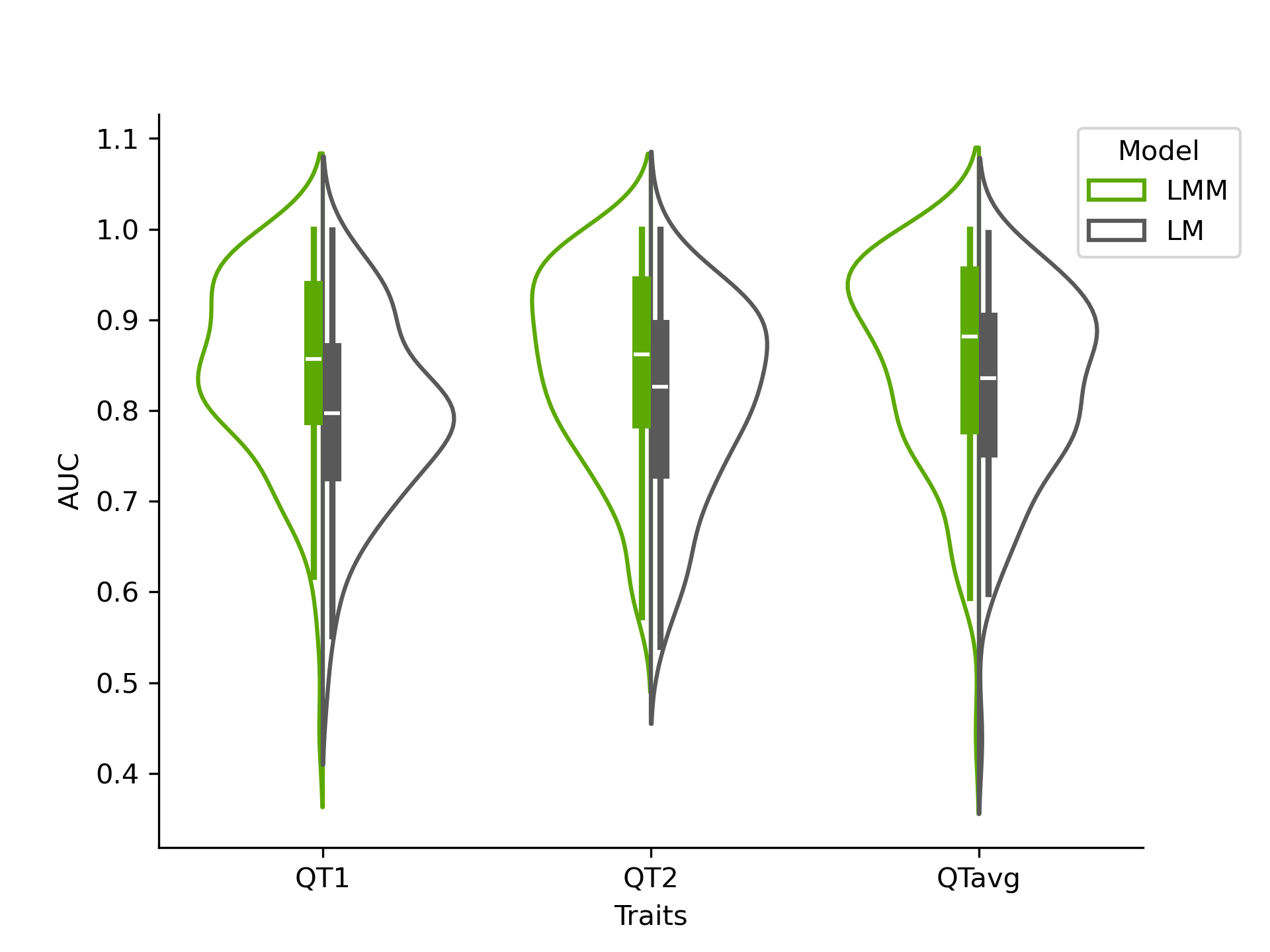}
    \label{subfig:case3_n=100_average_p}
    \subcaption{based on mcmc {\textit{P-}value}}
  \end{subfigure}
\par
\begin{subfigure}{0.49\linewidth}
    \includegraphics[width=\linewidth]{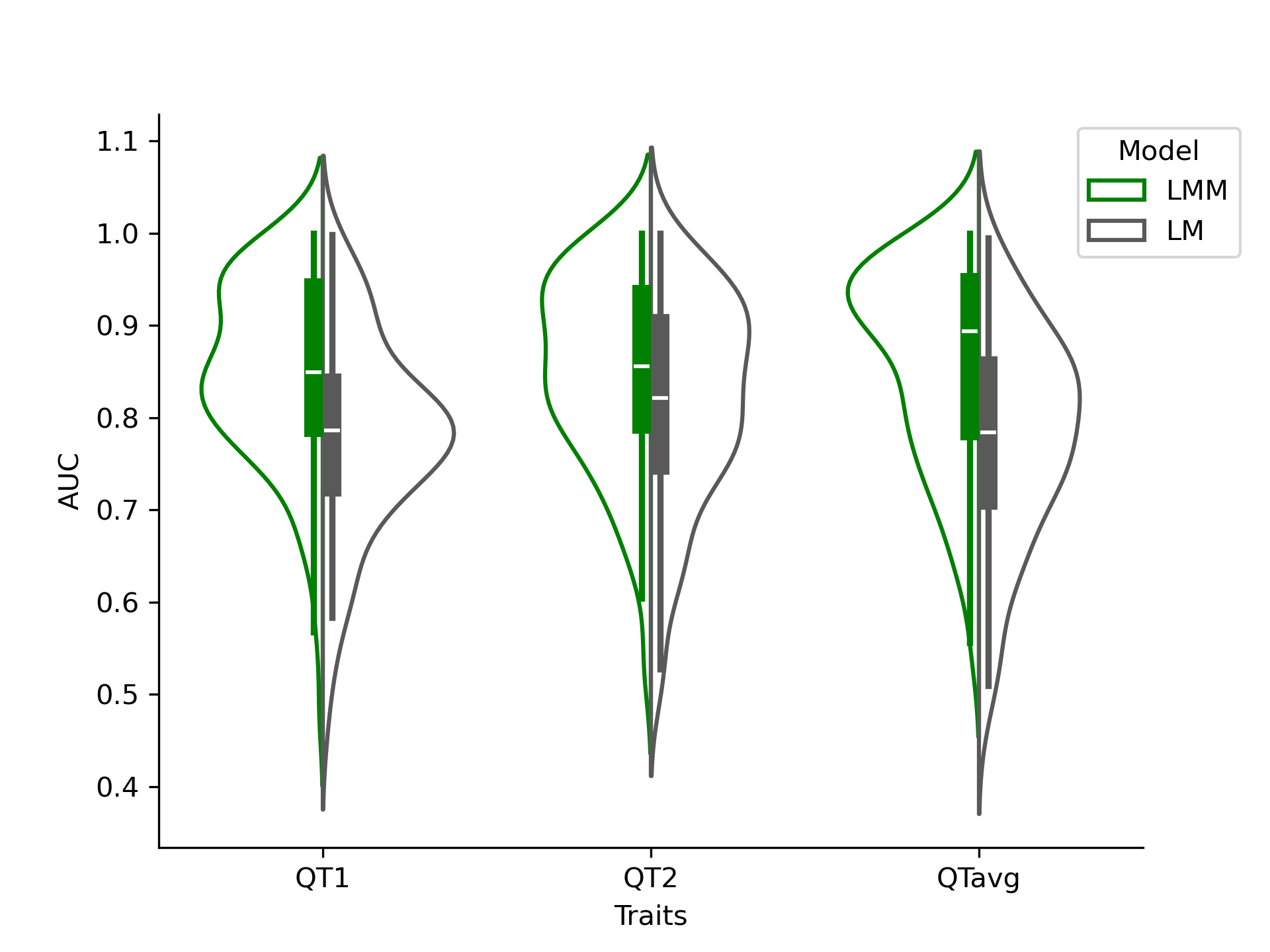}
    \label{subfig:case4_n=100_average_i}
    \subcaption{based on credible interval}
  \end{subfigure}
  \hfill
  \begin{subfigure}{0.49\linewidth}
    \includegraphics[width=\linewidth]{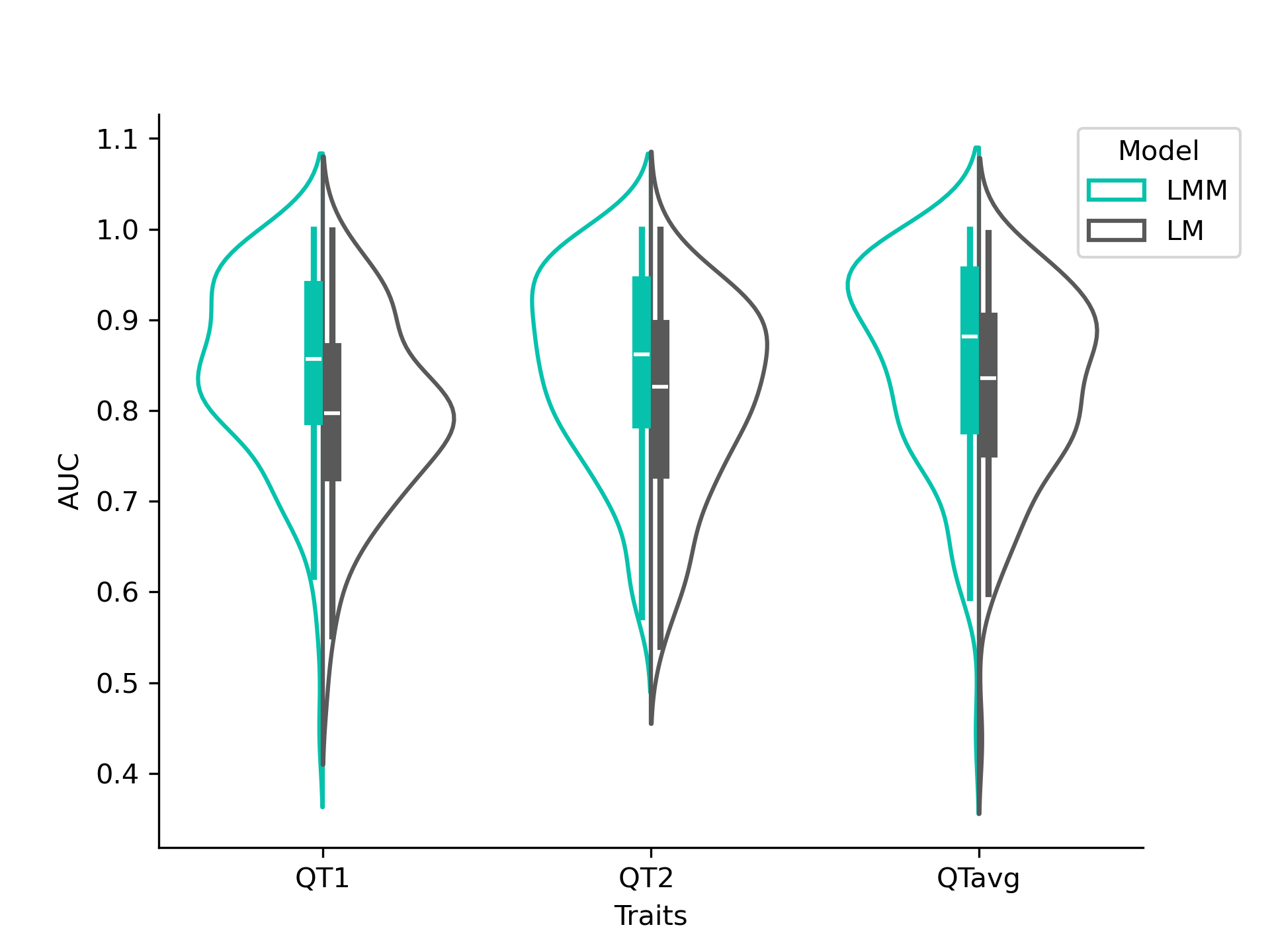}
    \label{subfig:case4_n=100_average_p}
    \subcaption{based on mcmc {\textit{P-}value}}
  \end{subfigure}
\caption{Cases 4 and 5. Violin plots of averaged AUCs over QTs when correlations among phenotypes were weak (a,b) and strong (c,d) with sample size $n=100$.}
  \label{fig:spc3}
\end{figure}

The AUC results were shown in violin plots in Figure \ref{fig:spc3}, indicating that the LMM performed better than LM with respect to metric AUC. Tables  \ref{tab:table1}, \ref{tab:table2} summarized the mean and standard deviation of the AUCs. We also performed the Wilcoxon Rank Sum and Signed Rank tests to identify if there existed a significant difference concerning metric AUC between the LMM and LM, and the result was reported in Tables \ref{tab:table3}, \ref{tab:table4}. As Tables  \ref{tab:table3}, \ref{tab:table4} showed, the LMM performed significantly better than the LM with respect to metric AUC based on both the CCT and credible intervals. {The computation time for each setting of the simulations was reported in Table 1 in Appendix B.}

{
In addition, we explored the performance of the LMM and LM when the true underlying distributions of the mixed effect term and error term were not normally distributed but heavy tailed, i.e., followed a multivariate t-distribution.  We repeated the data simulating process for \textbf{Case 1} to \textbf{5} according to Eq. \ref{con: sim}, all the data generating procedures were the same as before except that 
$$\bm{\epsilon}_i \sim Multi-t(v, \sigma_e^2 I_p),$$ 
$$\textbf{h} \sim Multi-t(v, \sigma_p^2 G).$$ 
Here we set the degree parameter $v =3$. The results were shown in Figures 1-5 in Appendix C . We did not observe a significant decrease in performance regarding the AUC metric when the model was misspecified for the LMM and LM, compared to the correctly specified cases. Again, improved performance of the LMM over the LM in terms of the metric AUC based on both the credible intervals and aggregated \textit{P-}values was observed. }

\begin{figure}
  \centering
  \begin{subfigure}{0.45\linewidth}
    \includegraphics[width=\linewidth]{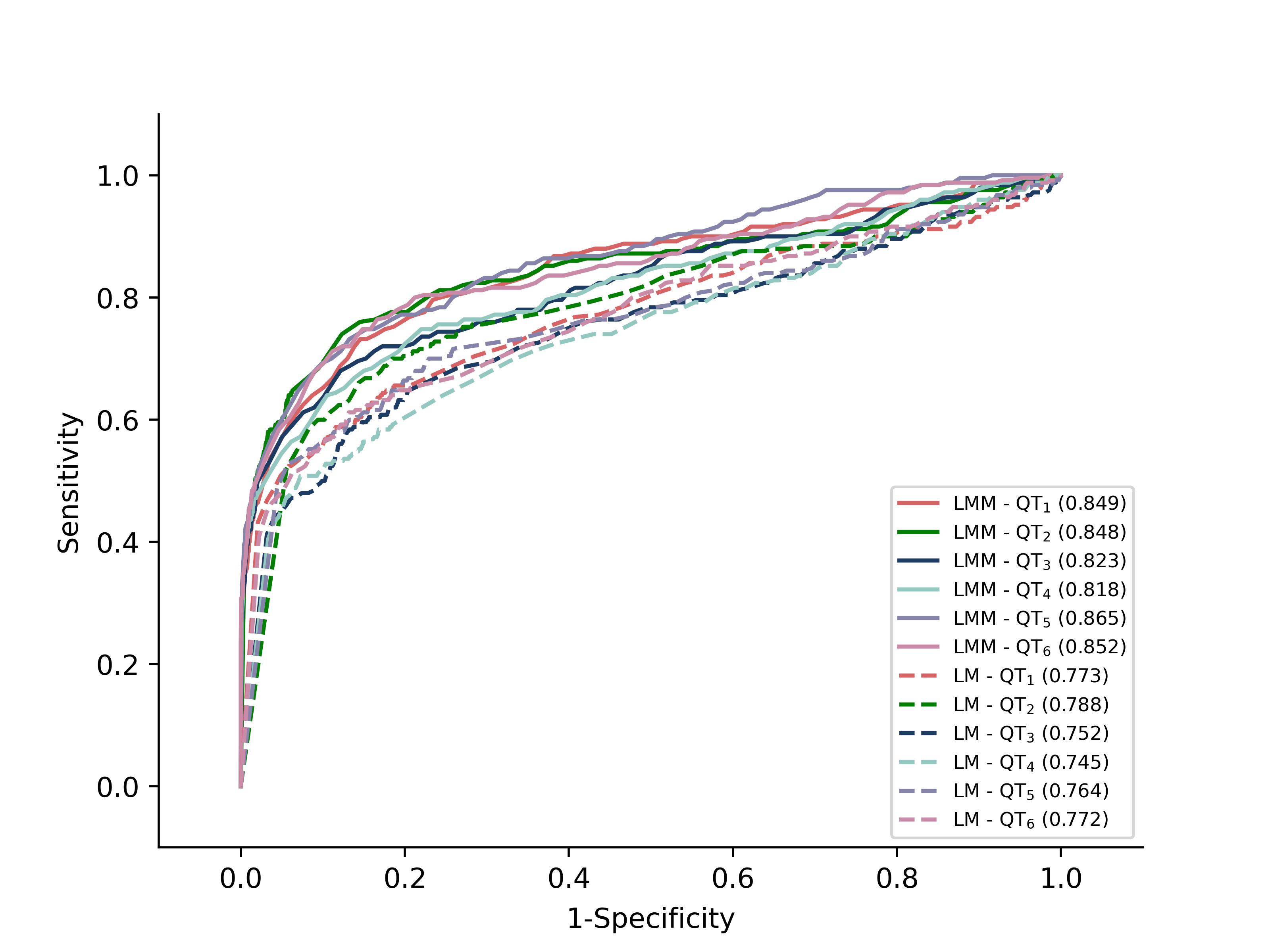}
    \label{subfig:i_n=50_average_adj}
    \subcaption{$n=50$}
  \end{subfigure}
  \hfill
  \begin{subfigure}{0.45\linewidth}
    \includegraphics[width=\linewidth]{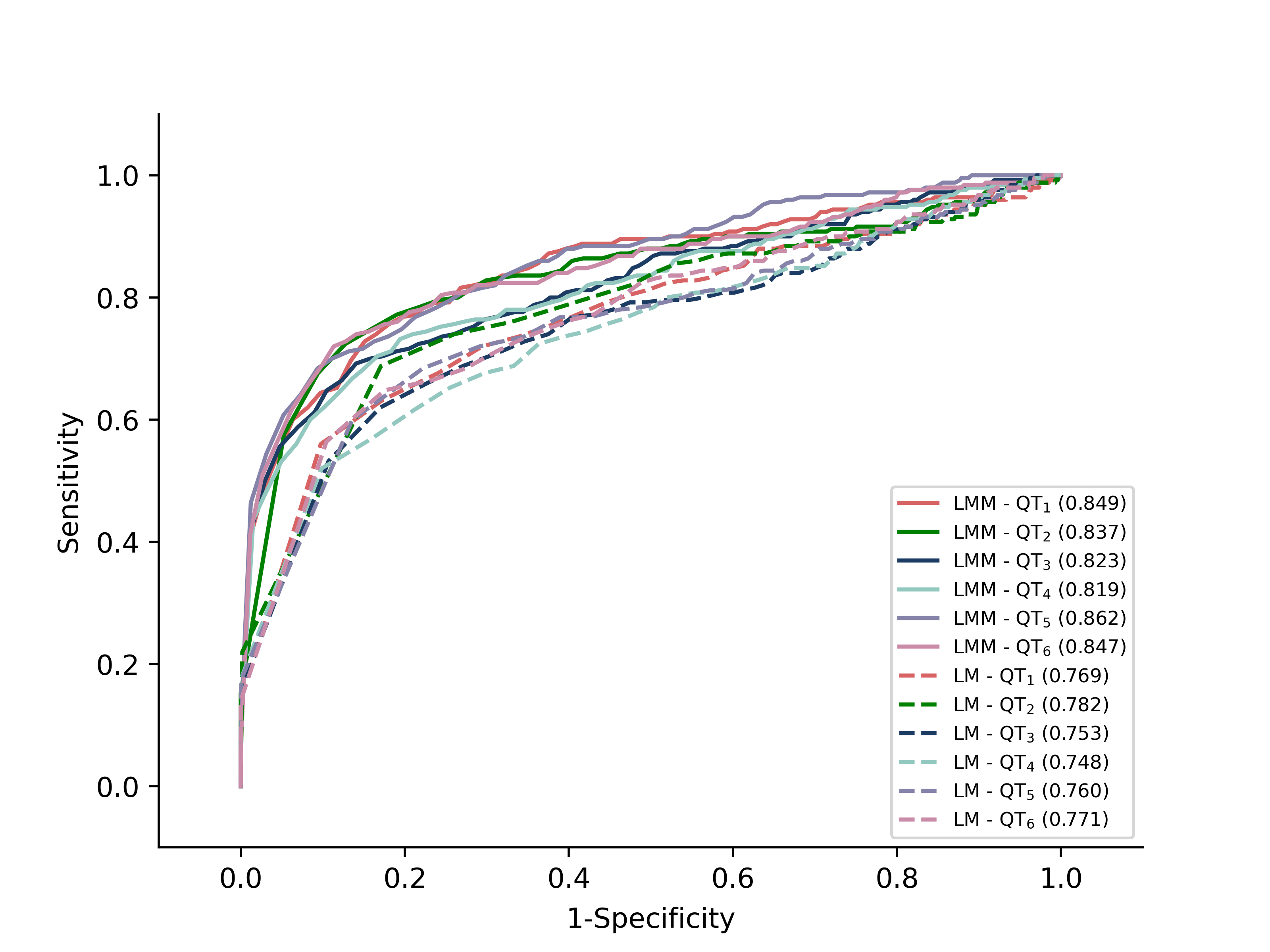}
    \label{subfig:p_n=50_average_adj}
    \subcaption{$n=50$}
  \end{subfigure}
\par
 \begin{subfigure}{0.45\linewidth}
    \includegraphics[width=\linewidth]{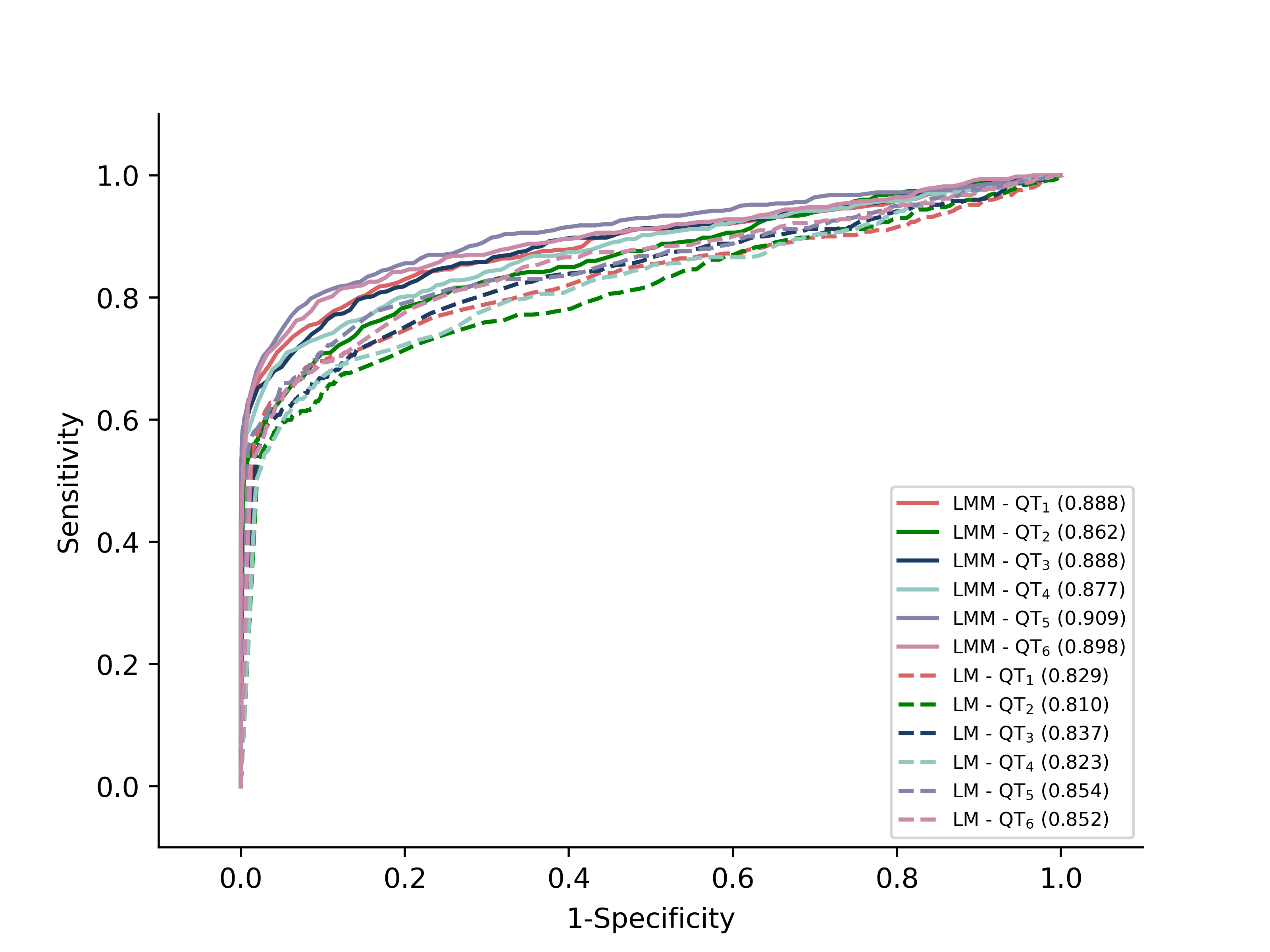}
    \label{subfig:i_n=100_average_adj}
    \subcaption{$n=100$}
  \end{subfigure}
  \hfill
  \begin{subfigure}{0.45\linewidth}
    \includegraphics[width=\linewidth]{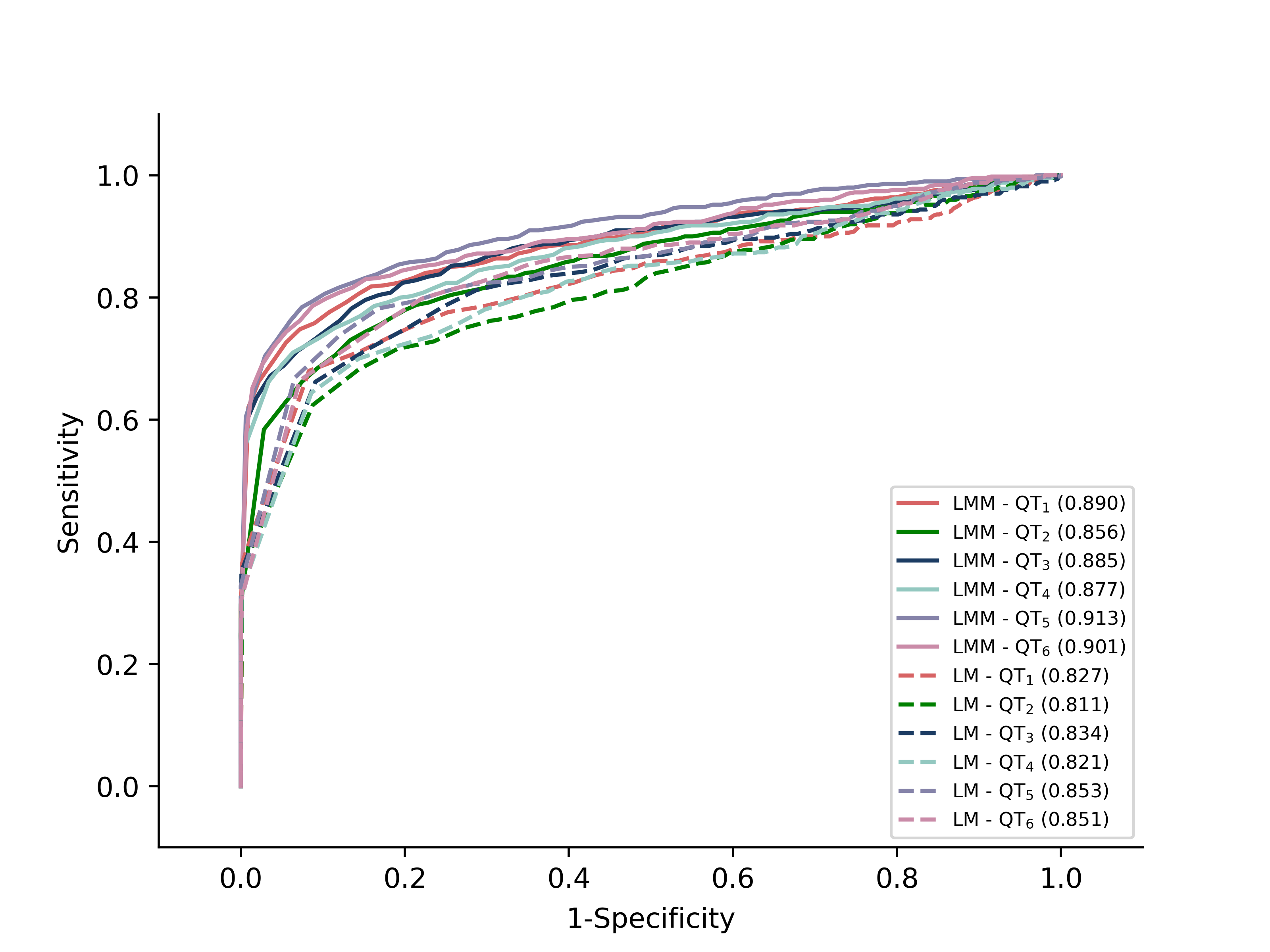}
    \label{subfig:p_n=100_average_adj}
    \subcaption{$n=100$}
 \end{subfigure}
    \caption{{Case 6. ROC curves of the LMM and LM based on credible intervals (a,c) and the CCT (b,d) when the covariance between phenotypes was specified by an adjacent matrix with element $1$ indicating the existence of dependence. Curves of QT$_1$, QT$_2$ represented the ROC curves concerning QT$_1$ and QT$_2$.  The figures in brackets indicated the corresponding AUC values for each curve. } }  
  \label{fig:spc6}
\end{figure}

{Moreover, in \textbf{Case 6}, we considered the scenarios where the covariance matrix was specified by an adjacent matrix. In this case, $G$ was set to $A + \rho I_{p}$, where the first term $A = \begin{bmatrix}
     1 & 1 & 0 & 0 & 0 & 0\\
     1 & 1 & 1 & 1 & 0 & 0\\
     0 & 1 & 1 & 0 & 0 & 0\\
     0 & 1 & 0 & 1 & 0 & 0\\
     0 & 0 & 0 & 0 & 1 & 1\\
     0 & 0 & 0 & 0 & 1 & 1\\
    \end{bmatrix}$ was an adjacency matrix with element $1$ indicating the existence of dependency between phenotypes. $\rho$ was a small bias (satisfying $\rho > -\lambda$ where $\lambda$ ranges over the eigenvalues of $A$, and we set it as $0.8$) to ensure $G$ was positive definite \citep{lehmann2021characterising}. We repeated this experiment $100$ times and computed the averaged point-wise sensitivity and specificity, and obtained the ROC curves as shown in Figure \ref{fig:spc6}, which indicated the LMM outperformed the LM  concerning the metric AUC when neighbourhood structures of QTs were presented.}

In summary, the above simulation studies showed that LMM performed better than the LM concerning metric AUC when there was a dependency structure among the phenotypes, and their performance was similar when there was no spatial correlation among QTs. In addition, Figures \ref{fig:sp1}-{\ref{fig:spc6}} and Tables \ref{tab:table1}-\ref{tab:table4} indicated that the association test results based on CCT and the credible intervals were consistent. 

\begin{table}
  \begin{center}
    \caption{Mean and standard deviation of AUC values of the 100 repeated experiments for both the LM and LMM based on the credible interval. Values of $QT_1, QT_2$, $QT_{avg}$ correspond to the values of averaged AUC over $100$ experiments concerning QT$_1$, QT$_2$ and their average. } 
    \label{tab:table1}
    \begin{tabular}{cccccccc}
    \hline
          & & \multicolumn{3}{|c}{{LMM}}   &\multicolumn{3}{|c}{{LM}} 
       \\
           \hline 
      & & \multicolumn{1}{|c}{$QT_1$} & \multicolumn{1}{c}{$QT_2$} & \multicolumn{1}{c|}{$QT_{avg}$} &\multicolumn{1}{|c}{$QT_1$} &
       \multicolumn{1}{c}{$QT_2$} &
       \multicolumn{1}{c}{$QT_{avg}$} \\
    \hline
    \multirow{2}{*}{\textbf{Case 4}} & \multicolumn{1}{c|}{mean} & 0.84 & 0.84 & \multicolumn{1}{c|}{0.86} & 0.80 & 0.83 & 0.84 \\
    & \multicolumn{1}{c|}{sd} & 0.11 & 0.11 & \multicolumn{1}{c|}{0.11} & 0.11 & 0.10 & 0.11 \\
    \hline 
    \multirow{2}{*}{\textbf{Case 5}} & \multicolumn{1}{c|}{mean}& 0.85 & 0.85 & \multicolumn{1}{c|}{0.86} & 0.78 & 0.81 & \multicolumn{1}{c}{0.78}\\
    & \multicolumn{1}{c|}{sd} & 0.10 & 0.11 & \multicolumn{1}{c|}{0.11} & 0.11 & 0.12 & 0.12\\
    \hline 
    \end{tabular} 
  \end{center}
\end{table}

\begin{table}
  \begin{center}
    \caption{Mean and standard deviation of AUC values of the $100$ repeated experiments for both the LM and LMM based on based on the CCT. Values of $QT_1, QT_2$, $QT_{avg}$ correspond to the values of averaged AUC over $100$ experiments concerning QT$_1$, QT$_2$ and their average. }
    \label{tab:table2}
    \begin{tabular}{cccccccc}
       \hline
          & & \multicolumn{3}{|c}{{LMM}}   &\multicolumn{3}{|c}{{LM}} 
       \\
           \hline 
      & & \multicolumn{1}{|c}{$QT_1$} & \multicolumn{1}{c}{$QT_2$} & \multicolumn{1}{c|}{$QT_{avg}$} &\multicolumn{1}{|c}{$QT_1$} &
       \multicolumn{1}{c}{$QT_2$} &
       \multicolumn{1}{c}{$QT_{avg}$} \\
    \hline
    \multirow{2}{*}{\textbf{Case 4}} & \multicolumn{1}{c|}{mean} & 0.84 & 0.84 & \multicolumn{1}{c|}{0.86} & 0.80 & 0.82 & 0.84 \\
    & \multicolumn{1}{c|}{sd} & 0.11 & 0.11 & \multicolumn{1}{c|}{0.10} & 0.09 & 0.10 & 0.10 \\
    \hline 
    \multirow{2}{*}{\textbf{Case 5}} & \multicolumn{1}{c|}{mean}& 0.85 & 0.85 & \multicolumn{1}{c|}{0.86} & 0.80 & 0.81 & \multicolumn{1}{c}{0.83}\\
    & \multicolumn{1}{c|}{sd} & 0.10 & 0.10 & \multicolumn{1}{c|}{0.11} & 0.10 & 0.11 & 0.10\\
    \hline 
    \end{tabular} 
  \end{center}
\end{table}

\begin{table}
  \begin{center}
    \caption{{\textit{P-}values} of the Wilcoxon Rank Sum and Signed Rank test of AUC of the LM v.s. LMM across the 100 repeated experiments based on the credible interval. The alternative hypothesis was the AUC of LMM was larger than that of the LM. {\textit{P-}values} of $QT_1, QT_2$, $QT_{avg}$ correspond to the values of concerning AUC of QT$_1$, QT$_2$ and their average.} 
    \label{tab:table3}
    \begin{tabular}{ccccccccccccc}
    \hline
      &\multicolumn{4}{|c}{$QT_1$} & \multicolumn{4}{|c}{$QT_2$} & \multicolumn{4}{|c}{$QT_{avg}$} \\
    \hline
    \multicolumn{1}{c|}{\textbf{Case 4}} &
    \multicolumn{4}{|c}{$4.46 \times 10^{-5}$} & 
    \multicolumn{4}{|c}{$3.00 \times 10^{-2}$} & 
    \multicolumn{4}{|c}{$9.20\times 10^{-9}$} \\
    \hline 
    \multicolumn{1}{c|}{\textbf{Case 5}} &
    \multicolumn{4}{|c}{$2.21\times 10^{-9}$} & 
    \multicolumn{4}{|c}{$1.37 \times 10^{-5}$} & 
    \multicolumn{4}{|c}{$4.27 \times 10^{-9}$} \\
    \hline 
    \end{tabular} 
  \end{center}
\end{table}

\begin{table}
  \begin{center}
    \caption{{\textit{P-}values} of the Wilcoxon Rank Sum and Signed Rank test of AUC of the LM v.s. LMM across the 100 repeated experiments based on the CCT. The alternative hypothesis was the AUC of LMM was larger than that of the LM. {\textit{P-}values} of $QT_1, QT_2$, $QT_{avg}$ correspond to the values of concerning AUC of QT$_1$, QT$_2$ and their average. 
    } 
    \label{tab:table4}
    \begin{tabular}{ccccccccccccc}
    \hline
      &\multicolumn{4}{|c}{$QT_1$} & \multicolumn{4}{|c}{$QT_2$} & \multicolumn{4}{|c}{$QT_{avg}$} \\
    \hline
    \multicolumn{1}{c|}{\textbf{Case 4}} &
    \multicolumn{4}{|c}{$5.51\times 10^{-6}$} & 
    \multicolumn{4}{|c}{$1.98\times 10^{-2}$} & 
    \multicolumn{4}{|c}{$1.38\times 10^{-2}$} \\
    \hline 
    \multicolumn{1}{c|}{\textbf{Case 5}} &
    \multicolumn{4}{|c}{$3.20 \times 10^{-9}$} & 
    \multicolumn{4}{|c}{$8.76 \times 10^{-7}$} & 
    \multicolumn{4}{|c}{$3.12 \times 10^{-5}$} \\
    \hline 
    \end{tabular} 
  \end{center}
\end{table}

\section{Application}

We applied our proposed method to an imaging genetics dataset collected
from the ADNI-1 database\footnote{Data used in the preparation of this article were obtained from the Alzheimer’s Disease Neuroimaging
Initiative (ADNI) database (adni.loni.usc.edu). The ADNI was launched in 2003 as a public-private partnership, led by Principal Investigator Michael W. Weiner, MD. The primary goal of ADNI has been to
test whether serial magnetic resonance imaging (MRI), positron emission tomography (PET), other biological markers, and clinical and neuropsychological assessment can be combined to measure the
progression of mild cognitive impairment (MCI) and early Alzheimer’s disease (AD).}. The imaging genetic data contained 632 individuals. After quality control and imputation, we included 486 SNPs from 33 of the top 40 Alzheimer’s Disease candidate genes listed on the AlzGene database as of June 10, 2010 \citep{song2022bayesian}. The imaging derived QTs included in this application study are thickness of the Supramarginal gyrus (Supramarg) and the Superior Temporal gyrus (SupTemporal) on both the left and right hemispheres. The average correlation between Supramarg and SupTemporal was used to describe the relationship between pairs, and the average correlation between the left and right Supramarg and SupTemporal was used to represent the correlation within each pair. Their Kronecker product was used to describe the dependency among these two QT pairs. To mimic the scenario where the number of SNPs was greater than the sample size, we randomly selected $100$ individuals from the $632$ individuals. {The total computational time for the real data application was $1.08$ hours with $5000$ MCMC iterations, running with a single core (3.20-GHz AMD Ryzen 7 7735H) on a computing cluster with 16GB of RAM.}

Table \ref{tab:table5} listed the 19 SNPs selected from the proposed Bayesian spatial LMM based on the CCT at $\alpha=0.05$ with a Bonferroni Correction. The SNP ID in bold indicated that it was also reported in existing literature and the corresponding references were listed in the fourth column in Table \ref{tab:table5}. For example, the genetic marker  $rs16871157$ was also identified in relative studies, such as \cite{song2022bayesian}, \cite{kundu2016semiparametric} and \cite{choi2019contribution}. Our model also identified several new genetic sites not reported in previous studies,  such as $rs212515$ and $rs1251753$, which may provide new insights into Alzheimer's disease. {Among the four phenotypes we used, the SupTemporal (left and right) mainly involved in the production, interpretation and self-monitoring of language, while dysfunction of this region might cause auditory hallucinations and thought disorder \citep{sun2009superior}. The Supramarg (left and right) played an active role in phonological processing during both language and verbal working memory tasks \citep{deschamps2014role}. Among these 19 SNPs, $rs2025935$ located in CR1 was the key significant molecular factor to modulate tau pathology and cause reductions in cortical thickness of the superior temporal gyrus. Thus, it was recognized as an important risk locus for late-onset Alzheimer’s disease \citep{chibnik2011cr1, hazrati2012genetic, zhu2015cr1}. DAPK1 was detected as a significant mediator of cell death and synaptic damage in central nervous system, thus was related to AD \citep{li2006dapk1, hazrati2012genetic, chen2019death}. The detected SNPs $rs1014306$, $rs10780849$, $rs1105384$ and $rs1473180$ indicated that DAPK1 could affect brain regions in the right hemisphere more significantly than the left part. Besides, SNPs we found located in ECE1, like $rs212515$, $rs213023$, $rs213025$ and $rs471359$,  showed significant effect on the left SupTemporal and Supramarg, indicating that these mutations might play a more important role in the left brain hemisphere than the right. This indication in term of spatial differences across the hemispheres was also mentioned in \cite{hoshi2023decreased}. There was little evidence supporting the relatedness between these genetic mutations and AD in literature, these new findings might warrant further investigation. The protein encoded by NEDD9 was identified as one of the signalling proteins in Alzheimer's disease \citep{xing2011nedd9, beck2014adaptors}, and SORCS1 was found genetically associated with hippocampal volume or grey matter density changes accounting for APOE \citep{xu2013genetic}. }{As shown by the results,  $rs3739784$ in gene DAPK1, $rs12758257$ in gene NEDD9, $rs10787010$ in gene SORCS1 were significantly correlated with the SupTemporal phenotype (left and right).  Additionally, it suggested non-significant differences in the genetic effects across hemispheres for these mutations.
}

\begin{table}[h!]
  \begin{center}
    \caption{ADNI-1 Study: Selected SNPs and the Corresponding Regions of Interest}
    \label{tab:table5}
    \begin{tabular}{cccc}
    \hline
      \multicolumn{1}{c}{SNP}
      &\multicolumn{1}{|c}{Gene} & \multicolumn{1}{|c}{Phenotype (hemisphere)} & \multicolumn{1}{|c}{Reference}\\
    \hline
    \multicolumn{1}{c}{\textbf{rs2025935}} & 
    \multicolumn{1}{|c}{CR1} & 
    \multicolumn{1}{|c}{SupTemporal (right)} & 
    \multicolumn{1}{|c}
    {
     \makecell[c]{
     \cite{greenlaw2017bayesian} \\ 
     \cite{zhu2017effect} \\
     \cite{song2022bayesian} 
    }
    }
    \\
    \multicolumn{1}{c}{\textbf{rs1014306}} & 
    \multicolumn{1}{|c}{DAPK1} & 
    \multicolumn{1}{|c}{Supramarg (right)} & 
    \multicolumn{1}{|c}{
    \makecell[c]{
     \cite{phillips2013longitudinal} \\ 
     \cite{asensio2022association} 
    }}
    \\

    \multicolumn{1}{c}{\textbf{rs10780849}} & 
    \multicolumn{1}{|c}{DAPK1} & 
    \multicolumn{1}{|c}{SupTemporal (right)} &
    \multicolumn{1}{|c}{    
    \makecell[c]{
     \cite{phillips2013longitudinal} \\ 
     \cite{song2022bayesian} 
    }}
    \\

    \multicolumn{1}{c}{\textbf{rs1105384}} & 
    \multicolumn{1}{|c}{DAPK1} & 
    \multicolumn{1}{|c}{SupTemporal (right)} &
    \multicolumn{1}{|c}{    
    \makecell[c]{
     \cite{kundu2016semiparametric} \\ 
     \cite{song2022bayesian} 
    }}
    \\

    \multicolumn{1}{c}{\textbf{rs12378686}} & 
    \multicolumn{1}{|c}{DAPK1} & 
    \multicolumn{1}{|c}{Supramarg (right)} &
    \multicolumn{1}{|c}{    
    \makecell[c]{
     \cite{watza2020copd} \\ 
     \cite{choi2019contribution} 
    }}
    \\

    \multicolumn{1}{c}{\textbf{rs1473180}} & 
    \multicolumn{1}{|c}{DAPK1} & 
    \multicolumn{1}{|c}{SupTemporal (left)} &
    \multicolumn{1}{|c}{\makecell[c]{
     \cite{greenlaw2017bayesian} \\ 
     \cite{song2022bayesian} 
    }}
    \\

    \multicolumn{1}{c}{\textbf{rs1558889} }& 
    \multicolumn{1}{|c}{DAPK1} & 
    \multicolumn{1}{|c}{Supramarg (left)} &
    \multicolumn{1}{|c}{\makecell[c]{
     \cite{laumet2010systematic}
    }}
    \\

    \multicolumn{1}{c}{rs3028} & 
    \multicolumn{1}{|c}{DAPK1} & 
    \multicolumn{1}{|c}{SupTemporal (right)} &
    \multicolumn{1}{|c}{-}
    \\

    \multicolumn{1}{c}{\textbf{rs3739784}} & 
    \multicolumn{1}{|c}{DAPK1} & 
    \multicolumn{1}{|c}{SupTemporal (left, right)} &
    \multicolumn{1}{|c}{\makecell[c]{
     \cite{choi2019contribution} \\
    \cite{beaulac2023neuroimaging}
    }}
    \\

    \multicolumn{1}{c}{\textbf{rs12758257}} & 
    \multicolumn{1}{|c}{ECE1} & 
    \multicolumn{1}{|c}{SupTemporal (left, right)} &
    \multicolumn{1}{|c}{\cite{beaulac2023neuroimaging}}
    \\

    \multicolumn{1}{c}{{rs212515}} & 
    \multicolumn{1}{|c}{ECE1} & 
    \multicolumn{1}{|c}{SupTemporal (left)} &
    \multicolumn{1}{|c}{-}
    \\

    \multicolumn{1}{c}{{rs213023}} & 
    \multicolumn{1}{|c}{ECE1} & 
    \multicolumn{1}{|c}{SupTemporal (left)} &
    \multicolumn{1}{|c}{-}
    \\

    \multicolumn{1}{c}{{rs213025}} & 
    \multicolumn{1}{|c}{ECE1} & 
    \multicolumn{1}{|c}{SupTemporal (left)} &
    \multicolumn{1}{|c}{-}
    \\

    \multicolumn{1}{c}{{rs471359}} & 
    \multicolumn{1}{|c}{ECE1} & 
    \multicolumn{1}{|c}{Supramarg (left)}&
    \multicolumn{1}{|c}{-}
    \\

    \multicolumn{1}{c}{\textbf{rs10947021}} & 
    \multicolumn{1}{|c}{NEDD9} & 
    \multicolumn{1}{|c}{SupTemporal (right)} &
    \multicolumn{1}{|c}{\cite{laumet2010systematic}}
    \\

    \multicolumn{1}{c}{\textbf{rs16871157}} & 
    \multicolumn{1}{|c}{NEDD9} & 
    \multicolumn{1}{|c}{Supramarg (left)} &
    \multicolumn{1}{|c}{\makecell[c]{
    \cite{kundu2016semiparametric} \\
    \cite{choi2019contribution} \\
    \cite{song2022bayesian} 
    }}
    \\

    \multicolumn{1}{c}{\textbf{rs6912916}} & 
    \multicolumn{1}{|c}{NEDD9} & 
    \multicolumn{1}{|c}{SupTemporal (right)} &
    \multicolumn{1}{|c}{\cite{laumet2010systematic}}
    \\

    \multicolumn{1}{c}{\textbf{rs10787010}} & 
    \multicolumn{1}{|c}{SORCS1} & 
    \multicolumn{1}{|c}{SupTemporal (left, right)}&
    \multicolumn{1}{|c}{\makecell[c]{
    \cite{greenlaw2017bayesian} \\
    \cite{song2022bayesian} 
    }}
    \\

    \multicolumn{1}{c}{rs1251753} & 
    \multicolumn{1}{|c}{SORCS1} & 
    \multicolumn{1}{|c}{Supramarg (right)} &
    \multicolumn{1}{|c}{-}
    \\
    \hline
    \end{tabular} 
  \end{center}
\end{table}

\section{Discussion}
We proposed a spatial-correlated multitask LMM to uncover potential gene markers associated with multiple correlated imaging QTs simultaneously in the Bayesian framework. The mixed-effects term accounted for the population-level dependency among QTs and enabled the model to make use of the spatial information among QTs and further boost its statistical power in association studies. In this work, we introduced the population-level dependency among QTs to avoid the unidentifiable issue usually triggered by the individual-level mixed-effects term. This might sacrifice some spatial information at the individual level, and how to model the individual-level dependency properly would be one of the directions of our future work. {We considered three forms of covariance to depict the population-level dependency, a vanilla form represented by a positive definite matrix, a Kronecker product form accounting for both correlations between phenotypes between pairs and cross pairs and an adjacency matrix which only uses element $0$ and $1$ to represent if there exists correlations between one pair of phenotypes. The Gaussian kernel has been also popular in covariance construction and it can be considered in the future work. Besides the Gaussian assumption for random errors, we also simulated phenotypes based on a multivariate t-distribution to represent the scenarios where the distribution of phenotypes was miss assumed. The results indicated that even the model was misspecified, the LMM performance was comparable to the cases when the model was not misspecified concerning the metric AUC, and the LMM outperformed the LM. In addition, applying a transformation to the phenotypes, for example, a log transformation maps the positive-valued QTs to $(-\infty, \infty)$, is a common practice. However, the resulting $\ log(QTs)$ might violate the Gaussian assumption heavily. How to model the transformed phenotypes accurately is a changing work. A possible approach might be the generalization of LMM inferring the transformation function directly from the data \citep{fusi2014warped}. }
\\

{ In addition to the standard LM, we also compared the LMM with the Bayesian Group Sparse Multi-Task Regression (BGSMTR, \citealp{song2022bayesian}). The BGSMTR was an innovative approach that explicitly modelled phenotypic correlations both within and across cerebral hemispheres using a bivariate conditional autoregressive (CAR) process. This allowed for considering spatial dependencies of imaging QTs. In addition, the BGSMTR tested grouped SNPs, by encouraging sparsity between and within SNP groupings. While this simultaneous estimation could enhance statistical power when SNPs are strongly correlated (e.g. exhibit strong linkage disequilibrium (LD) within groups) \citep{greenlaw2017bayesian}, it could also incur significant computational overhead. We conducted a direct comparison to the BGSMTR in Case 3, the results are given in Appendix D. The results (Figure 6 in Appendix D) indicated that the LMM outperformed the BGSMTR concerning the metric AUC when the SNPs were independently simulated. Meanwhile, compared to the LM, significant performance improvements were observed for both the LMM and BGSMTR, as the LM ignores the inter-trait dependencies. The computational complexity shows (see Table 2 in Appendix D), when $mn < p$, the two methods exhibit comparable theoretical costs. However, in real-world imaging-genetic studies, where $p \ll mn$ in general, the LMM would demonstrate significant efficiency. This was empirically validated in Case 3 (Figure 7 in Appendix D). In conclusion, the BGSMTR represented a valuable advanced approach to model structured genetic effects and spatial phenotype dependencies, whereas our LMM provided a balanced trade-off between statistical performance and computational feasibility for high-dimensional imaging-genetic applications. }\\

 Besides, Variational Bayes (VB) is also a widely acknowledged technique in Bayesian inference, which often serves as an alternative to MCMC sampling methods \citep{tran2021practical}.  In the future work, we will consider incorporating the VB results into the model inference to improve the computational efficiency.

\section*{Acknowledgments}
Data used in preparation of this article were obtained from the Alzheimer’s Disease Neuroimaging Initiative
(ADNI) database (adni.loni.usc.edu). As such, the investigators within the ADNI contributed to the design
and implementation of ADNI and/or provided data but did not participate in analysis or writing of this report.
A complete listing of ADNI investigators can be found at:\url{http://adni.loni.usc.edu/wp-content/uploads/how_to_apply/ADNI_Acknowledgement_List.pdf}

Data collection and sharing for the Alzheimer's Disease Neuroimaging Initiative (ADNI) is funded by the National
Institute on Aging (National Institutes of Health Grant U19 AG024904). The grantee organization is the Northern
California Institute for Research and Education. In the past, ADNI has also received funding from the National
Institute of Biomedical Imaging and Bioengineering, the Canadian Institutes of Health Research, and private
sector contributions through the Foundation for the National Institutes of Health (FNIH) including generous
contributions from the following: AbbVie, Alzheimer’s Association; Alzheimer’s Drug Discovery Foundation;
Araclon Biotech; BioClinica, Inc.; Biogen; Bristol-Myers Squibb Company; CereSpir, Inc.; Cogstate; Eisai Inc.;
Elan Pharmaceuticals, Inc.; Eli Lilly and Company; EuroImmun; F. Hoffmann-La Roche Ltd and its affiliated
company Genentech, Inc.; Fujirebio; GE Healthcare; IXICO Ltd.; Janssen Alzheimer Immunotherapy Research \&
Development, LLC.; Johnson \& Johnson Pharmaceutical Research \&Development LLC.; Lumosity; Lundbeck;
Merck \& Co., Inc.; Meso Scale Diagnostics, LLC.; NeuroRx Research; Neurotrack Technologies; Novartis
Pharmaceuticals Corporation; Pfizer Inc.; Piramal Imaging; Servier; Takeda Pharmaceutical Company; and Transition Therapeutics. 

The authors thank ShanghaiTech University for supporting this work through the
startup fund and the HPC Platform.

This article has been accepted for publication in \textit{Journal of Computational Biology}, published by Mary Ann Liebert, Inc..

\section*{Authors' contributions}
Z.P. - Conceptualization, Data curation, Formal analysis, Methodology, Software, Writing—original draft. S.G. - Conceptualization, Data Curation, Formal Analysis, Methodology, Software, Project Administration, Supervision, Writing - Review and Editing.  All authors read and approved the final manuscript.



\section*{Conflict of Interests}
The authors declare that they have no competing interests.

\section*{Funding information}
This project was supported by the Shanghai Science and Technology Program (No. 21010502500), the National Natural Science Foundation of China (12401383), the startup fund of ShanghaiTech University. and the HPC Platform of ShanghaiTech University.

\newpage

\nocite*{}
\bibliographystyle{plainnat-revised}
\bibliography{sample}

\end{document}


\maketitle 

\section*{Appendix A.  Gibbs sampling}
\label{app_gibbs}

\subsection*{A1. Derivation of the full conditional distributions of unknown parameters in the LM}
\par
Followed by the LM model describe in Section 2.1,  the joint posterior distribution of ${\bm{\beta}_1}$, ${\bm{\beta}_0}$ and $\sigma^2$ is:
\begin{equation}
\begin{aligned}
p({\bm{\beta}_1}, {\bm{\beta}_0}, \sigma^2| {\mathbf{y}_1}, \cdots, {\mathbf{y}_n}) 
&\propto \prod_{i=1}^n p({\mathbf{y}_i}|{\bm{\beta}_1}, {\bm{\beta}_0}, \sigma^2)p({\bm{\beta}_1})p(\sigma^2)p({\bm{\beta}_0})\\
&\propto \prod_{i=1}^n |2\pi\sigma^2I_p|^{-1/2} 
\exp\{-\frac{1}{2\sigma^2}({\mathbf{y}_i}-x_i{\bm{\beta}_1}-{\bm{\beta}_0})^T({\mathbf{y}_i}-x_i{\bm{\beta}_1}-{\bm{\beta}_0})\} \\
&\times |2\pi\sigma^2I_p|^{-1/2}\exp\{-\frac{1}{2}{\bm{\beta}_1}^T{\bm{\beta}_1}\} \\
&\times (\sigma^2)^{-a-1}e^{-\frac{b}{\sigma^2}} \\
&\times \exp \{ -\frac{1}{2}({\bm{\beta}_0}-{\bm{\mu}_0})^T({\bm{\beta}_0}-{\bm{\mu}_0})\}.
\end{aligned}
\end{equation}
The full conditional distributions of $\sigma^2, {\bm{\beta}_1}$ and ${\bm{\beta}_0}$ are derived as follows.
\begin{equation}
\begin{aligned}
p(\sigma^2| {\bm{\beta}_1}, {\bm{\beta}_0}, {\mathbf{y}_1}, \cdots, {\mathbf{y}_n}) 
&\propto \prod_{i=1}^n |2\pi\sigma^2I_p|^{-1/2} 
\exp\{-\frac{1}{2\sigma^2}({\mathbf{y}_i}-x_i{\bm{\beta}_1}-{\bm{\beta}_0})^T({\mathbf{y}_i}-x_i{\bm{\beta}_1}-{\bm{\beta}_0})\} \\
&\times (\sigma^2)^{-a-1}e^{-\frac{b}{\sigma^2}} \\
& = (\sigma^2)^{-a-1-np/2}
\exp\{-\frac{\frac{1}{2}\sum_{i=1}^n({\mathbf{y}_i}-x_i{\bm{\beta}_1}-{\bm{\beta}_0})^T({\mathbf{y}_i}-x_i{\bm{\beta}_1}-{\bm{\beta}_0})+b}{\sigma^2}\}.  \label{con: lm1sigma}
\end{aligned}
\end{equation}

\begin{equation}
\begin{aligned}
p({\bm{\beta}_1}|{\bm{\beta}_0}, \sigma^2, {\mathbf{y}_1}, \cdots, {\mathbf{y}_n}) 
&\propto \prod_{i=1}^n |2\pi\sigma^2I_p|^{-1/2} 
\exp\{-\frac{1}{2\sigma^2}(y_i-x_i{\bm{\beta}_1}-{\bm{\beta}_0})^T(y_i-x_i{\bm{\beta}_1}-{\bm{\beta}_0})\} \\
&\times |2\pi\sigma^2I_p|^{-1/2}\exp\{-\frac{1}{2}{\bm{\beta}_1}^T{\bm{\beta}_1}\} \\
&\propto \exp\{-\frac{1}{2}(\frac{\sum_ix_i^2}{\sigma^2}+1)
{\bm{\beta}_1}^T{\bm{\beta}_1} + \frac{1}{\sigma^2}\sum_ix_i({\mathbf{y}_i}-{\bm{\beta}_0})^T{\bm{\beta}_1}\}. \label{con: lm2beta1}
\end{aligned}
\end{equation}
and 
\begin{equation}
\begin{aligned}
p({\bm{\beta}_0}|{\bm{\beta}_1}, \sigma^2, {\mathbf{y}_1}, \cdots, {\mathbf{y}_n}) 
&\propto \prod_{i=1}^n |2\pi\sigma^2I_p|^{-1/2} 
\exp\{-\frac{1}{2\sigma^2}({\mathbf{y}_i}-x_i{\bm{\beta}_1}-{\bm{\beta}_0})^T({\mathbf{y}_i}-x_i{\bm{\beta}_1}-{\bm{\beta}_0})\} \\
&\times \exp \{ -\frac{1}{2}({\bm{\beta}_0}-{\bm{\mu}_0})^T({\bm{\beta}_0}-{\bm{\mu}_0})\}\\
&\propto \exp\{-\frac{1}{2}(\frac{n}{\sigma^2}+1)
{\bm{\beta}_0}^T{\bm{\beta}_0} + (\frac{1}{\sigma^2}\sum_i({\mathbf{y}_i}-x_i{\bm{\beta}_1})+ {\bm{\mu}_0})^T{\bm{\beta}_0}\}. \label{con: lm3beta0}
\end{aligned}
\end{equation}

In summary, the full conditional distribution of each parameter is given as follows:
\begin{align}
    \sigma^2| rest &\sim IG(a+\frac{np}{2}, \frac{1}{2}\sum_{i=1}^n||{\mathbf{y}_i}-x_i{\bm{\beta}_1}-
    {\bm{\beta}_0}||^2+b), \label{con: lm1}\\
    {\bm{\beta}_1}| rest &\sim 
    MVN({\bm{\mu_{\beta_1}}}, \Sigma_{{\bm{\beta}_1}}),
    \label{con: lm2}\\
    {\bm{\beta}_0}| rest &\sim 
    MVN({\bm{\mu_{\beta_0}}}
    , \Sigma_{{\bm{\beta}_0}}),
    \label{con: lm3}
\end{align}
where ${\bm{\mu_{\beta_1}}}=(\frac{\sum_i x_i^2}{\sigma^2}+1)^{-1}\frac{\sum_i x_i({\mathbf{y}_i}-{\bm{\beta}_0})}{\sigma^2}$, $\Sigma_{{\bm{\beta}_1}}= (\frac{\sum_ix_i^2}{\sigma^2}+1)^{-1}I_p$, ${\bm{\mu_{\beta_0}}}=(\frac{n}{\sigma^2}+1)^{-1}(\frac{\sum_i({\mathbf{y}_i}-x_i{\bm{\beta}_1})}{\sigma^2}+ {\bm{\mu}_0})
    $, and $\Sigma_{{\bm{\beta}_0}}=(\frac{n}{\sigma_e^2}+1)^{-1}I_p$.

Algorithm \ref{alg:lm} depicts the Gibbs sampling algorithm of the LM.
\begin{algorithm}[h]
    \caption{Gibbs sampling algorithm for the LM}
    \label{alg:lm}
  {\bfseries Input:} $\textbf{X}=(x_{ij}) \in \mathbb{R}^{n \times d}$
        , $\textbf{Y}=\{\textbf{y}_1, \textbf{y}_2, \cdots, \textbf{y}_n\} \in \mathbb{R}^{p \times n}$, total number of iteration $m$. \\
 {\bfseries Output:} Gibbs Samples \{$\sigma^{2(t)}$, ${\bm{\beta}_1}^{(t)}$, ${\bm{\beta}_0}^{(t)}$$\}_{t=1}^{m}$.\\
  {\bfseries Initialization:}  Set $t \leftarrow 0$, and initialize $\sigma^{2(0)}$, ${\bm{\beta}_1}^{(0)}$, ${\bm{\beta}_0}^{(0)}$.\\
\While{$t \le m$} {
 Set $t \leftarrow t + 1$.\\
    Sample $\sigma^{2(t+1)}$ according to Eq. (\ref{con: lm1})      conditional on ${\bm{\beta}_1}^{(t)}$, ${\bm{\beta}_0}^{(t)}$.\\
  Sample ${\bm{\beta}_1}^{(t+1)}$ according to Eq. (\ref{con: lm2}) conditional on  $\sigma^{2(t+1)}$, ${\bm{\beta}_0}^{(t)}$.\\
Sample ${\bm{\beta}_0}^{(t+1)}$ according to Eq. (\ref{con: lm3}) conditional on 
        $\sigma^{2(t+1)}$, ${\bm{\beta}_1}^{(t+1)}$.\\
} 
\end{algorithm}

\subsection*{A2. Derivation of the full conditional distributions of unknown parameters in the LMM}
\par
Followed by the LMM model describe in Section 2.1, the joint posterior distribution of ${\bm{\beta}_1}$, ${\bm{\beta}_0}$, $\sigma_e^2$, $\sigma_p^2$ and ${\mathbf{h}}$ is 
\begin{equation}
\begin{aligned}
p({\bm{\beta}_1}, \sigma_e^2, \sigma_p^2, {\bm{\beta}_0}, {\mathbf{h}}| {\mathbf{y}_1}, \cdots, {\mathbf{y}_n}) 
&\propto \prod_{i=1}^n p({\mathbf{y}_i}| {\bm{\beta}_1},  {\bm{\beta}_0},\sigma_e^2, \sigma_p^2, {\mathbf{h}})p({\bm{\beta}_1})p({\mathbf{h}}|\sigma_p^2)p(\sigma_e^2)p({\bm{\beta}_0})
p(\sigma_p^2)\\
&\propto \prod_{i=1}^n |\sigma_e^2I_p|^{-1/2} 
\exp\{-\frac{1}{2\sigma_e^2}||{\mathbf{y}_i}-x_i{\bm{\beta}_1}-{\mathbf{h}}-{\bm{\beta}_0}||^2\} \\
&\times \exp\{-\frac{1}{2}{\bm{\beta}_1}^T{\bm{\beta}_1}\} \\
&\times |\sigma_p^2G|^{-1/2} 
\exp\{-\frac{1}{2\sigma_p^2}{\mathbf{h}}^TG^{-1}{\mathbf{h}}\} \\
&\times (\sigma_p^2)^{-a-1}e^{-\frac{b}{\sigma_p^2}}\\
&\times (\sigma_e^2)^{-c-1}e^{-\frac{d}{\sigma_e^2}} \\
&\times \exp \{ -\frac{1}{2}({\bm{\beta}_0}-{\bm{\mu}_0})^T({\bm{\beta}_0}-{\bm{\mu}_0})\}. \\
\end{aligned}
\end{equation}
Denote $p(\cdot | rest)$ the full conditional distribution of one parameter is conditioned on the rest parameters.

Followed by the joint distribution, the full conditional distributions of $\sigma_e^2$, ${\bm{\beta}_1}$, $\sigma_p^2$, ${\mathbf{h}}$ and ${\bm{\beta}_0}$ are derived as follows.
\begin{equation}
\begin{aligned}
    p(\sigma_e^2 | rest) 
    &\propto (\sigma_e^2)^{-\frac{np}{2}} \exp \{-\frac{1}{2\sigma_e^2}\sum_i||{\mathbf{y}_i}-x_i{\bm{\beta}_1} -{\mathbf{h}}-{\bm{\beta}_0}||^2 \}\\
    &\times (\sigma^2_e)^{-c-1}\exp \{-\frac{d}{\sigma_e^2}\} \\
    &\propto (\sigma^2_e)^{-c-\frac{np}{2}-1}
    \exp \{-\frac{\frac{1}{2}\sum_i||y_i-x_i{\bm{\beta}_1} -{\mathbf{h}}-{\bm{\beta}_0}||^2+d}{\sigma^2_e}\}.
\end{aligned}
\end{equation}
\begin{equation}
\begin{aligned}
    p({\bm{\beta}_1} | rest) 
    &\propto (\sigma_e^2)^{-\frac{np}{2}} \exp \{-\frac{1}{2\sigma_e^2}\sum_i||{\mathbf{y}_i}-x_i{\bm{\beta}_1} -{\mathbf{h}}-{\bm{\beta}_0}||^2 \}\\
    &\times \exp\{-\frac{1}{2}{\bm{\beta}_1}^T{\bm{\beta}_1} \} \\
    &\propto \exp \{-\frac{1}{2}(\frac{1}{\sigma_e^2}
    \sum_{i=1}^nx_i^2+1){\bm{\beta}_1}^T{\bm{\beta}_1} + 
    \frac{1}{\sigma_e^2}{\bm{\beta}_1}^T\sum_{i=1}^n({\mathbf{y}_i}-{\mathbf{h}}-{\bm{\beta}_0})x_i \}.
\end{aligned}
\end{equation},

\begin{equation}
\begin{aligned}
    p(\sigma_p^2 | rest) 
    &\propto (\sigma_p^2)^{-\frac{p}{2}} \exp \{-\frac{1}{2\sigma_p^2}{\mathbf{h}}^TG^{-1}{\mathbf{h}} \}\\
    &\times (\sigma_p)^{-a-1}\exp \{-\frac{b}{\sigma_p^2}\} \\
    &\propto (\sigma_p^2)^{-a-\frac{p}{2}-1} \exp 
    \{-\frac{\frac{1}{2}{\mathbf{h}}^TG^{-1}{\mathbf{h}}+b}{\sigma_p^2}\}.
\end{aligned}
\end{equation},

\begin{equation}
\begin{aligned}
    p({\mathbf{h}} | rest) 
    &\propto \exp \{-\frac{1}{2\sigma_e^2}\sum_i||{\mathbf{y}_i}-x_i{\bm{\beta}_1} -{\mathbf{h}}-{\bm{\beta}_0}||^2 \}\\
    &\times \exp \{-\frac{1}{2\sigma_p^2}{\mathbf{h}}^TG^{-1}{\mathbf{h}} \}\\
    &\propto \exp \{-\frac{1}{2}{\mathbf{h}}^T(\frac{n}{\sigma_e^2}I_p+\frac{1}{\sigma_p^2}G^{-1}){\mathbf{h}}
    + \frac{1}{\sigma_e^2}{\mathbf{h}}^T\sum_i(y_i-x_i{\bm{\beta}_1}-{\bm{\beta}_0})\}.
\end{aligned}
\end{equation}
and 
\begin{equation}
\begin{aligned}
    p({\bm{\beta}_0} | rest) 
    &\propto \exp \{-\frac{1}{2\sigma_e^2}\sum_i||{\mathbf{y}_i}-x_i{\bm{\beta}_1} -{\mathbf{h}}-{\bm{\beta}_0}||^2 \}\\
    &\times \exp \{ -\frac{1}{2}({\bm{\beta}_0}-{\bm{\mu}_0})^T({\bm{\beta}_0}-{\bm{\mu}_0})\} \\
    &\propto \exp \{-\frac{1}{2}(\frac{n}{\sigma_e^2}+1){\bm{\beta}_0}^T{\bm{\beta}_0}
    + (\frac{1}{\sigma_e^2}\sum_i({\mathbf{y}_i}-x_i{\bm{\beta}_1}-{\mathbf{h}})+{\bm{\mu}_0})^T{\bm{\beta}_0}\}.
\end{aligned}
\end{equation}

In summary, the full conditional distribution of each parameter is derived as follows:
\begin{align}
    \sigma_e^2| rest &\sim IG(c+\frac{np}{2}, \frac{1}{2}\sum_{i=1}^n||{\mathbf{y}_i}-x_i{\bm{\beta}_1} -{\mathbf{h}}-{\bm{\beta}_0}||^2+d),\\
        \sigma_p^2| rest &\sim IG(a+\frac{p}{2}, \frac{1}{2}{\mathbf{h}}^TG^{-1}{\mathbf{h}}+b),\\
    {\bm{\beta}_1}| rest &\sim 
    MVN({\bm{\mu_{\beta_1}}}, \Sigma_{{\bm{\beta}_1}}), \\
    {\bm{\beta}_0}| rest &\sim 
    MVN({\bm{\mu_{\beta_0}}}
    , \Sigma_{{\bm{\beta}_0}}),\\
    {\mathbf{h}}| rest &\sim 
    MVN({\bm{\mu}_{{\mathbf{h}}}}, \Sigma_{{\mathbf{h}}}). 
\end{align}
where $\mu_h=(\frac{n}{\sigma_e^2}I_p+\frac{1}{\sigma_p^2}G^{-1})^{-1}\frac{\sum_i({\mathbf{y}_i}-x_i{\bm{\beta}_1}-
    {\bm{\beta}_0})}{\sigma_e^2}$, $\Sigma_{{\mathbf{h}}}=(\frac{n}{\sigma_e^2}I_p+\frac{1}{\sigma_p^2}G^{-1})^{-1}$, ${\bm{\mu_{\beta_1}}}=\frac{\frac{1}{\sigma_e^2}\sum_i({\mathbf{y}_i}-{\mathbf{h}}-
    {\bm{\beta_0}})x_i}{\frac{1}{\sigma_e^2}\sum_ix_i^2+1} $, $\Sigma_{{\bm{\beta_1}}}= (\frac{\sum_ix_i^2}{\sigma_e^2}+1)^{-1}I_p$, ${\bm{\mu_{\beta_0}}}=(\frac{n}{\sigma_e^2}+1)^{-1}(\frac{\sum_i({\mathbf{y}_i}-x_i{\bm{\beta_1}}- {\mathbf{h}})}{\sigma_e^2}+{\bm{\mu_0}})
    $, and $\Sigma_{{\bm{\beta_0}}}=(\frac{n}{\sigma_e^2}+1)^{-1}I_p$.

\section*{Appendix B. Computational time for the simulations}
\begin{table}[h!]
  \begin{center}
    \caption{{Mean and standard deviation of computational times (in seconds) of each single job for $100$ repeated experiments for both the LM and LMM. The total number of MCMC iterations was $5000$. Each experiment was run with a single core ($3.20$-GHz AMD Ryzen $7$ $7735$H) with $16$GB of RAM.}} 
    \label{tab:table_neo}
    \begin{tabular}{cccccc}
    \hline
          & & \multicolumn{2}{|c}{{LMM}}   &\multicolumn{2}{|c}{{LM}} 
       \\
           \hline 
      & & \multicolumn{1}{|c}{$n = 50$} & \multicolumn{1}{c}{$n = 100$} & \multicolumn{1}{|c}{$n = 50$} &
       \multicolumn{1}{c}{$n = 100$} 
       \\
    \hline
    \multirow{2}{*}{\textbf{Case 1}} & \multicolumn{1}{c|}{mean} & 4.174 & \multicolumn{1}{c|}{6.139} & 3.744 & 5.118 \\
    & \multicolumn{1}{c|}{sd} & 0.046 & \multicolumn{1}{c|}{0.043} & 1.177 & 0.970\\
    \hline 
    \multirow{2}{*}{\textbf{Case 2}} & \multicolumn{1}{c|}{mean}& 4.098 & \multicolumn{1}{c|}{6.155} & 3.215 & 4.169 \\
    & \multicolumn{1}{c|}{sd} & 0.067 & \multicolumn{1}{c|}{0.059} & 0.060 & 0.047\\
    \hline 
    \multirow{2}{*}{\textbf{Case 3}} & \multicolumn{1}{c|}{mean}& 4.291 & \multicolumn{1}{c|}{6.300} & 3.881 & 5.351 \\
    & \multicolumn{1}{c|}{sd} & 0.723 & \multicolumn{1}{c|}{0.168} & 1.168 & 1.598\\
    \hline 
    \multirow{2}{*}{\textbf{Case 4}} & \multicolumn{1}{c|}{mean}& 4.082 & \multicolumn{1}{c|}{6.149} & 5.771 & 8.147 \\
    & \multicolumn{1}{c|}{sd} & 0.035 & \multicolumn{1}{c|}{0.028} & 1.361 & 2.407\\
    \hline
    \multirow{2}{*}{\textbf{Case 5}} & \multicolumn{1}{c|}{mean}& 4.119 & \multicolumn{1}{c|}{6.773} & 3.845 & 5.179 \\
    & \multicolumn{1}{c|}{sd} & 0.046 & \multicolumn{1}{c|}{3.585} & 1.042 & 0.428\\
    \hline 
    \end{tabular} 
  \end{center}
\end{table}

\section*{Appendix C. Simulation studies with model misspecification}
\par
\begin{figure}
  \centering
  \begin{subfigure}{0.45\linewidth}
    \includegraphics[width=\linewidth]{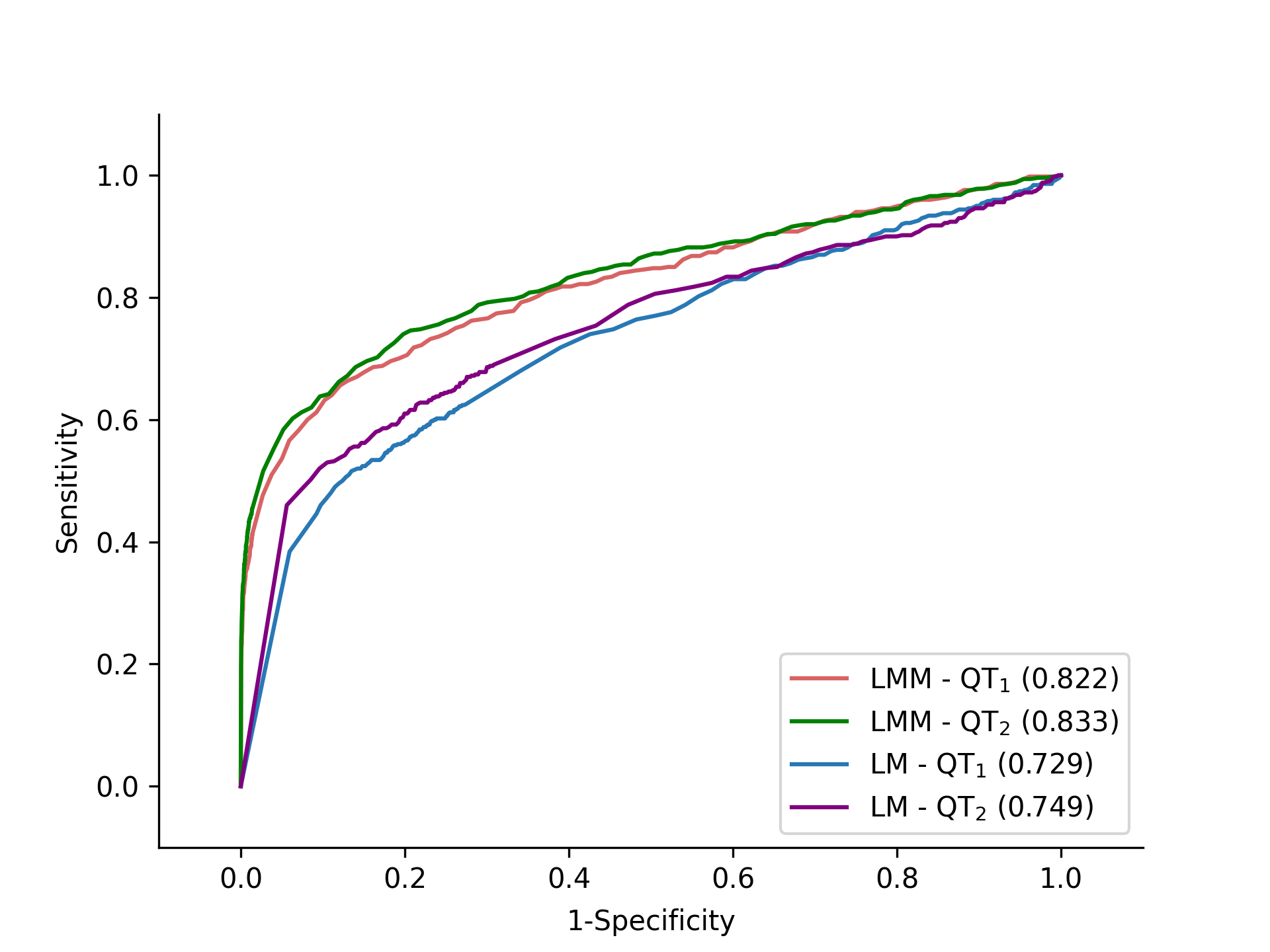}
    \label{subfig:t_50_i1}
    \subcaption{$n=50$}
  \end{subfigure}
  \hfill
  \begin{subfigure}{0.45\linewidth}
    \includegraphics[width=\linewidth]{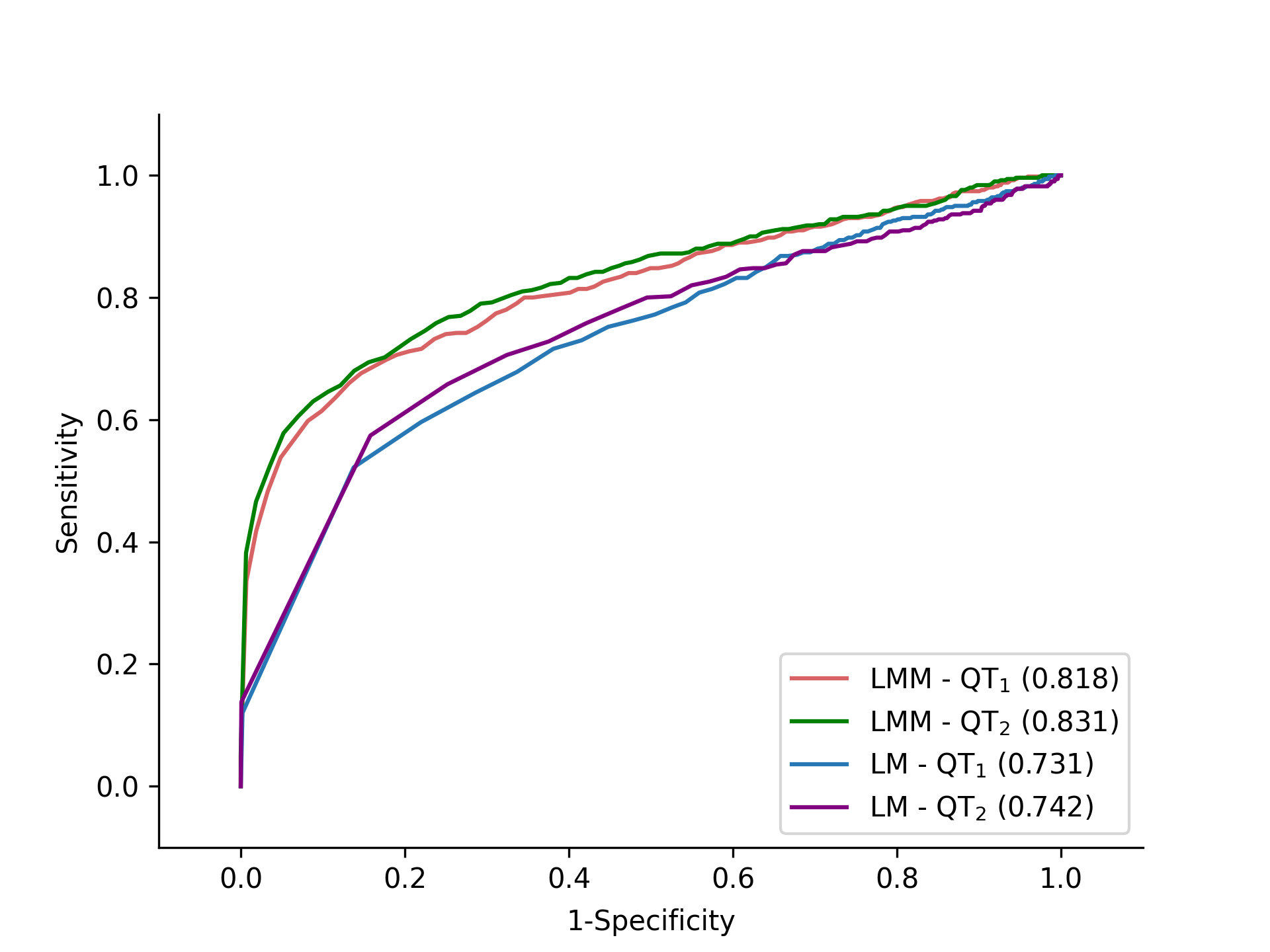}
    \label{subfig:t_50_p1}
    \subcaption{$n=50$}
  \end{subfigure}
\par
\begin{subfigure}{0.45\linewidth}
    \includegraphics[width=\linewidth]{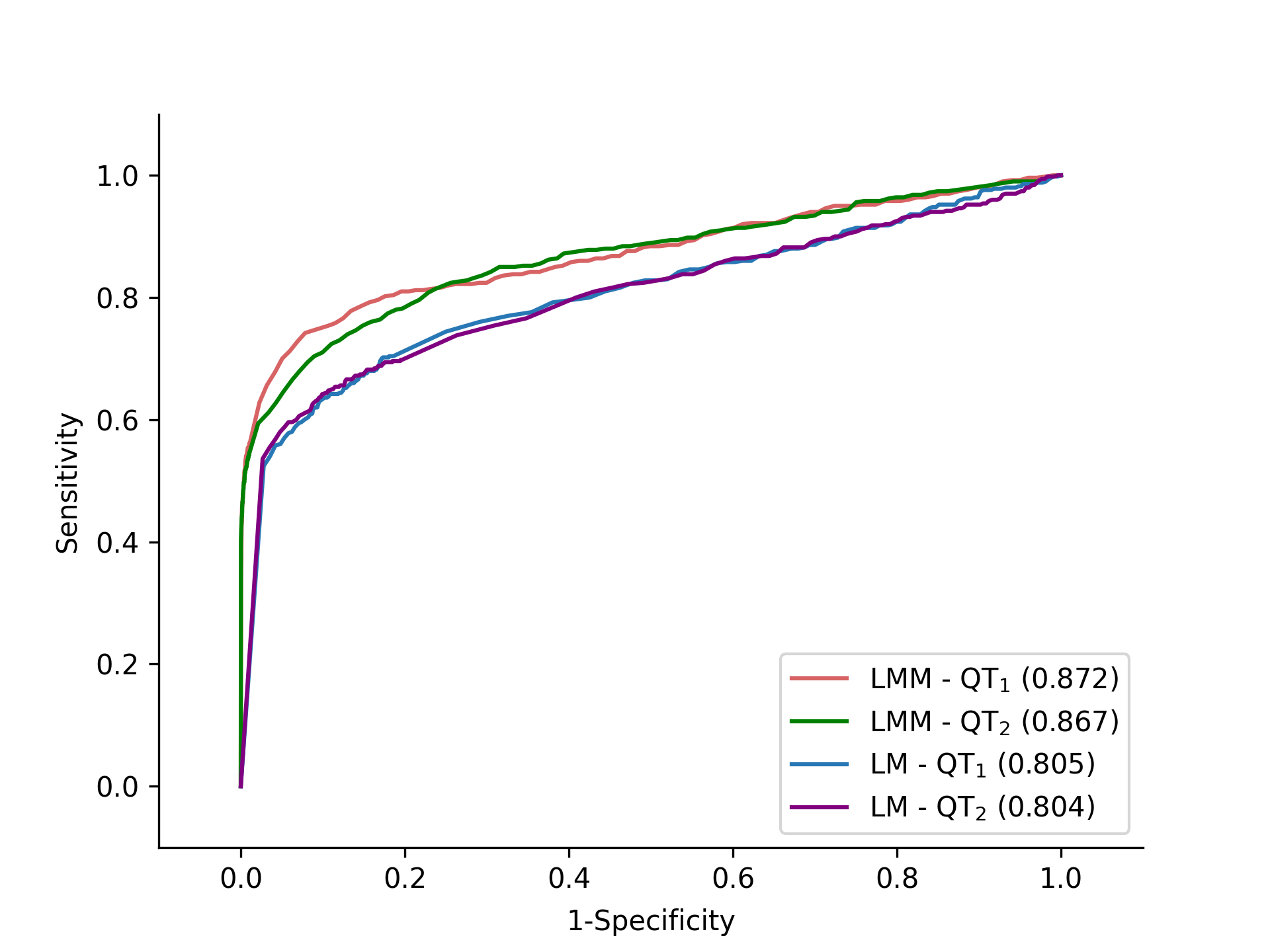}
    \label{subfig:t_100_i1}
    \subcaption{$n=100$}
  \end{subfigure}
  \hfill
  \begin{subfigure}{0.45\linewidth}
    \includegraphics[width=\linewidth]{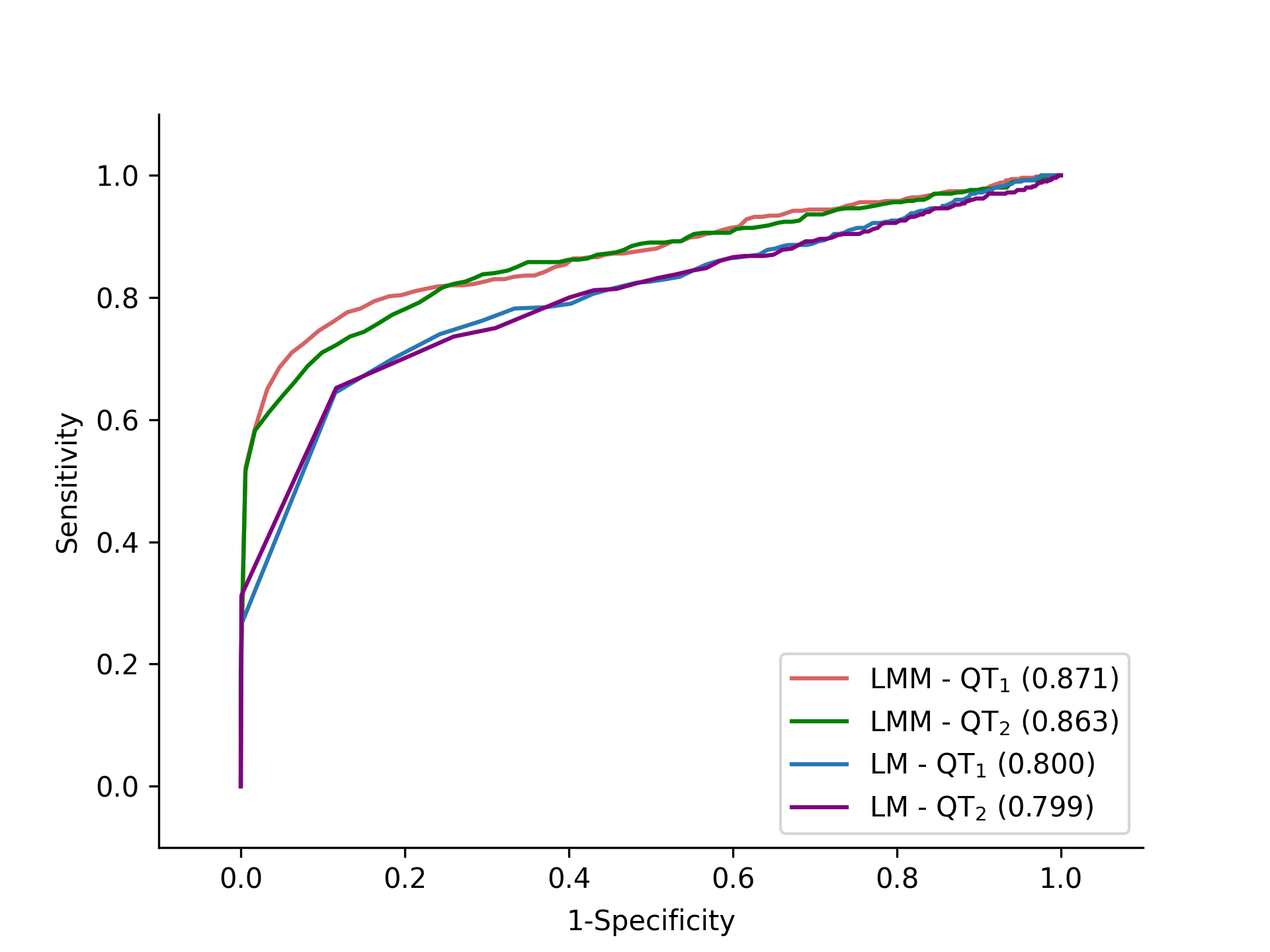}
    \label{subfig:t_100_p1}
    \subcaption{$n=100$}
  \end{subfigure}
    \caption{{Case 1 with model misspecification. ROC curves of the LMM and LM based on credible intervals (a,c) and the CCT (b,d) when there was a moderate dependency presented among phenotypes with sample size varies.  Curves of QT$_1$, QT$_2$ represent the ROC curves concerning QT$_1$ and QT$_2$. The figures in brackets indicated the corresponding AUC for each curve.} }  
  \label{fig:t_1}
\end{figure}
\begin{figure}
  \centering
  \begin{subfigure}{0.45\linewidth}
    \includegraphics[width=\linewidth]{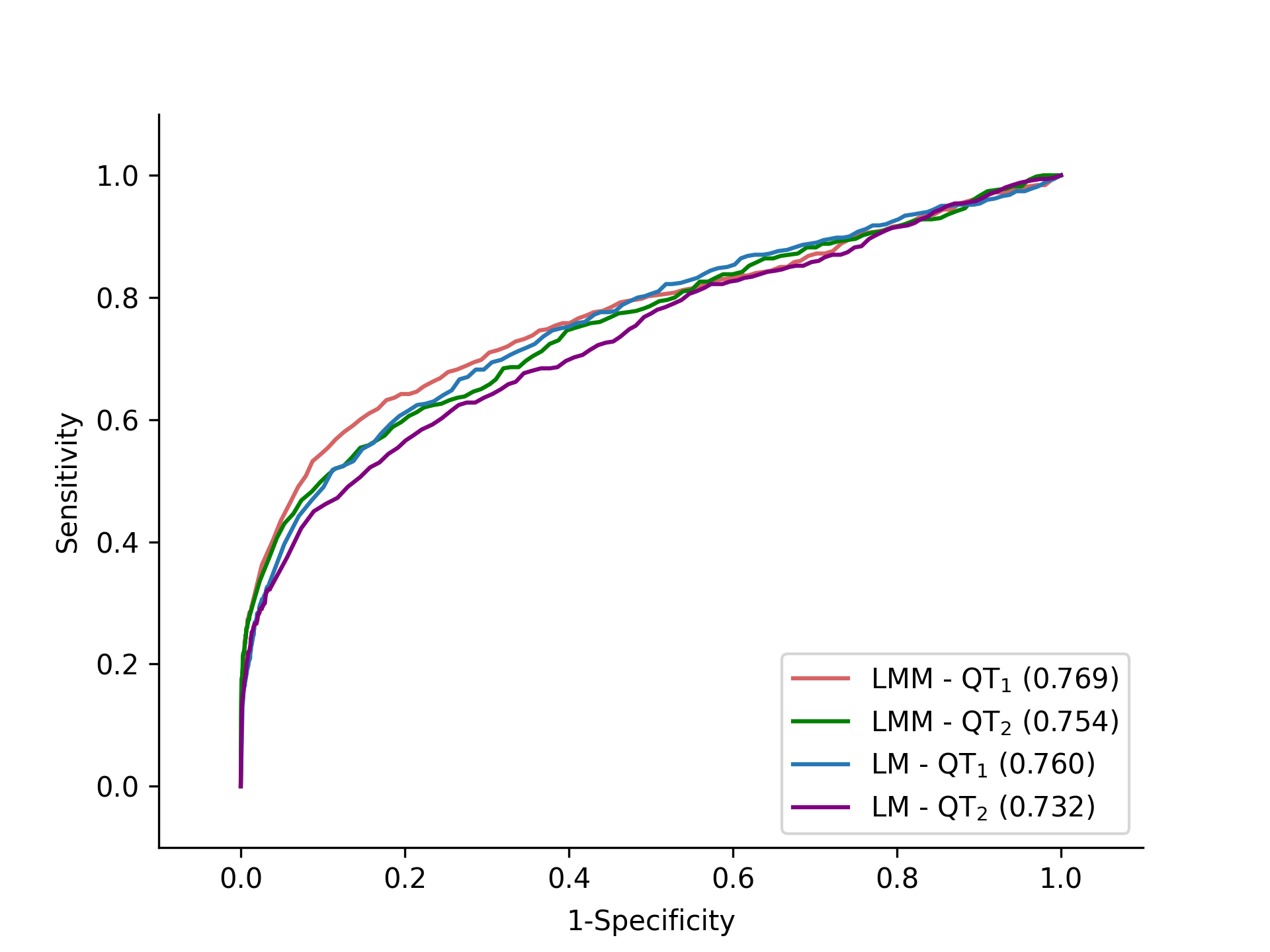}
    \label{subfig:t_50_i2}
    \subcaption{$n=50$}
  \end{subfigure}
  \hfill
  \begin{subfigure}{0.45\linewidth}
    \includegraphics[width=\linewidth]{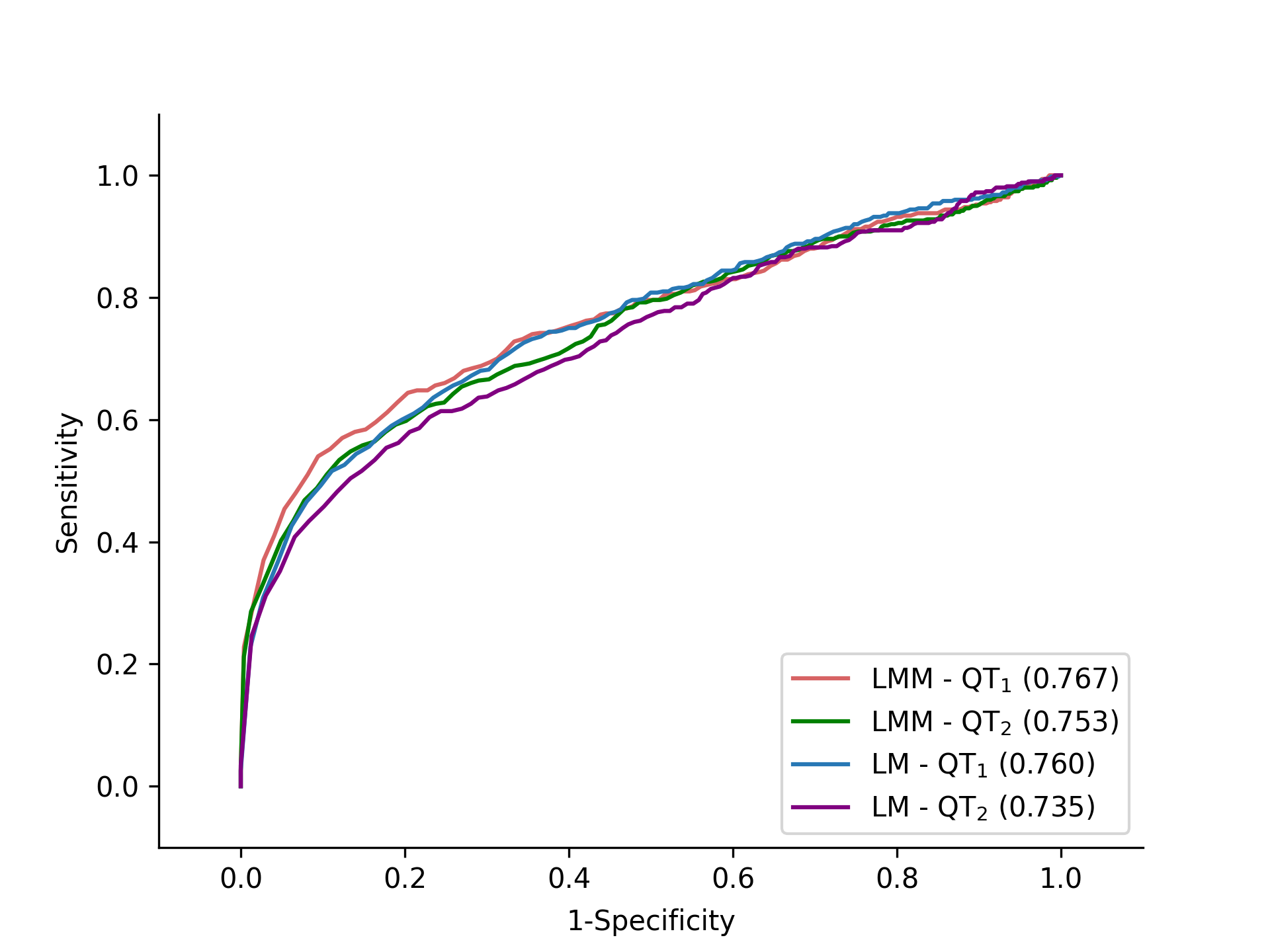}
    \label{subfig:t_50_p2}
    \subcaption{$n=50$}
  \end{subfigure}
\par
\begin{subfigure}{0.45\linewidth}
    \includegraphics[width=\linewidth]{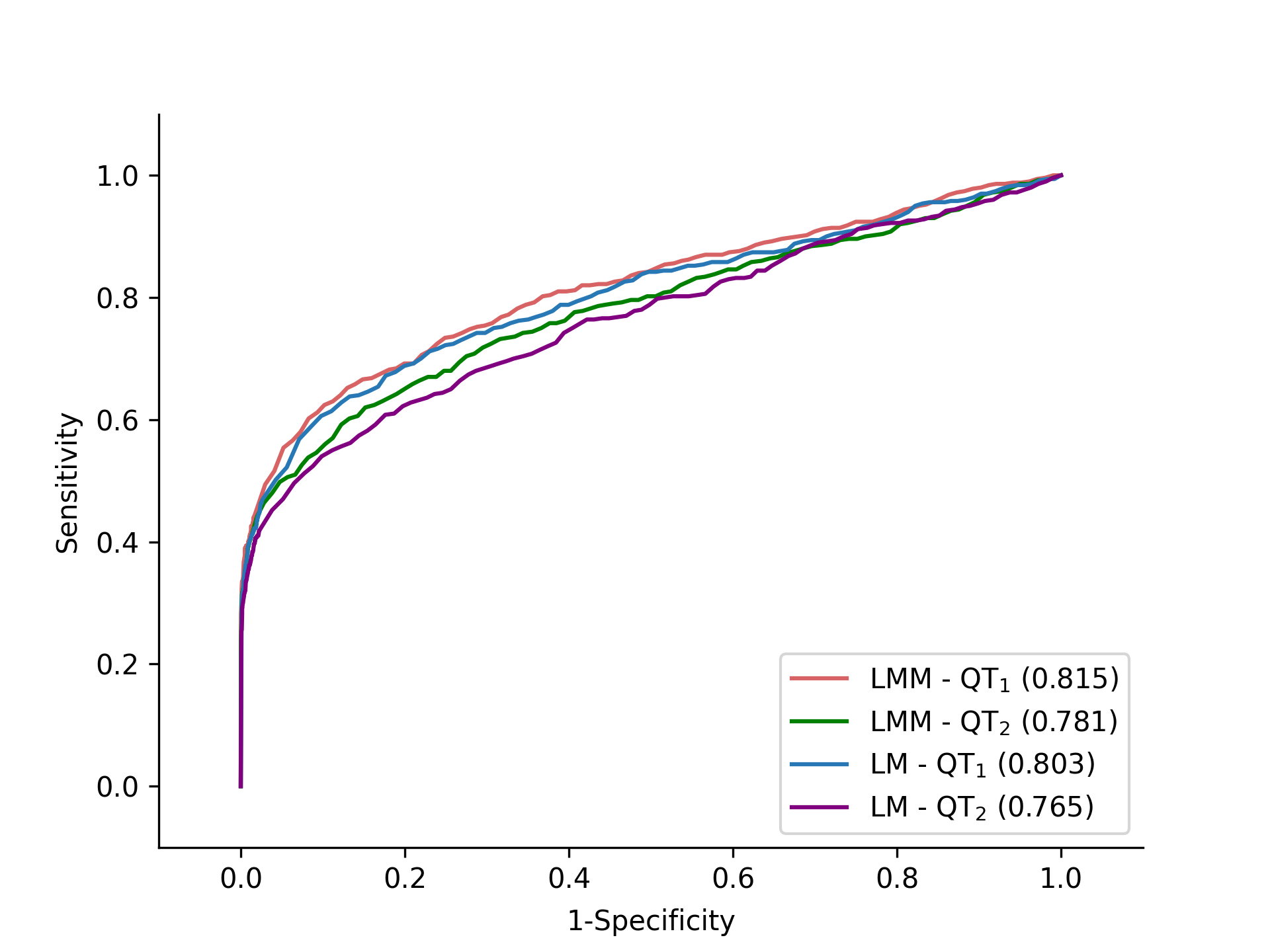}
    \label{subfig:t_100_i2}
    \subcaption{$n=100$}
  \end{subfigure}
  \hfill
  \begin{subfigure}{0.45\linewidth}
    \includegraphics[width=\linewidth]{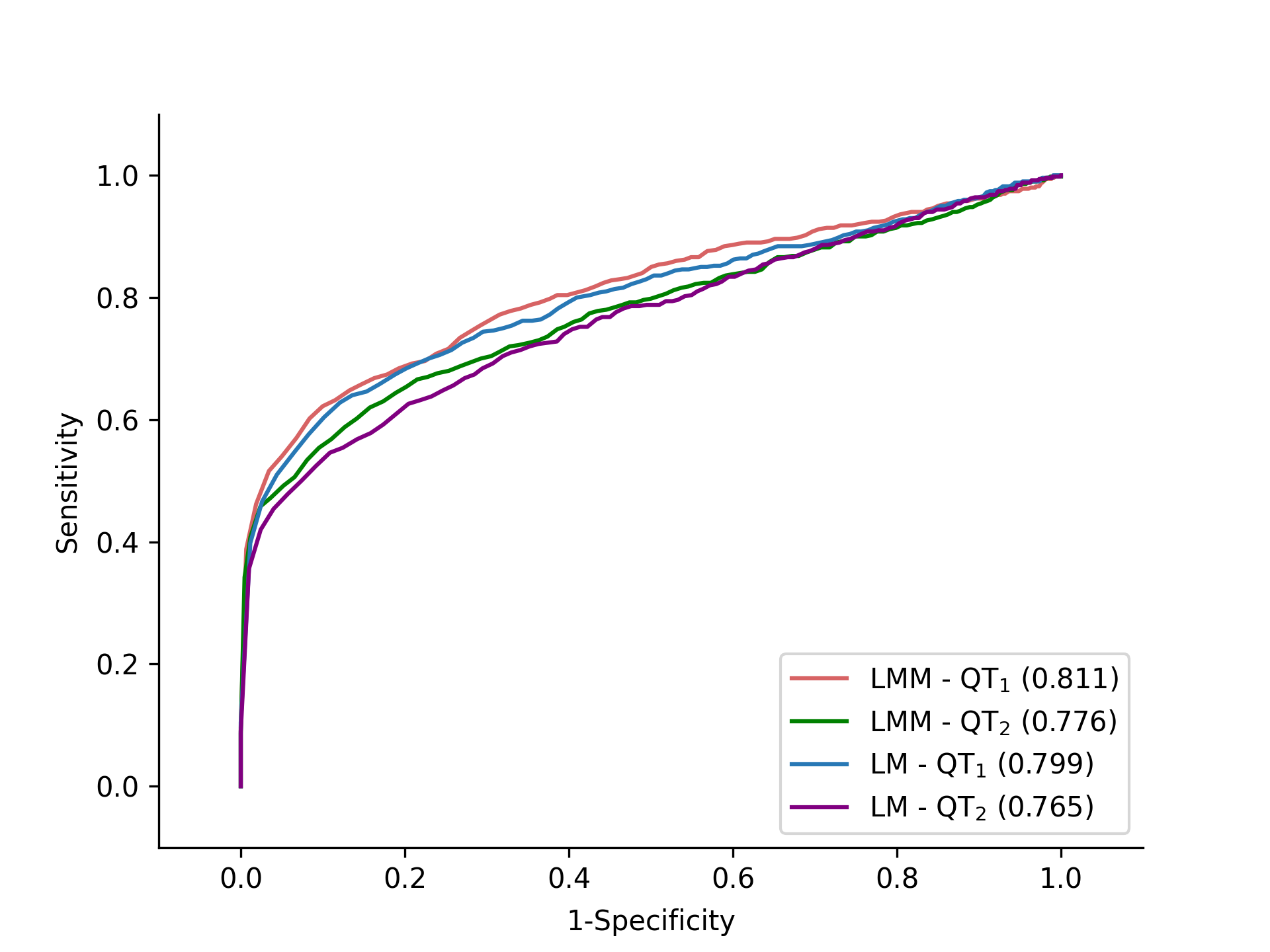}
    \label{subfig:t_100_p2}
    \subcaption{$n=100$}
  \end{subfigure}
    \caption{{Case 2 with model misspecification. ROC curves of the LMM and LM based on credible intervals (a,c) and the CCT (b,d) when there was no dependence presented among phenotypes with sample size varies. Curves of QT$_1$, QT$_2$ represent the ROC curves concerning QT$_1$ and QT$_2$.  The figures in  brackets indicated the corresponding AUC for each curve.} }  
  \label{fig:t_2}
\end{figure}
\begin{figure}
  \centering
  \begin{subfigure}{0.45\linewidth}
    \includegraphics[width=\linewidth]{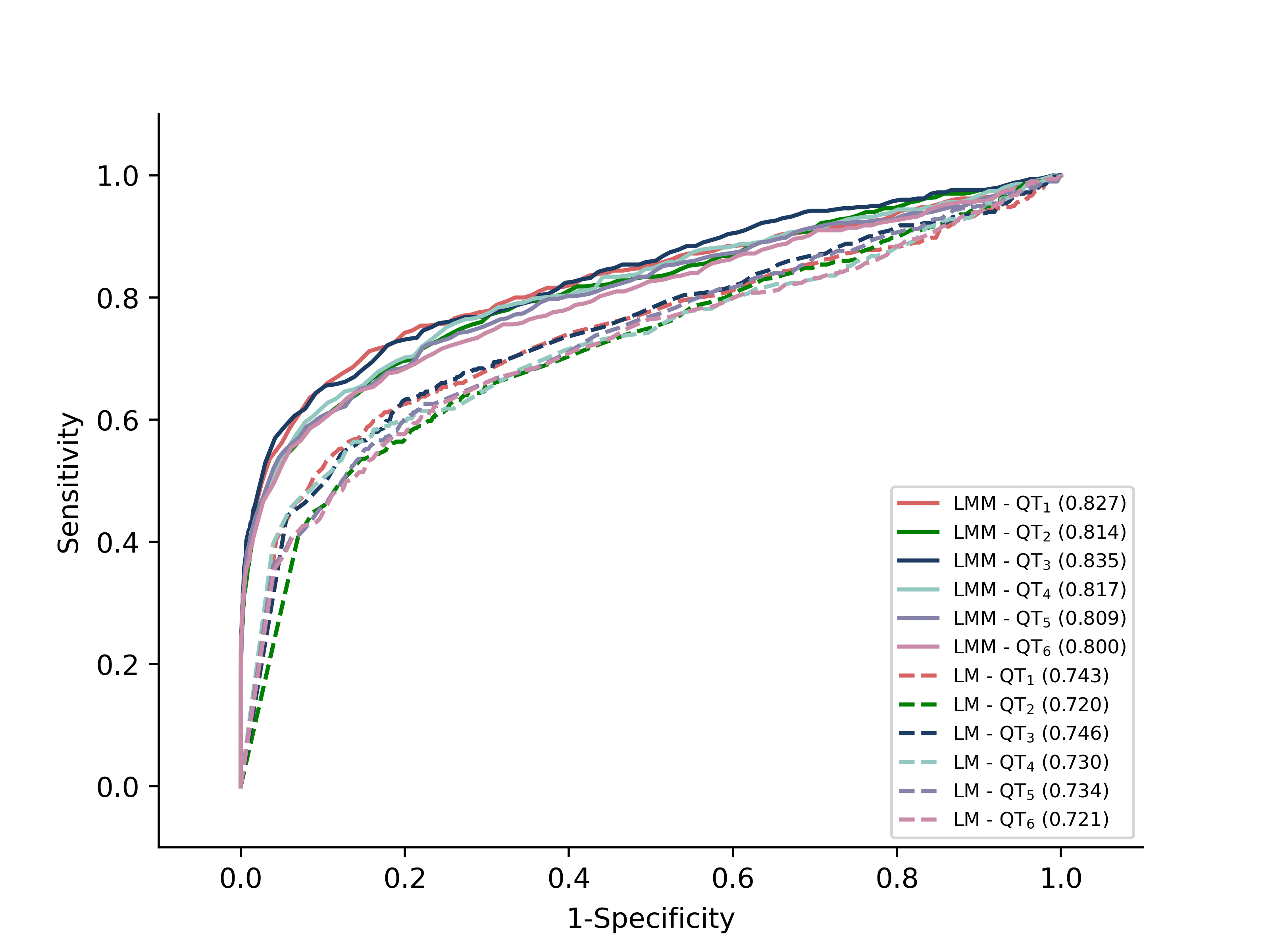}
    \label{subfig:t_50_i3}
    \subcaption{$n=50$}
  \end{subfigure}
  \hfill
  \begin{subfigure}{0.45\linewidth}
    \includegraphics[width=\linewidth]{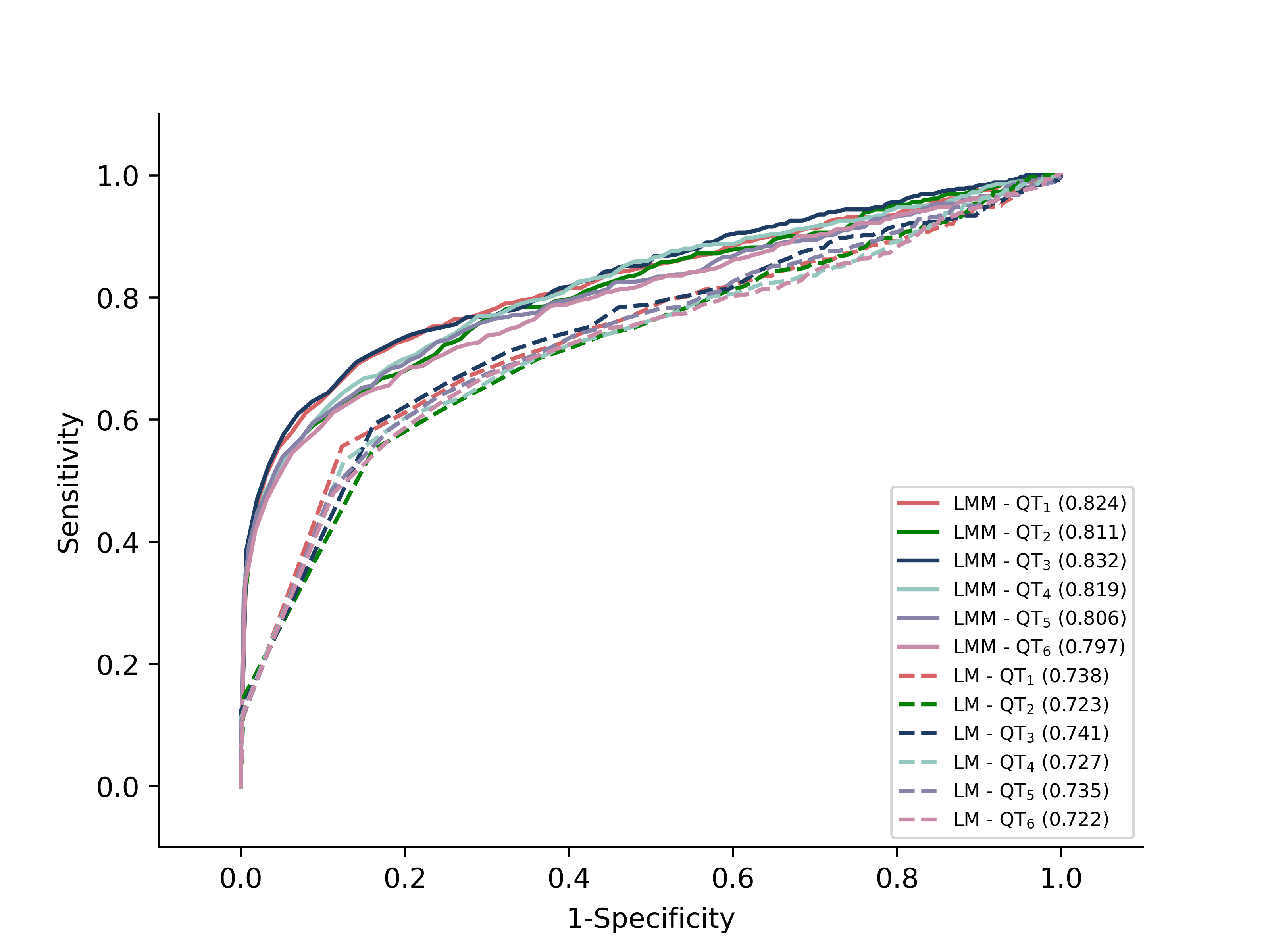}
    \label{subfig:t_50_p3}
    \subcaption{$n=50$}
  \end{subfigure}
\par
\begin{subfigure}{0.45\linewidth}
    \includegraphics[width=\linewidth]{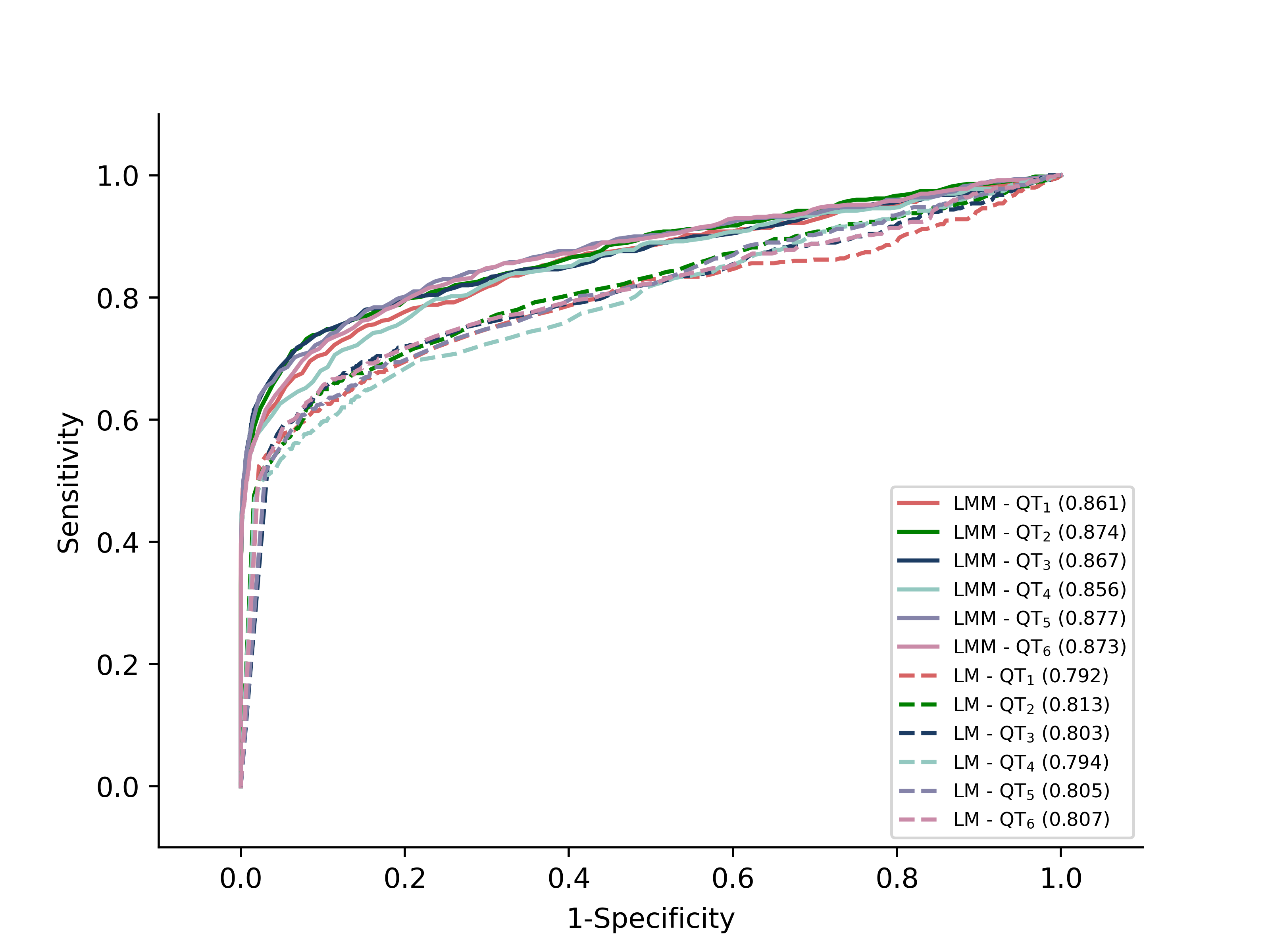}
    \label{subfig:t_100_i3}
    \subcaption{$n=100$}
  \end{subfigure}
  \hfill
  \begin{subfigure}{0.45\linewidth}
    \includegraphics[width=\linewidth]{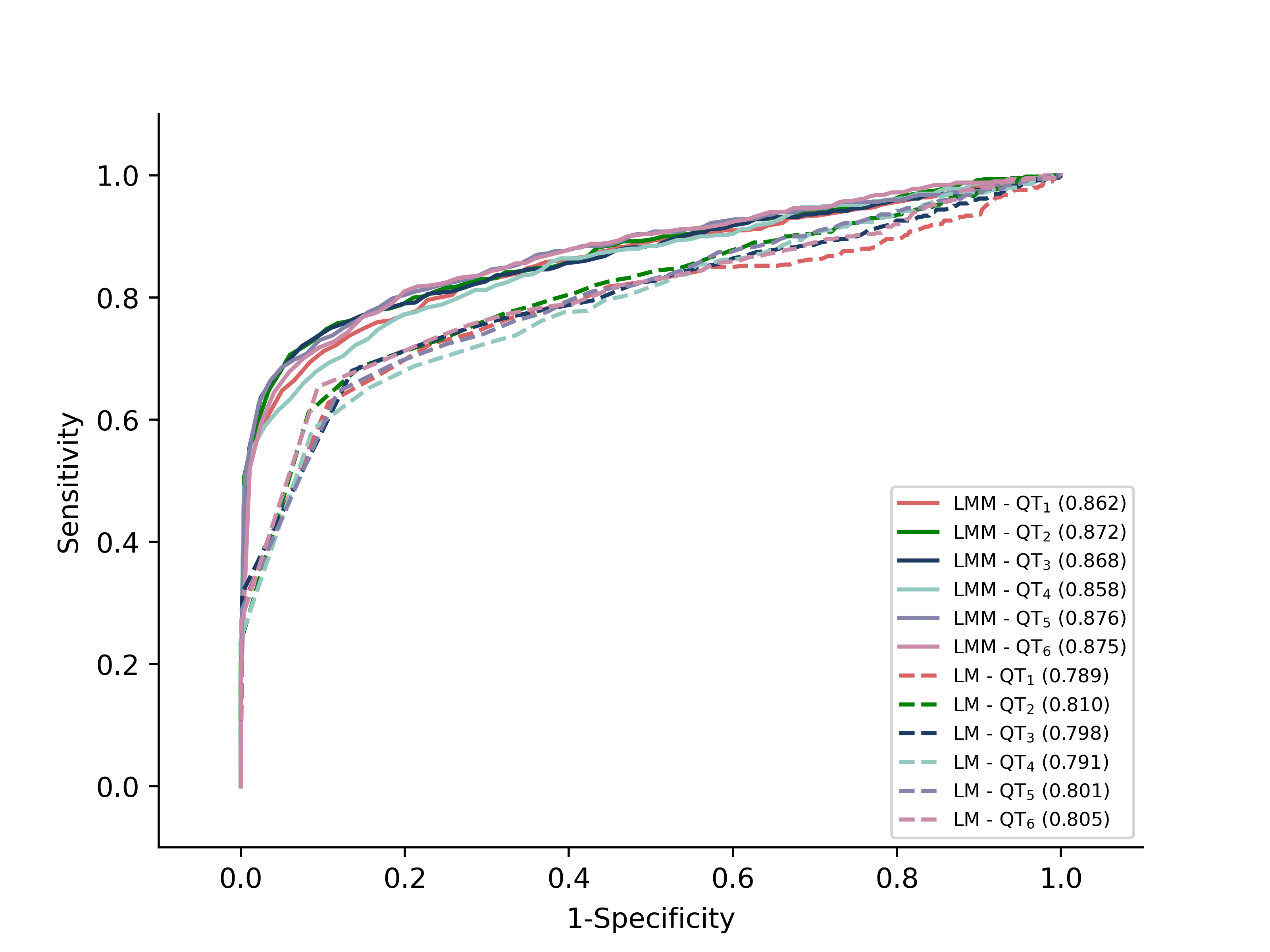}
    \label{subfig:t_100_p3}
    \subcaption{$n=100$}
  \end{subfigure}
    \caption{{Case 3 with model misspecification. ROC curves of the LMM and LM based on credible intervals (a,c) and the CCT (b,d) when there was a complex dependency presented among phenotypes with sample size varies. Curves of QT$_1,\ldots$, QT$_6$ represents the ROC curves concerning QT$_1, \ldots$, QT$_6$. The figures in  brackets indicated the corresponding AUC for each curve.}}  
  \label{fig:t_3}
\end{figure}
\begin{figure}
  \centering
  \begin{subfigure}{0.45\linewidth}
    \includegraphics[width=\linewidth]{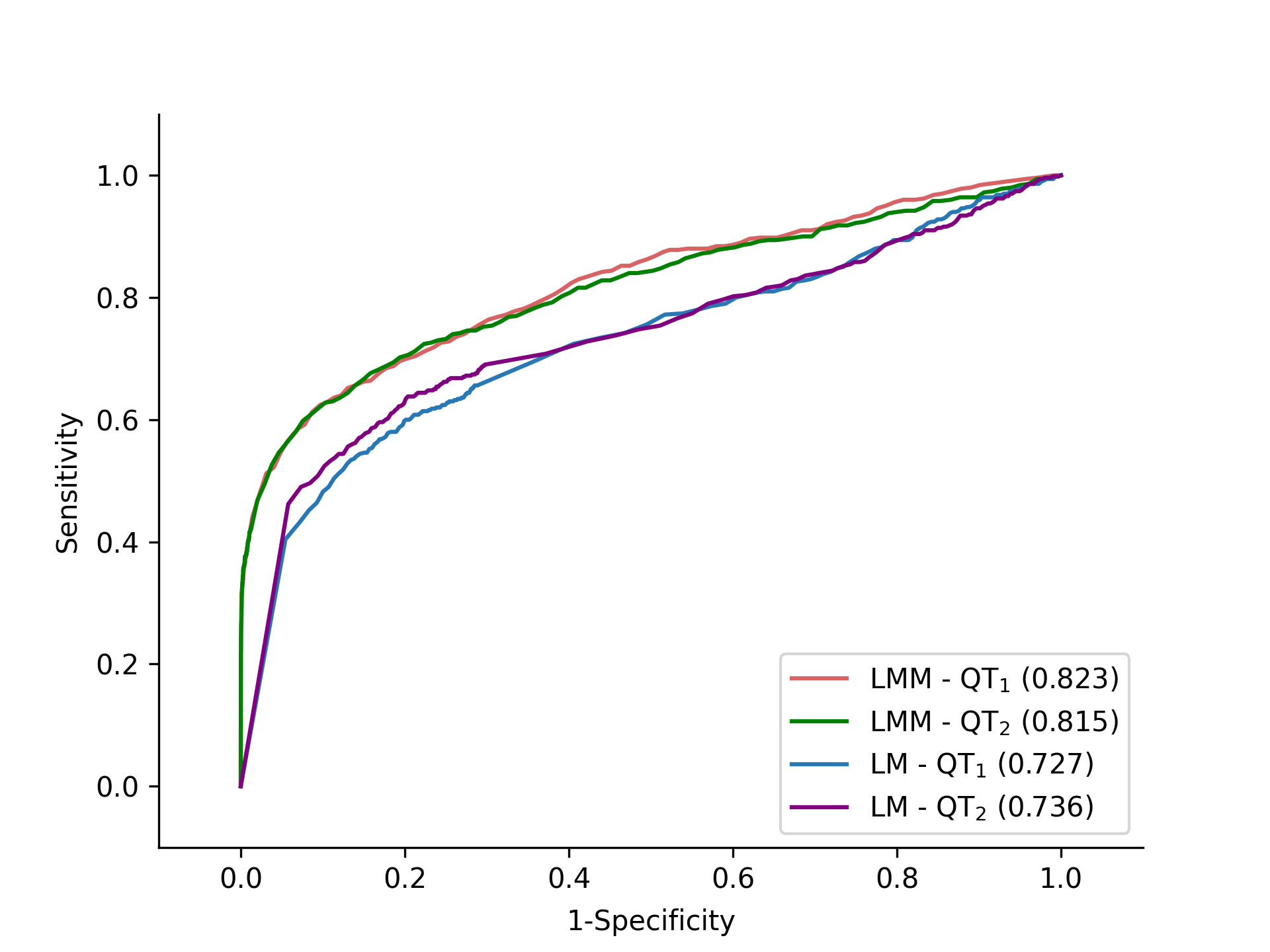}
    \label{subfig:t_50_i4}
    \subcaption{$n=50$}
  \end{subfigure}
  \hfill
  \begin{subfigure}{0.45\linewidth}
    \includegraphics[width=\linewidth]{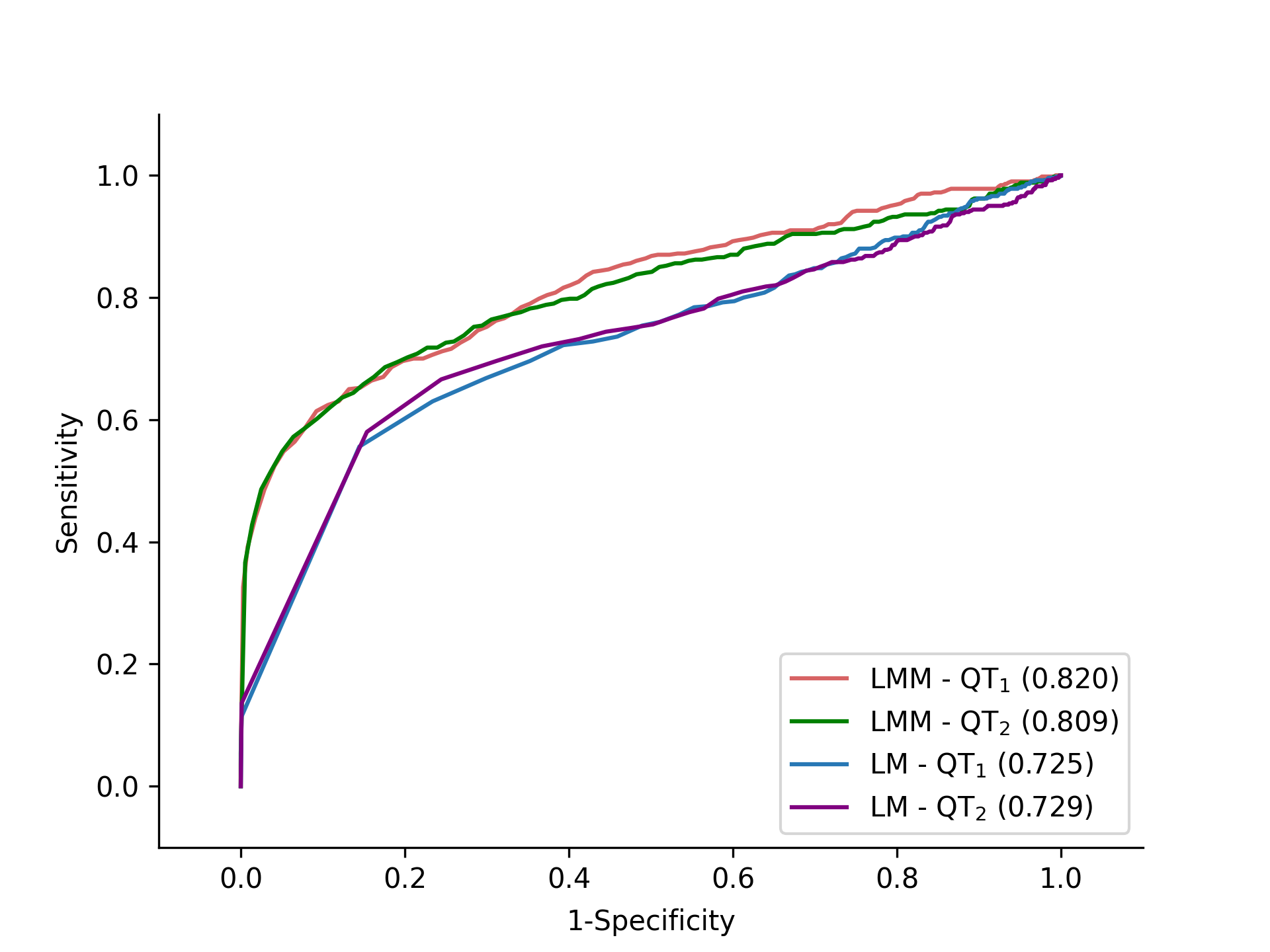}
    \label{subfig:t_50_p4}
    \subcaption{$n=50$}
  \end{subfigure}
\par
\begin{subfigure}{0.45\linewidth}
    \includegraphics[width=\linewidth]{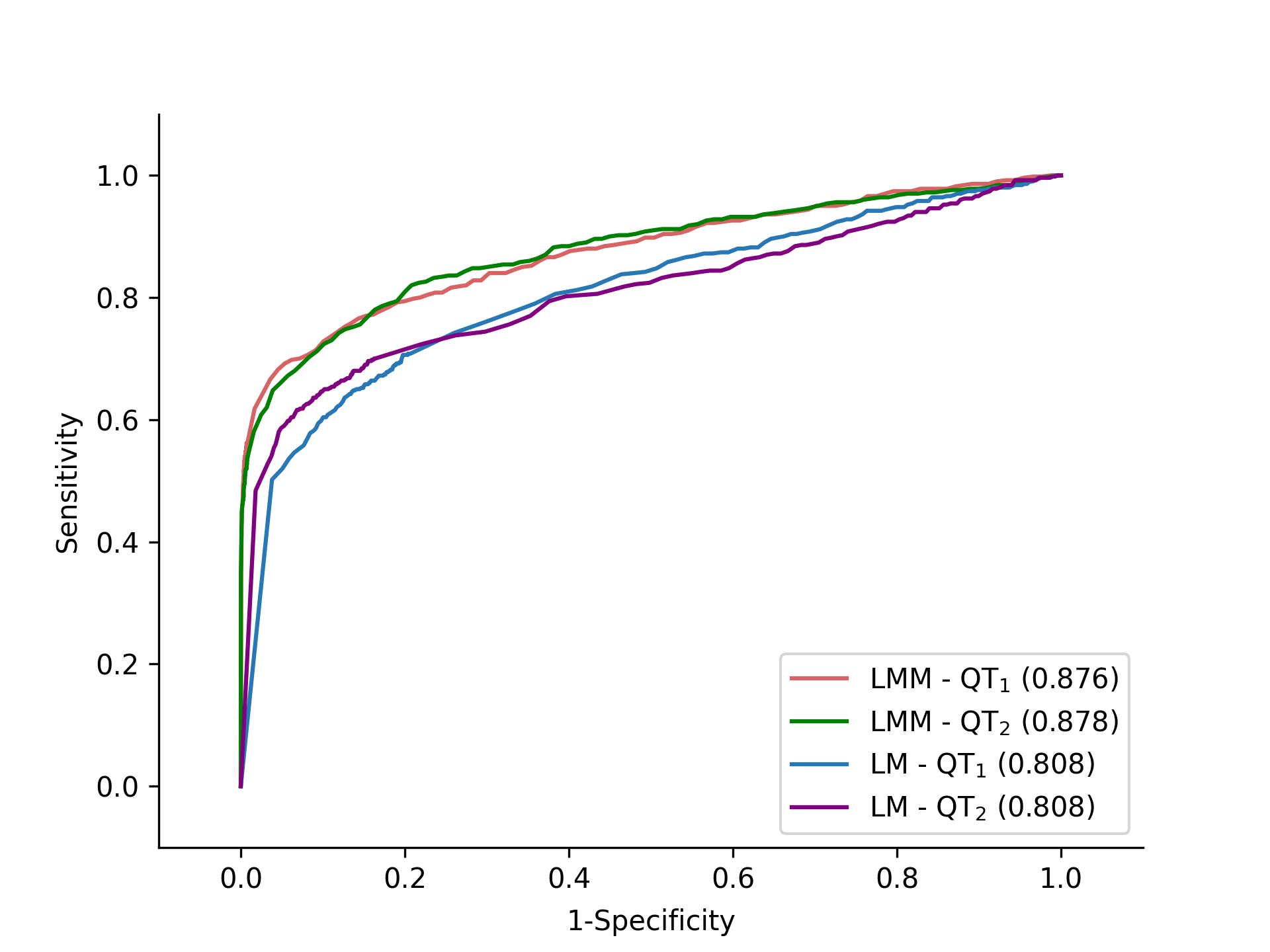}
    \label{subfig:t_100_i4}
    \subcaption{$n=100$}
  \end{subfigure}
  \hfill
  \begin{subfigure}{0.45\linewidth}
    \includegraphics[width=\linewidth]{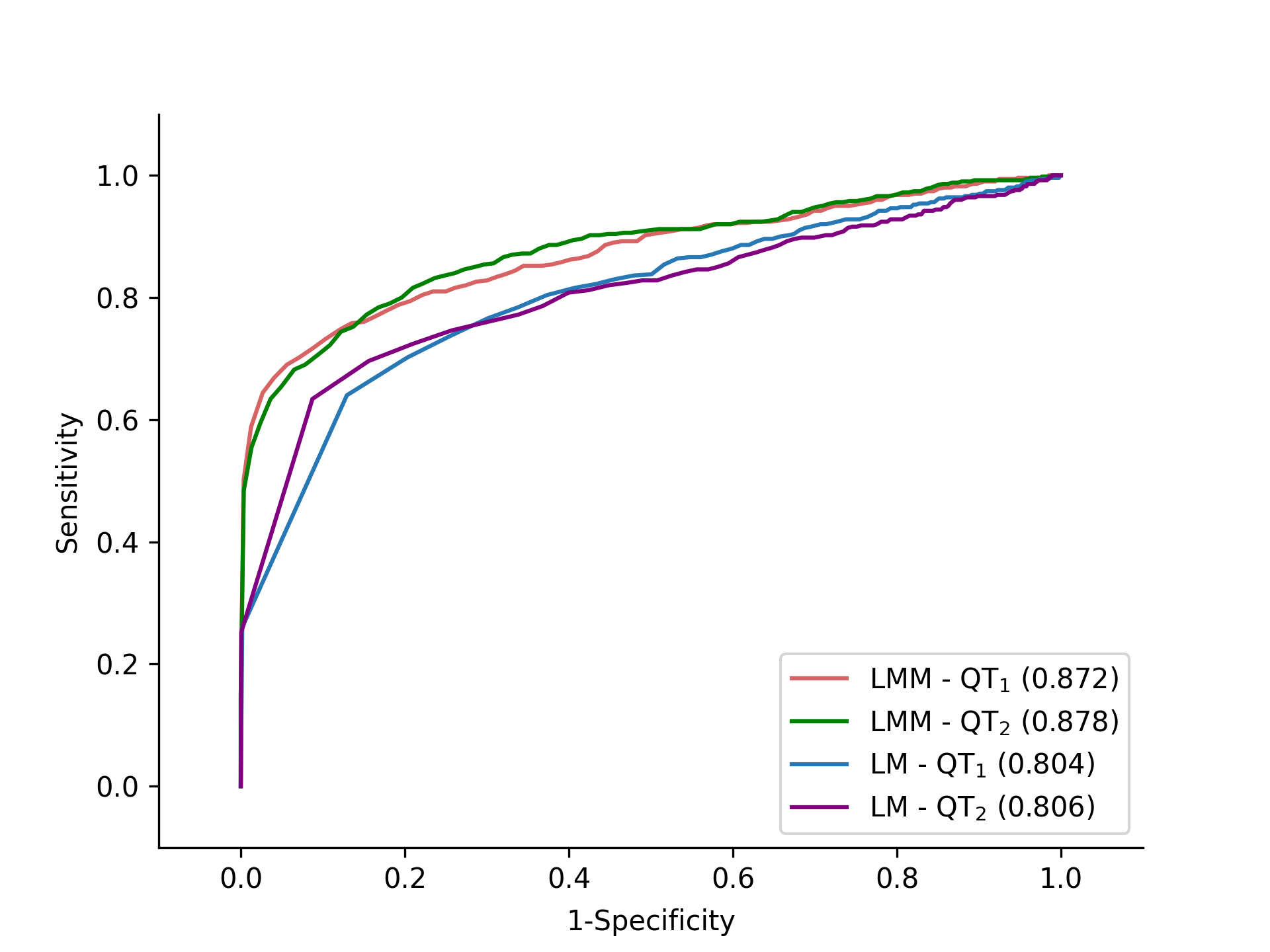}
    \label{subfig:t_100_p4}
    \subcaption{$n=100$}
  \end{subfigure}
    \caption{{Case 4 with model misspecification. ROC curves of the LMM and LM based on credible intervals (a,c) and the CCT (b,d) when there was weak dependence presented among phenotypes with sample size varies. Curves of QT$_1$, QT$_2$ represent the ROC curves concerning QT$_1$ and QT$_2$.  The figures in  brackets indicated the corresponding AUC for each curve. }}  
  \label{fig:t_4}
\end{figure}
\begin{figure}
  \centering
  \begin{subfigure}{0.45\linewidth}
    \includegraphics[width=\linewidth]{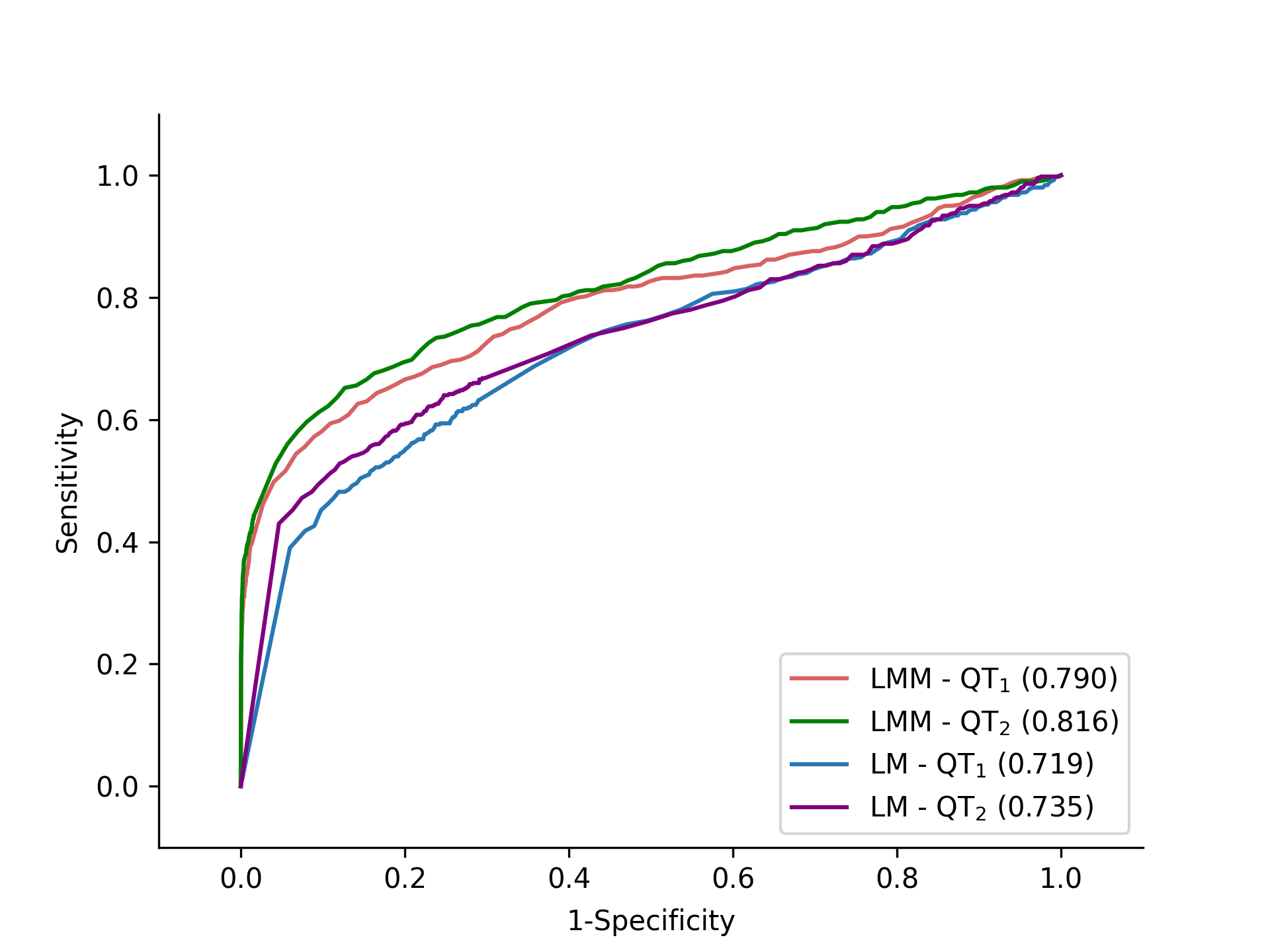}
    \label{subfig:t_50_i5}
    \subcaption{$n=50$}
  \end{subfigure}
  \hfill
  \begin{subfigure}{0.45\linewidth}
    \includegraphics[width=\linewidth]{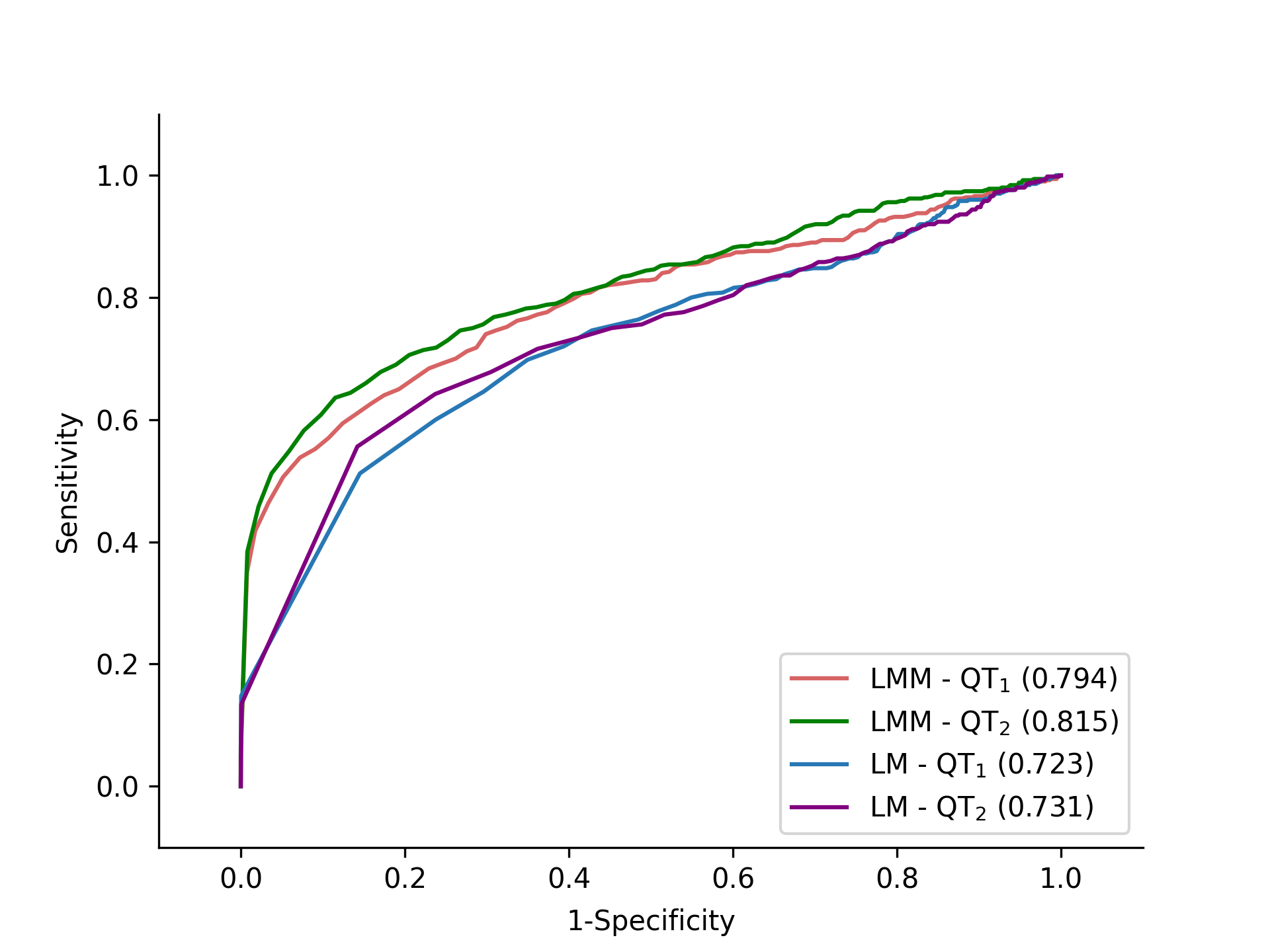}
    \label{subfig:t_50_p5}
    \subcaption{$n=50$}
  \end{subfigure}
\par
\begin{subfigure}{0.45\linewidth}
   \includegraphics[width=\linewidth]{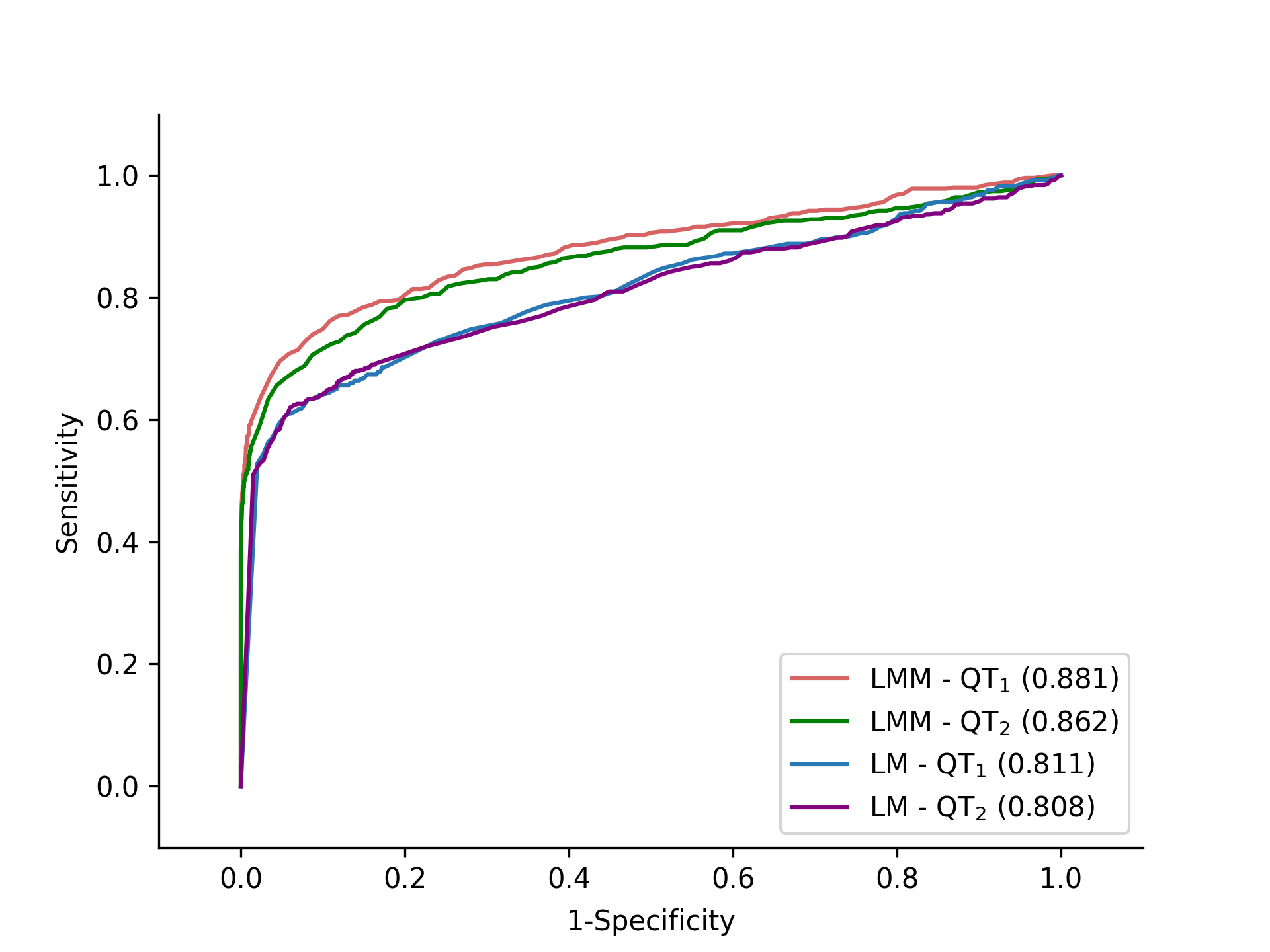}
   \label{subfig:t_100_i5}
   \subcaption{$n=100$}
 \end{subfigure}
 \hfill
 \begin{subfigure}{0.45\linewidth}
   \includegraphics[width=\linewidth]{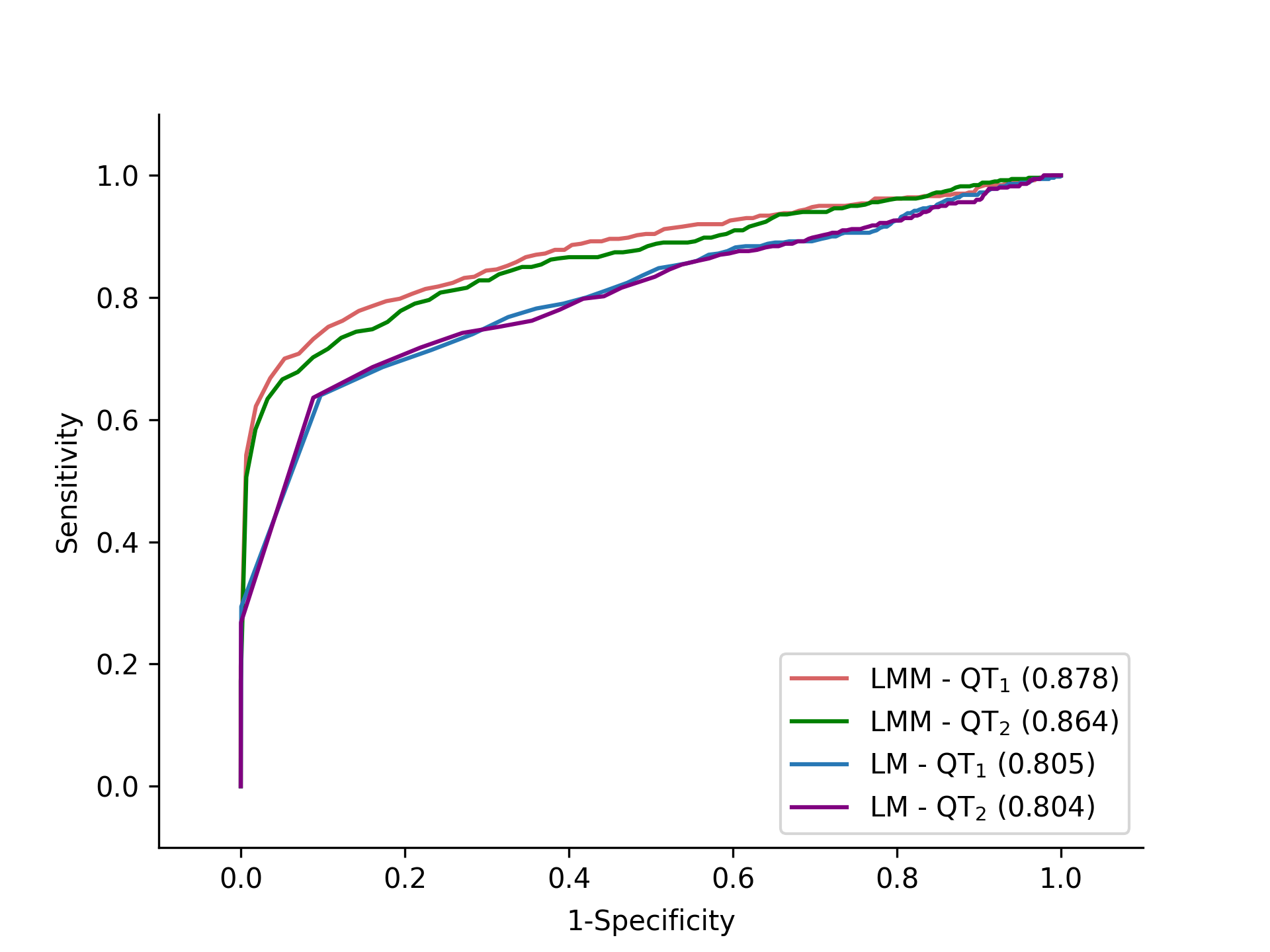}
   \label{subfig:t_100_p5}
   \subcaption{$n=100$}
\end{subfigure}
    \caption{{Case 5 with model misspecification. ROC curves of the LMM and LM based on credible intervals (a,c) and the CCT (b,d) when there was strong dependence presented among phenotypes with sample size varies. Curves of QT$_1$, QT$_2$ represent the ROC curves concerning QT$_1$ and QT$_2$.  The figures in  brackets indicated the corresponding AUC for each curve. }}  
  \label{fig:t_5}
\end{figure}

{
In this experiment, we explored the performance of the LMM and LM when the the true underlying distribution of the mixed effect term and error term were not normally distributed but heavy tailed, i.e., followed a multivariate t-distribution.  We repeated the data simulating process for \textbf{Case 1} to \textbf{5} according to Eq. (15),
 all the data generated procedure were same as before except that 
$$\bm{\epsilon}_i \sim Multi-t(v, \sigma_e^2 I_p),$$ 
$$\textbf{h} \sim Multi-t(v, \sigma_p^2 G).$$ 
Here we set the degree of freedom parameter $v =3$. The results were shown in Figure \ref{fig:t_1} to Figure \ref{fig:t_5}. We did not observe a significant decrease in performance regarding the AUC metric when the model was misspecified for the LMM and LM, compared to the correctly specified cases. Again, improved performance of the LMM over the LM in terms of the metric AUC based on both the credible intervals and aggregated \textit{P-}values was observed.  }

\newpage

\section*{Appendix D. Comparison between the LMM and Bayesian Group Sparse Multi-Task Regression \\(BGSMTR)}
\par

\begin{figure}[H]
  \centering
  \begin{subfigure}{0.45\linewidth}
    \includegraphics[width=\linewidth]{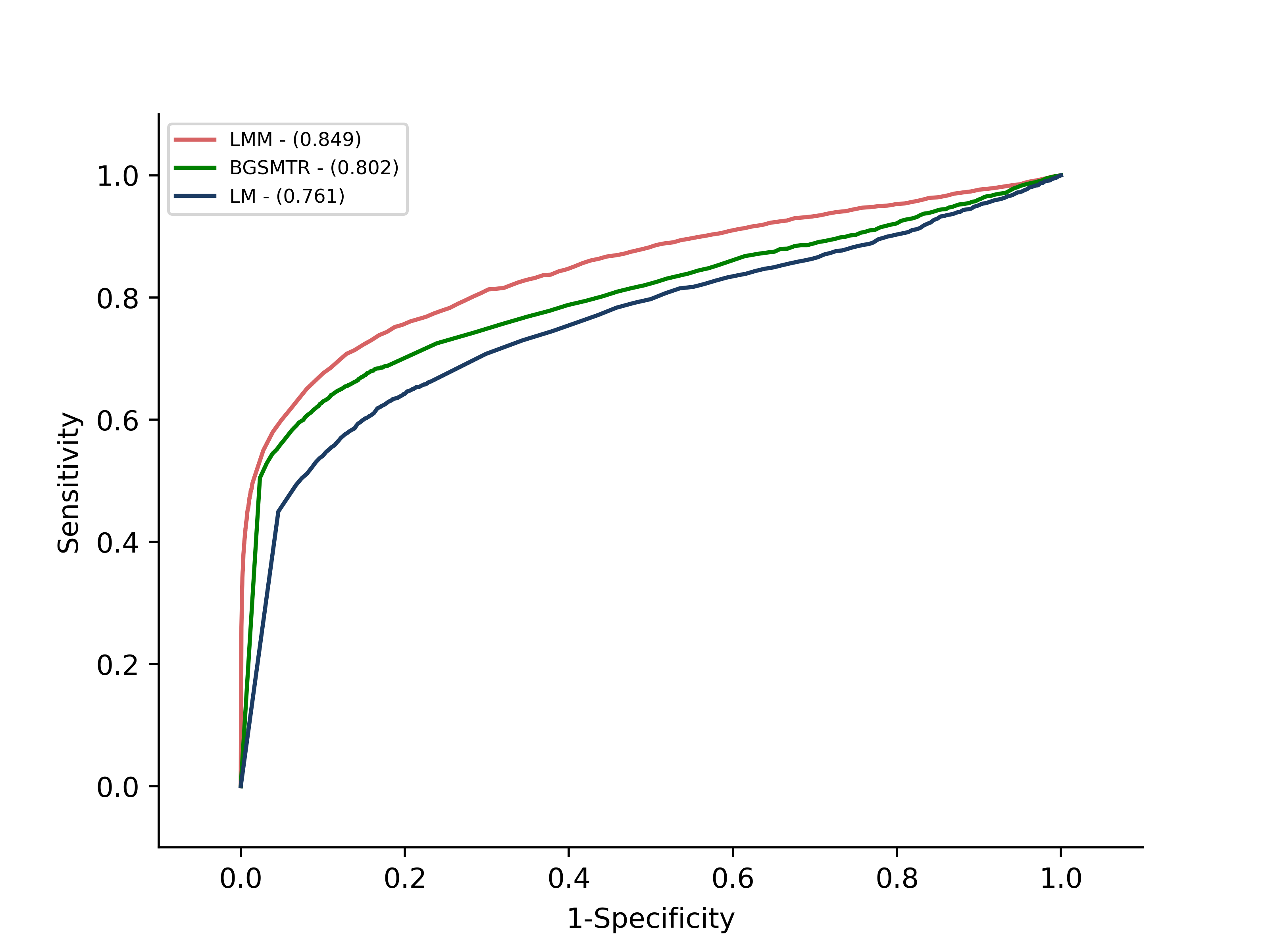}
    \label{subfig: ci_50_auc}
    \subcaption{$n=50$, CI}
  \end{subfigure}
  \hfill
  \begin{subfigure}{0.45\linewidth}
    \includegraphics[width=\linewidth]{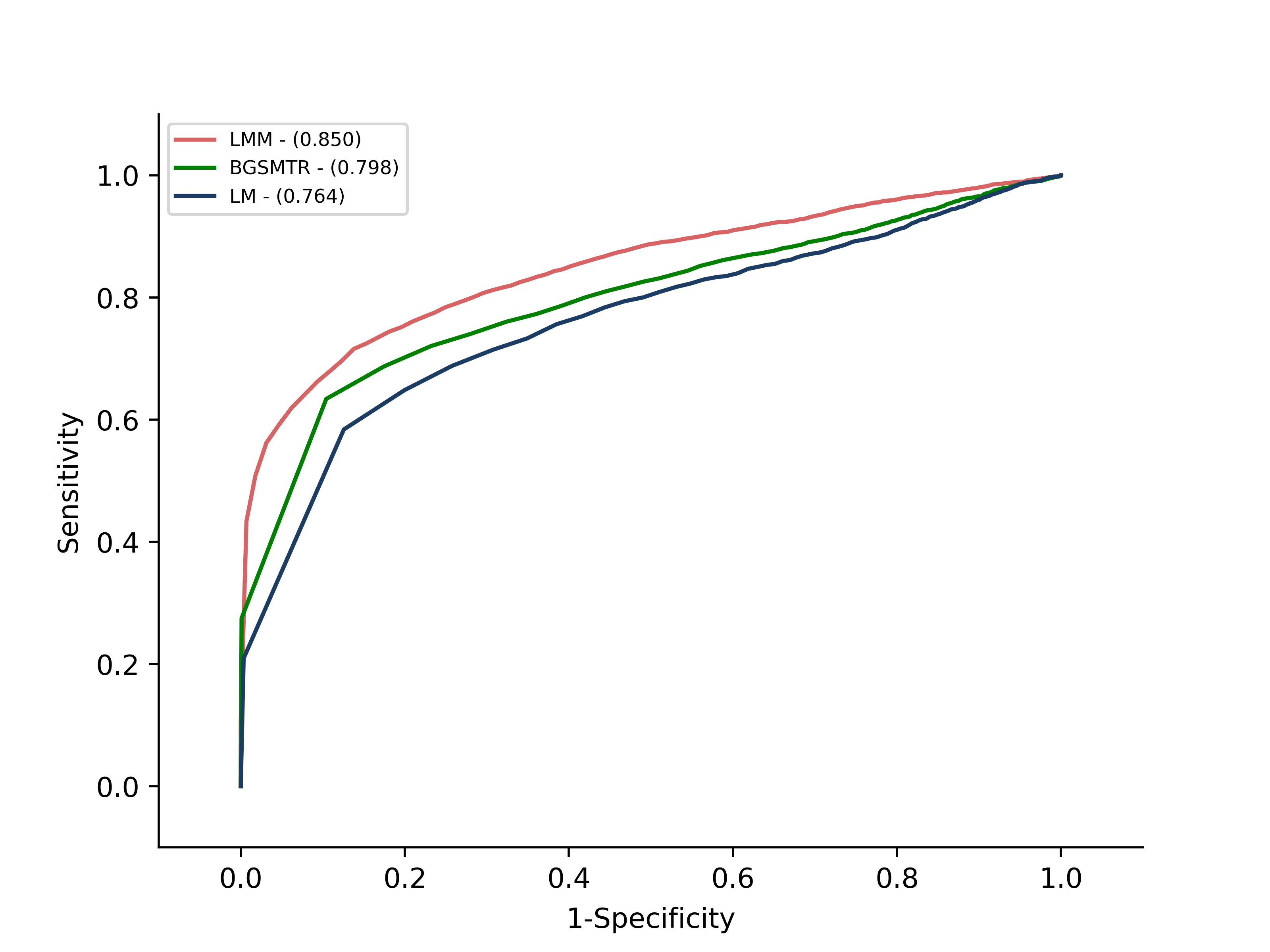}
    \label{subfig: pval_50_auc}
    \subcaption{$n=50$, CCT}
  \end{subfigure}
\par
\begin{subfigure}{0.45\linewidth}
\includegraphics[width=\linewidth]{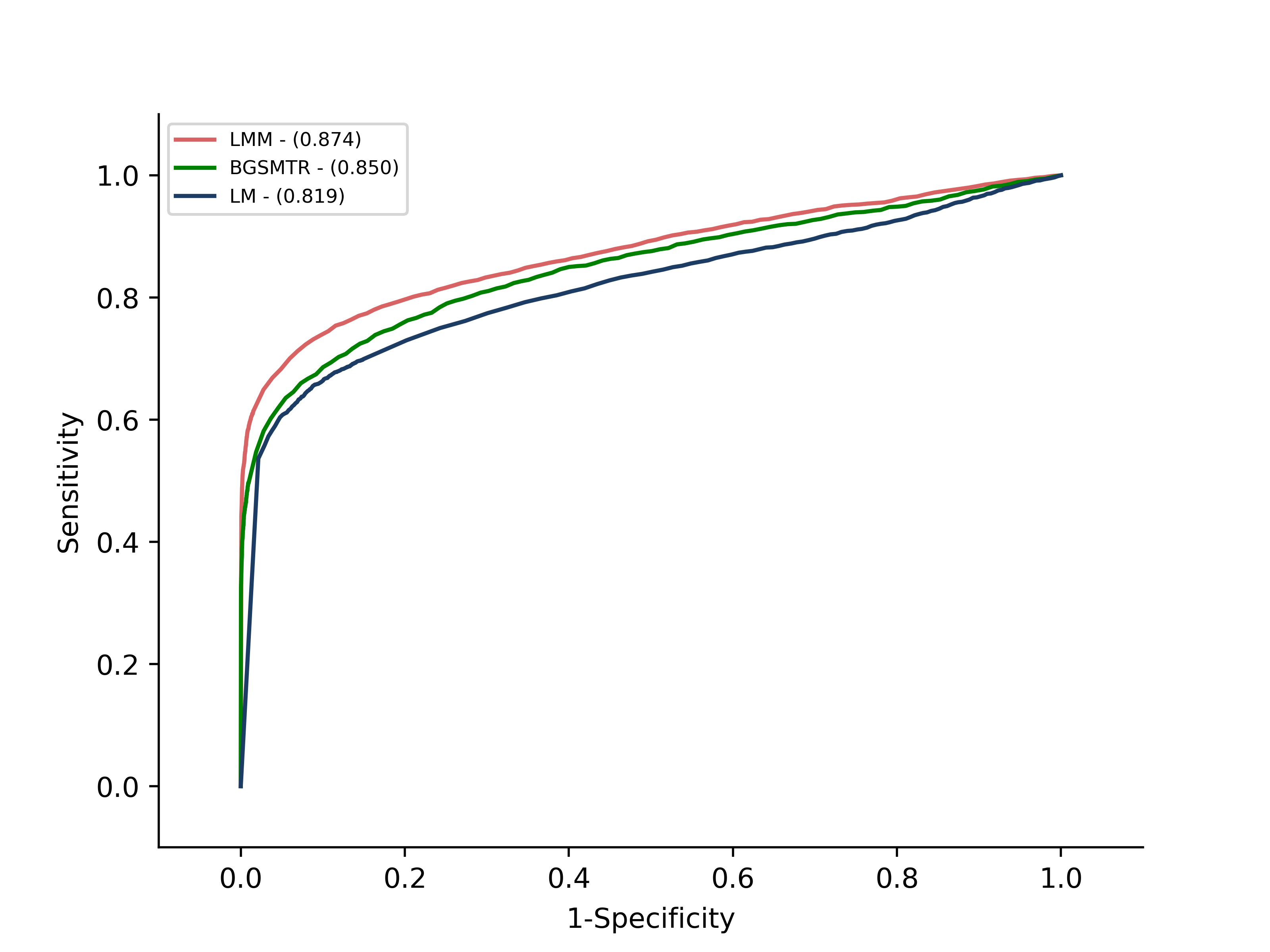}
   \label{subfig: ci_100_auc}
   \subcaption{$n=100$, CI}
 \end{subfigure}
 \hfill
  \begin{subfigure}{0.45\linewidth}
    \includegraphics[width=\linewidth]{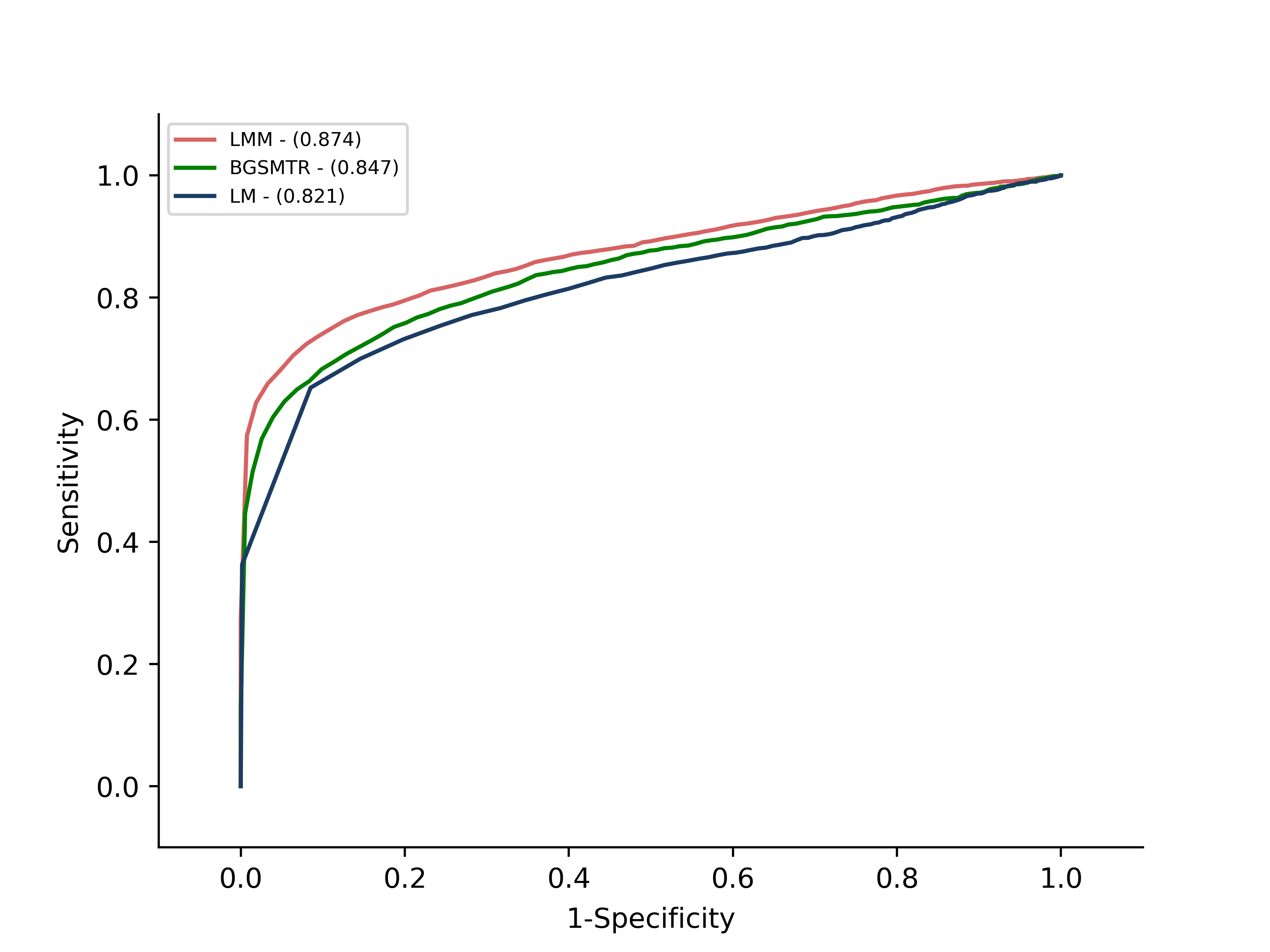}
    \label{subfig: pval_100_auc}
    \subcaption{$n=100$, CCT}
  \end{subfigure}
    \caption{Comparison of the LM, LMM and BGSMTR in Case 3. Mean ROC curves over all the traits of the LM, LMM and BGSMTR based on credible intervals (a,c) and the CCT (b,d) when complex dependence structures presented among phenotypes with sample size varying. The figures in brackets indicated the corresponding AUC for each curve. } 
  \label{fig: case7}
\end{figure}

We provided a direct comparison of the BGSMTR and our method under the settings of Case 3.  Figure \ref{fig: case7} indicated that the LMM outperformed the BGSMTR concerning the metric AUC. This difference might be due to the assumption of homogeneity in the BGSMTR, whereas the LMM allowed for heterogeneity among SNPs.  Meanwhile, compared to the LM, significant performance improvements were observed for both the LMM and BGSMTR, as the LM ignores the inter-trait dependencies.  Also, the BGSMTR used a whole set of SNPs to estimate the effect sizes simultaneously, instead of estimating the effect sizes for each SNP one by one. This might improve the statistical power when SNPs are strongly correlated (e.g. exhibit strong linkage disequilibrium (LD) within groups) \citep{greenlaw2017bayesian}, it could also incur significant computational overhead.

\begin{figure}[H]
  \centering
  \begin{subfigure}{0.32\linewidth}
    \includegraphics[width=\linewidth]{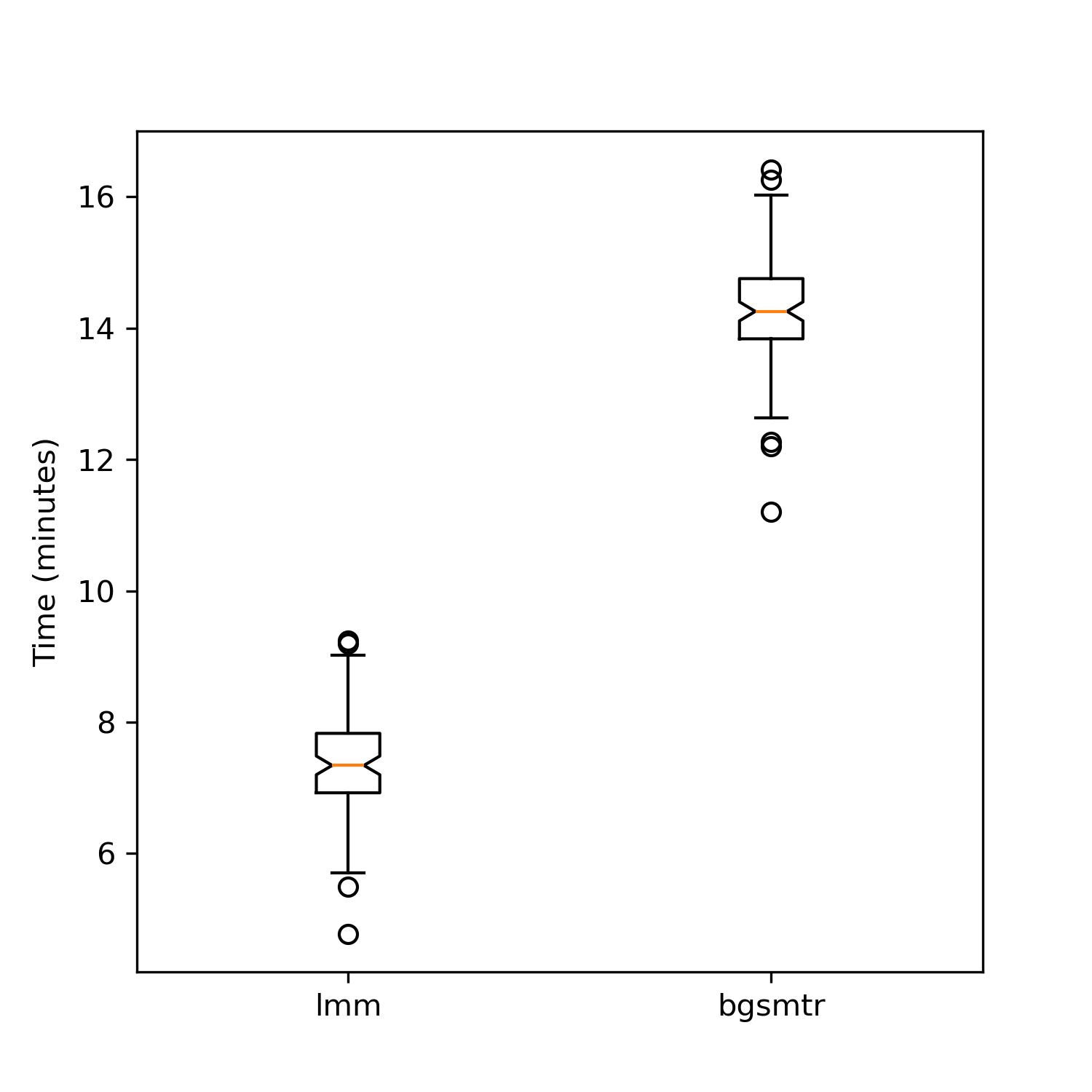}
    \label{subfig: 50_time}
    \subcaption{$n=50$}
  \end{subfigure}
  \hfill
  \begin{subfigure}{0.32\linewidth}
    \includegraphics[width=\linewidth]{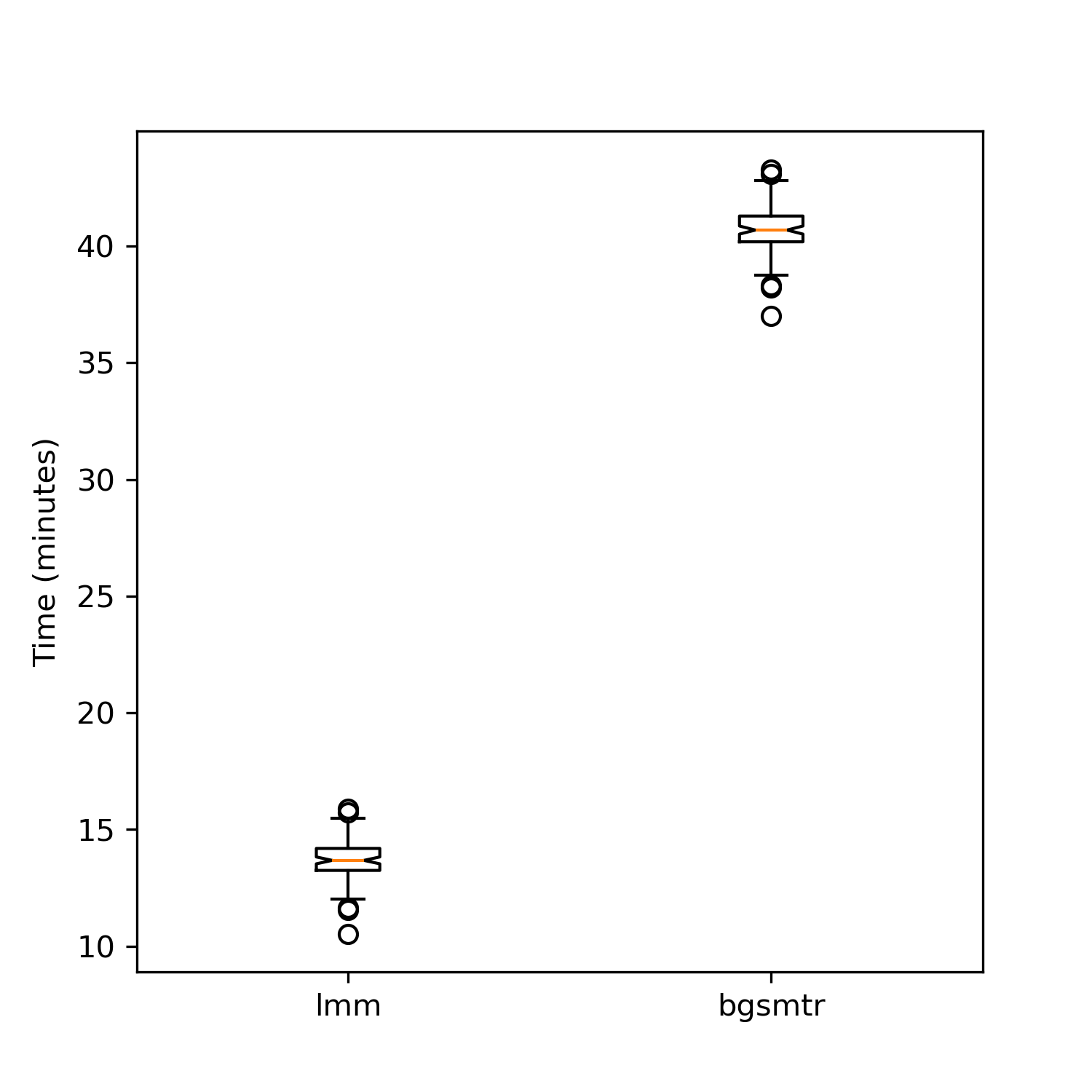}
    \label{subfig: time_100}
    \subcaption{$n=100$}
  \end{subfigure}
    \hfill
  \begin{subfigure}{0.32\linewidth}
    \includegraphics[width=\linewidth]{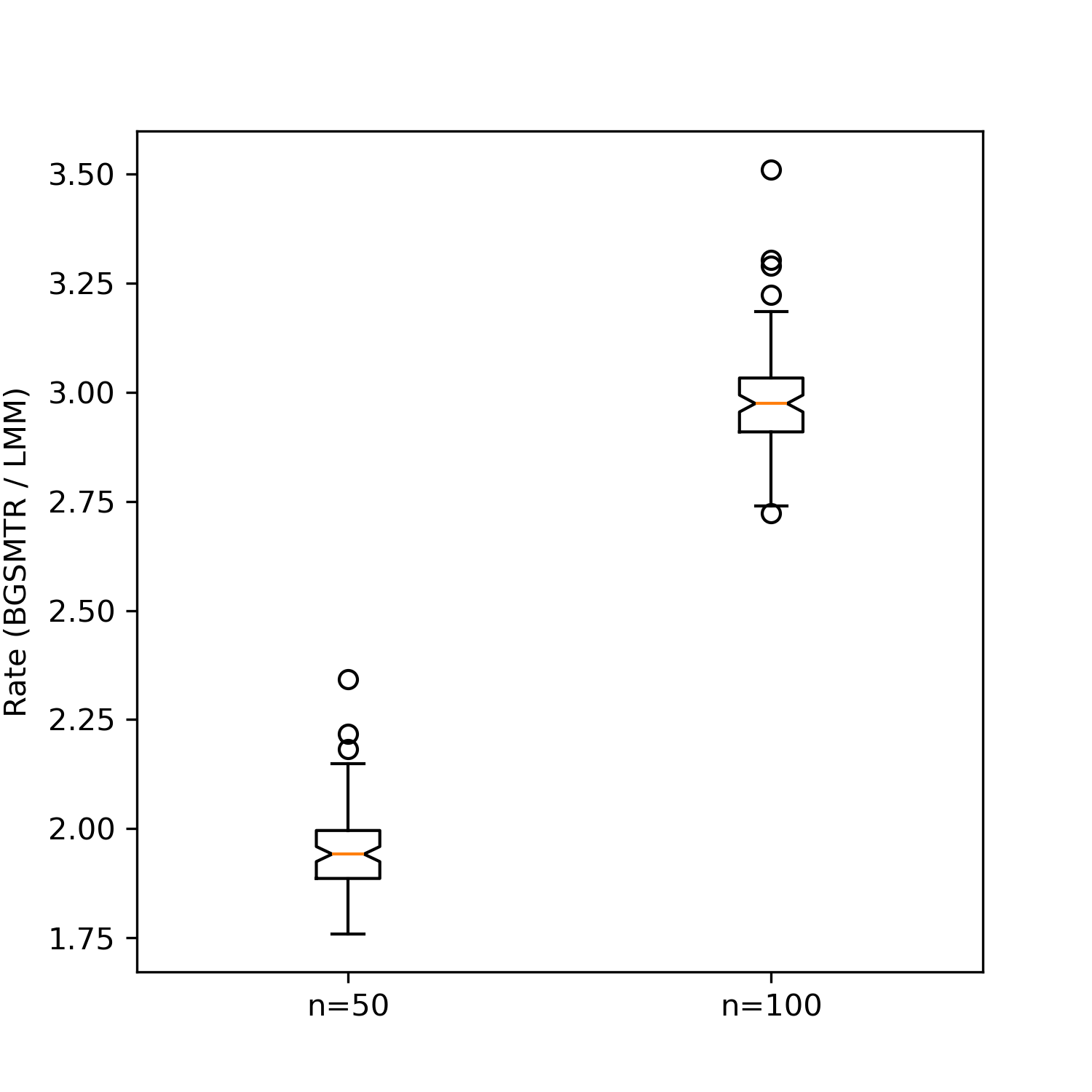}
    \label{subfig: rate}
    \subcaption{Rate}
  \end{subfigure}
    \caption{The computational time of the LMM and BGSMTR in Case 3. The box plots (a) and (b) showed the cost in $100$ repeated experiments for both the BGSMTR and LMM with the whole SNP set, when $n=50,~100$, respectively. (c) showed the odds of the computational cost in the experiments (BGSMTR/LMM). The total number of MCMC iterations was $5000$. Each experiment was run with a single core ($3.20$-GHz AMD Ryzen $7$ $7735$H, $16$GB of RAM) on a computing cluster.}  
  \label{fig: time}
\end{figure}

As shown in Table \ref{tab:table_time}, 
when $mn < p$, the two methods exhibited comparable theoretical costs. However, in real-world imaging genetic studies, where $p \ll mn$ in general, the LMM would demonstrate significant efficiency. This was empirically validated in Case 3 as shown in Figure \ref{fig: time}: when $n=50$ and $n=100$ in Case 3, the LMM required substantially less computation time than the BGSMTR.  For example,  fixing the number of MCMC iterations to $5000$, for a set of $100$ SNPs,  $6$ phenotypes,  with a single core ($3.20$-GHz AMD Ryzen $7$ $7735$H, $16$GB of RAM ) on a computing cluster,  when $n=50$, the averaged running time of the BGSMTR was $1.948$ times that of the LMM, and this number increased to $2.979$ when $n=100$.

\begin{table}[h]
  \begin{center}
    \caption{Computational complexity of the LMM and BGSMTR in a single iteration of MCMC (for all SNPs). $T_{inv}$ represented the time for computing the inverse of a ${p \times p}$ matrix. \cite{tveit2003complexity} showed that the lower bound of $T_{inv}$ was $O(p^2\log p)$, and $T_{inv}$ at most took $\mathcal{O}(p^3)$. }
    \label{tab:table_time}
    \begin{tabular}{cccc}
    \hline
    \multicolumn{2}{c|}{Method}  & \multicolumn{2}{|c}{Computational Complexity} \\
           \hline 
    \multicolumn{2}{c|}{LMM}  & \multicolumn{2}{|c}{$\max\{\mathcal{O}(nmp^2), \mathcal{O}(mT_{inv})\}$} \\
           \hline 
    \multicolumn{2}{c|}{BGSMTR}  & \multicolumn{2}{|c}{$\max\{\mathcal{O}(nm^2p^2), \mathcal{O}(mT_{inv})\}$} \\
           \hline 
    \end{tabular} 
  \end{center}
\end{table}

\newpage

\bibliographystyle{plainnat-revised}
\bibliography{sample}